\documentclass[aps,twocolumn,prb,floatfix,altaffilletter,superscriptaddress]{revtex4-2}

\usepackage{graphicx}
\usepackage{amsmath,amsfonts,amssymb,bm}
\usepackage{lmodern}
\usepackage{xcolor}
\usepackage[shortlabels]{enumitem}
\usepackage{float}
\usepackage[caption=false]{subfig}  
\usepackage{natbib}

\newcommand{\sectitle}[1]{\emph{#1}:}

\newcommand{\ket}[1]{\left|{#1}\right.\rangle}

\newcommand{\bs}[1]{\boldsymbol{#1}}
\newcommand{\beq}{\begin{equation}}
\newcommand{\eeq}{\end{equation}}
\newcommand{\la}{\left\langle}
\newcommand{\ra}{\right\rangle}

\newcommand{\element}[2]{\left\{ #1 \right\vert \left. #2 \right\}}
\newcommand{\sselement}[3]{\left[ #1 \rVert \left\{ #2 \lvert #3\right\} \right]}

\newcommand{\av}{{\bs{a}}}
\newcommand{\bv}{{\bs{b}}}
\newcommand{\cv}{{\bs{c}}}

\newcommand{\kv}{{\bs{k}}}

\newcommand{\sv}{{\bs{s}}}
\newcommand{\Ts}{{\mathcal{T}}}
\newcommand{\Ps}{{\mathcal{P}}}

\newcommand{\rv}{{\bs{r}}}

\makeatother
\begin{document}

\title{Spin-Space Groups and Magnon Band Topology}

\author{A.~Corticelli}
\affiliation{Max Planck Institute for the Physics of Complex Systems, N\"{o}thnitzer Str. 38, 01187 Dresden, Germany}
\author{R. Moessner}
\affiliation{Max Planck Institute for the Physics of Complex Systems, N\"{o}thnitzer Str. 38, 01187 Dresden, Germany}
\author{P.~A.~McClarty}
\affiliation{Max Planck Institute for the Physics of Complex Systems, N\"{o}thnitzer Str. 38, 01187 Dresden, Germany}

\begin{abstract}
Band topology is both constrained and enriched by the presence of symmetry. The importance of anti-unitary symmetries such as time reversal was recognized early on leading to the classification of topological band structures based on the ten-fold way. Since then, lattice point group and non-symmorphic symmetries have been seen to lead to a vast range of possible topologically nontrivial band structures many of which are realized in materials. In this paper we show that band topology is further enriched in many physically realizable instances where magnetic and lattice degrees of freedom are wholly or partially decoupled. The appropriate symmetry groups to describe general magnetic systems are the spin-space groups. Here we describe cases where spin-space groups are essential to understand the band topology in magnetic materials. We then focus on magnon band topology where the theory of spin-space groups has its simplest realization. We consider magnetic Hamiltonians with various types of coupling including Heisenberg and Kitaev couplings revealing a hierarchy of enhanced magnetic symmetry groups depending on the nature of the lattice and the couplings. We describe, in detail, the associated representation theory and compatibility relations thus characterizing symmetry-enforced constraints on the magnon bands revealing a proliferation of nodal points, lines, planes and volumes. 
\end{abstract}

\maketitle

\section{Introduction}

Most readers will be very familiar with the huge abundance and diversity with which crystalline solids occur in nature. Underlying the richness of the chemistry and details of the structure is the set of lattice symmetries which fall into one of $17$ wallpaper groups in 2D materials and $230$ space groups in 3D materials \cite{bradley2009mathematical}. The profusion of structures and symmetries grows when we focus on magnetic crystals as one is then forced to include the role of time reversal and its interplay with magnetic order.  Altogether there are $80$ magnetic space groups in 2D and $1651$ in 3D \cite{bradley2009mathematical}. These symmetry groups and the point groups on which they are based form one of the cornerstones of condensed matter physics as they place constraints on couplings, dispersion relations, wavefunctions and matrix elements \cite{dresselhaus2007group}.

Symmetry is essential also to understand the variety of possible topological band structures \cite{Chiu2016}. While band topology can be nontrivial in the complete absence of symmetry it is greatly enriched by its presence as may be appreciated by inspecting the ten-fold classification of band topology with anti-unitary symmetries \cite{Ryu_2010,Chiu2016}. The connectivity of bands in momentum space, including the presence or absence of band touchings, is highly constrained by lattice symmetries. Such constraints $-$ originating from space groups and their magnetic counterparts $-$ have been the subject of intense study in electronic band structures and underlie various partial classification schemes of band topology \cite{Bradlyn2017,Po2017,Cano2020}. In turn, these classification tables are important for identifying topological materials.

In strongly spin-orbit coupled magnetic materials, the magnetic moments are typically locked to the lattice so that transformations performed on the moments are performed also in real space. However, as was noted long ago by Brinkman and Elliott \cite{BrinkmanElliott1966,BrinkmanElliott1966b,Brinkman1967} there are physically natural settings where the spin and space transformations are wholly or partially decoupled. In this paper, we investigate these enhanced symmetry groups $-$ so-called {\it spin-space groups} $-$ in relation to band topology. Beginning with a brief introduction to the spin-space groups, we give an account of their importance to the understanding of physically relevant condensed matter systems. We then turn to the investigation of constraints on band topology arising from the spin-space symmetries.  

Our presentation concentrates on magnon band topology which is the simplest context for studying symmetries in magnetic materials, though many of our considerations carry over to the electronic band structures of itinerant magnetic materials. For magnons, in common with other bosonic excitations, particle-hole and chiral symmetries are subsumed by time reversal resulting in a three-fold rather than a ten-fold table \cite{xuthreefold}. For this reason, crystalline symmetries and, more generally, spin-space symmetries are especially important if magnons are to be imbued with interesting band topology.  

Magnon band structures emerge through symmetry breaking from an underlying lattice of interacting magnetic moments. As we show, the symmetry of the band structure of coherent single magnon excitations is tethered both to the nature of the magnetic exchange Hamiltonian as well as to the magnetic structure. Such excitations are routinely probed in bulk magnetic materials in energy-momentum over the entire Brillouin zone using inelastic neutron scattering. 

One notable case where spin-space groups arise is in the study of Kitaev-Heisenberg models \cite{Kimchi2014} that are of considerable current interest \cite{rau2016spin,hermanns2017physics,trebst2017kitaev,winter2017,winter2016challenges} owing to the rich phenomenology in such models and their relevance to materials \cite{jackeli2009mott,chaloupka2010kitaev,choi2012spin,plumb2014alpha,sears2015magnetic,majumder2015anisotropic,williams2016incommensurate,banerjee2016proximate,little2017antiferromagnetic,ponomaryov2017,ran2017spin,sears2017phase,wolter2017field}. Kitaev physics is known to range over quantum spin liquid phases \cite{kitaev2006anyons}, complex ordered magnetic structures \cite{chaloupka2010kitaev,williams2016incommensurate}, rich magnetic field induced phase diagrams \cite{janssen2016honeycomb}, nontrivial magnon band topology \cite{mcclarty2018,joshi2018} and unusual heat transport properties \cite{Kasahara2018}. In the following pages we show that spin-space groups are essential to understand the magnon band structures of such models and conversely can be used as a means of establishing the importance of Kitaev-Heisenberg terms in real materials.

Spin-space groups also arise in the study of other physically motivated magnetic models including Heisenberg models, certain single ion anisotropies and certain anti-symmetric exchange couplings as we discuss below. For various models with spin-space symmetry we show direct connections between the enhanced symmetry and a wide variety of richly degenerate band structures with Dirac points in 2D, Weyl points, 4-fold degenerate points, nodal lines and planes and 2-fold degenerate volumes across the Brillouin zone \cite{wan2011weyl,burkov2011,armitage2018,fang2016,wu2018nodalsurface}. In addition, in the course of this analysis we show in detail how to analyse band representations for spin-space groups including non-symmorphic groups.

\subsection{Outline of the paper}

Spin-space groups characterize the symmetries of a range of physically relevant interacting magnetic systems. In such cases, the standard magnetic space groups are insufficient to capture all the symmetries of the problem. 

In the next section, we give a number of examples where such enhanced magnetic symmetries arise. In fact, such models are very common as they include Heisenberg models that appear, to an excellent approximation, in many materials with weak spin-orbit coupling. In such models, the spin space transformations are completely decoupled from the lattice transformations and therefore lie at the extreme end of possible spin-space symmetry groups. We show that spin-space groups appear also in cases where spin-orbit coupling is important. For example, for various kinds of single ion anisotropy, for Kitaev-Heisenberg models and for certain kinds of anti-symmetric or Dzyaloshinskii-Moriya exchange.

In Section~\ref{sec:magnons} we briefly review spin wave theory and show that spin-space symmetries are inherited by magnons. Then, we give a short account of the band representation theory of spin-space groups that is the tool of choice to determine the symmetry constraints on the magnon band structure (Section~\ref{sec:BR}). Indeed, our analysis is grounded in the representation theory of spin-space groups and the associated compatibility relations. In contrast to the representation theory of magnetic space groups, this has not been worked out in detail and our study of band topology relies on calculations of band representations and their decomposition into irreducible representations of spin-space groups from first principles. 

Symmetry constraints on band topology have been extensively analysed for space groups and their magnetic analogues. Here we show that the further enhancement of symmetry in going to spin-space groups can lead to a proliferation of band degeneracies including nodal points, lines, planes and volumes. Here ``nodal" refers to points, curves, surfaces and volumes where pairs of bands become degenerate.

We illustrate this through a number of examples beginning with the particularly rich case of the Heisenberg-Kitaev model on a hyperhoneycomb lattice with a simple collinear antiferromagnetically ordered ground state. We show (Section~\ref{sec:overview}), by direct calculation, that the magnon band structure has a nodal plane that is punctured by a number of nodal lines and that these features survive a tuning of the Kitaev-Heisenberg couplings within this phase suggesting that they originate from the Hamiltonian symmetries. 

We then enumerate all the symmetries of the magnetically ordered state. These include a combination of non-symmorphic symmetry elements, anti-unitary elements and spin-space symmetries. The resulting group is not isomorphic to any of the $1651$ magnetic space groups and is instead a concrete example of a spin-space group. Section~\ref{sec:representationtheory} is a derivation of the nodal plane and other degeneracies in the richly featured band structure on the basis of the band representation theory of this group.

Having given one detailed calculation of the magnon band structure from a spin-space group over the entire Brillouin zone, we present a number of further examples. The first set of examples comes from Heisenberg models in Section~\ref{sec:heisenberg}. We make general observations about the nature of the spin-space symmetries of collinear ferromagnets and antiferromagnets and their effect on magnon bands. In the latter case, we describe how these symmetries can lead to nodal volumes. 

In Section~\ref{sec:honeycomb}, we focus on the honeycomb lattice $-$ one of the most symmetric lattices in two-dimensions. The recent literature on topological magnons includes the discovery of magnon Chern bands in the Kitaev-Heisenberg honeycomb model and in the case of the ferromagnet with second-neighbor anti-symmetric exchange. In these models, the band topology can arise in the complete absence of symmetry. We revisit them to show that their nontrivial spin-space symmetries allow one to tune the appearance of Dirac points in the magnon band structure. As a corollary, the symmetry analysis reveals that the known Chern band regimes have the necessary symmetry, or lack of it, to allow for nontrivial Berry curvature. 

Section~\ref{sec:hyperhoneycombfm} is a detailed study of the Kitaev-Heisenberg hyperhoneycomb ferromagnet. This model beautifully illustrates some important features of magnon band topology arising from spin-space groups because the symmetry of the model can be tuned by simply rotating the direction of the applied magnetic field. We give the full (spin-space) symmetry group corresponding to each of the symmetry-distinct moment directions revealing a hierarchy of magnetic symmetries. We also give the symmetry group that one would naively anticipate purely based on the invariance of the magnetic structure. The latter is, by definition, a magnetic space group and therefore generally has lower symmetry than the spin-space group for the same moment direction. 

We then enumerate all the magnon band degeneracies one would expect for each of these symmetry groups. This reveals, at a glance, that the spin-space group has various features including nodal lines and Weyl points that would be absent for the corresponding magnetic space group. For many of these cases we can give simple criteria or informal arguments for the appearance of nodal features thereby circumventing the detailed representation theory analysis.

We further show the surprising feature that the spin-space groups arising for this model are frequently isomorphic to some magnetic space group albeit of higher symmetry than the one that merely leaves the magnetic structure invariant. 

\begin{figure}[!htb]
\hbox to \linewidth{ \hss
\includegraphics[width=0.99\linewidth]{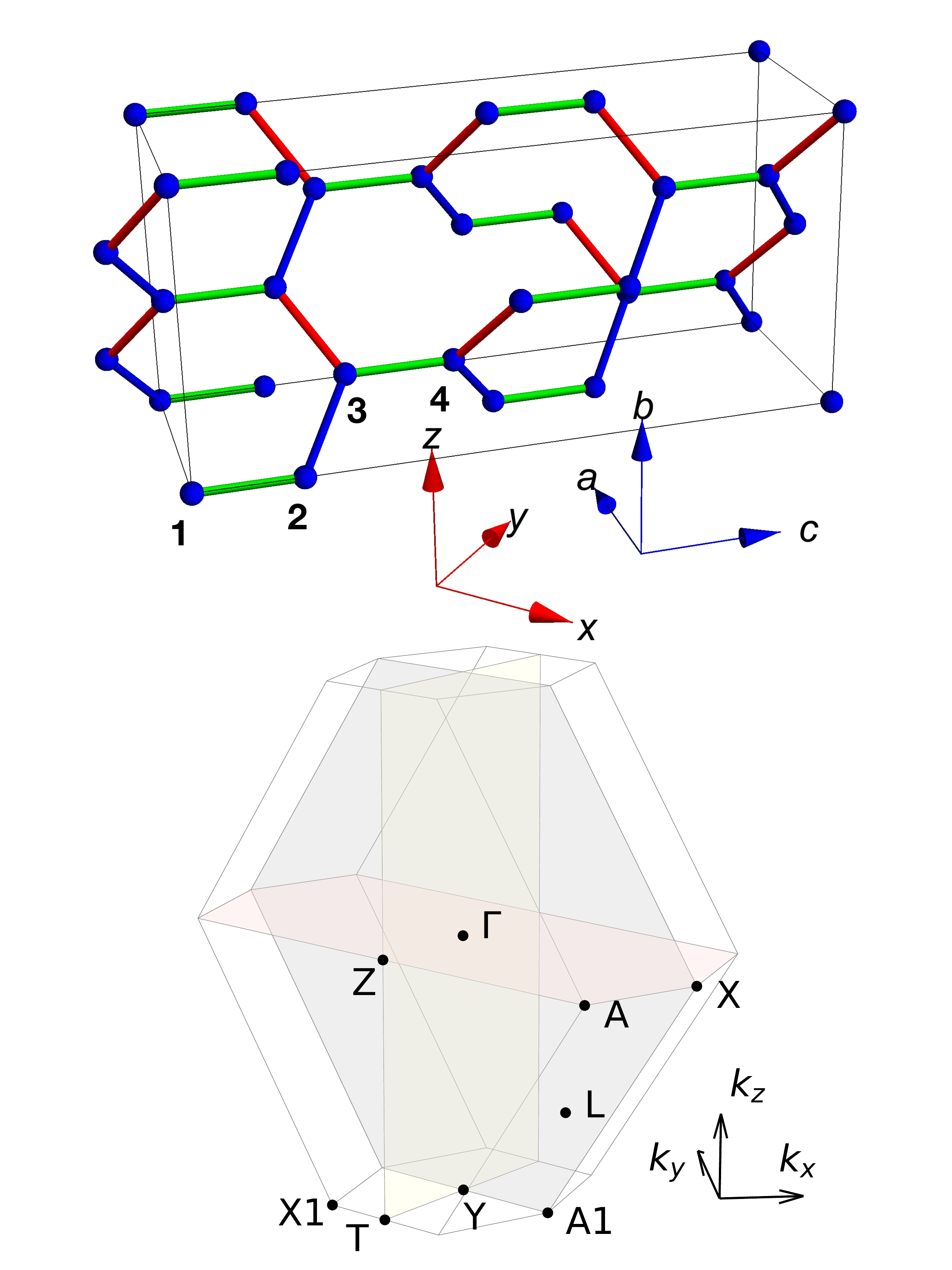}
\hss}
\caption{The upper figure shows the tri-coordinated network of sites on a hyperhoneycomb lattice. The primitive unit cell contains four sites  
labeled from 1 to 4. The primitive unit vectors for the 4-site unit cell are given by ${\bs a_i}$. The color of the bonds reflects the direction of the Kitaev interaction: $S^xS^x$ (red),  $S^yS^y$ (green), and $S^zS^z$ (blue) interactions respectively. The lower panel shows the Brillouin zone of the hyperhoneycomb lattice. The perpendicular planes are the mirror plane of glides: $d_1$ (red), $d_2$ (yellow), $d_3$ (grey). 
The high symmetry paths are: $\Gamma \rightarrow Y \rightarrow  T \rightarrow Z \rightarrow \Gamma 
\rightarrow X\rightarrow A_1 \rightarrow Y; ~ T \rightarrow X_1$; $X \rightarrow A \rightarrow Z$ and $\Gamma
 \rightarrow L$. The coordinates of the points are given in Appendix~\ref{app:lattice_data}.
}
\label{fig:Lattice}
\end{figure}

\section{Spin and Space Symmetries of the Magnetic Hamiltonian}
\label{sec:symmetry}

\subsection{Spin-Space Groups and Magnetic Couplings}
\label{sec:exchange}

Given a magnetic Hamiltonian $H$, we shall identify all symmetry operations that leave the Hamiltonian invariant. This parent group will be denoted $\mathbf{G}_H$. In general, the Hamiltonian will be invariant under lattice symmetries that form a space group $\mathbf{G}$ that includes the primitive translations $\mathbf{T}$ as a normal subgroup and the space group then admits a coset decomposition
\beq
\mathbf{G}=\bigcup_{\alpha}\element{R_\alpha}{\bs{t}_{\alpha}}\mathbf{T}
\eeq
where a general coset representative of the space group is conventionally denoted $\element{R}{\bs{t}}$ where $R$ is a point group element and $\bs{t}$ is a non-Bravais translation. These act on a general position $\mathbf{r}$ in real space as $\element{R}{\bs{t}}\mathbf{r}=R \mathbf{r} + \boldsymbol{t}$ and so:
\begin{align}
& \element{R_1}{\bs{t}_1}\element{R_2}{\bs{t}_2} = \element{R_1 R_2}{R_1 \bs{t}_2 + \bs{t}_1 }  \\
& \element{R}{\bs{t}}^{-1} = \element{R^{-1}}{- R^{-1}\bs{t}}.
\label{eq:sgmult}
\end{align}
The identity is denoted $\element{E}{\bs{0}}$. There are $17$ such groups in two dimensions $-$ usually called the wallpaper groups $-$ and $230$ in three dimensions. In the presence of time reversal symmetry, we allow for the possibility of anti-unitary elements $\hat{\Ts}\element{S}{\bs{w}}$ leading to magnetic space groups $\mathbf{M}$ that have coset decomposition
\beq
\mathbf{M} = \bigcup_\alpha \element{R_\alpha}{\bs{t}_{\alpha}}\mathbf{T} +  \hat{\Ts}\bigcup_{\alpha'} \element{S_{\alpha'}}{\bs{w}_{\alpha'}}\mathbf{T}.
\eeq
There are $80$ magnetic space groups in two dimensions and $1651$ in three dimensions. Magnetic Hamiltonians may have higher symmetry still: in general $\mathbf{G}_H$ is a direct product of a magnetic space group and a group acting only in spin space.

For example, consider the canonical Heisenberg model on some lattice with, for concreteness, nearest neighbor couplings 
\beq
\hat{H} = J \sum_{\la i,j \ra} \hat{\boldsymbol{S}}_i \cdot \hat{\boldsymbol{S}}_j.
\eeq
The symmetries of this Hamiltonian include:
\begin{enumerate}
\item the primitive lattice translations forming group $\mathbf{T}$,
\item  the symmetry elements denoted $\element{\alpha}{\bs{t}_{\alpha}}$ for $\alpha= 1,\ldots,\vert \mathbf{G}\vert$ belonging to space group $\mathbf{G}$,
\item the time reversal symmetry operator $\hat{\Ts}$: the operator acts on spin as $\hat{S}^\mu_i \xrightarrow[]{\hat{\Ts}} -\hat{S}^\mu_i$
\item and the group of global spin rotations $\mathbf{R}\cong SO(3)$.
\end{enumerate}
and combinations of these. The parent group is therefore $\mathbf{G}_H = \left( \mathbf{G}\oplus \hat{\Ts}\mathbf{G}\right) \otimes \mathbf{R}$ where $\mathbf{G}\oplus \hat{\Ts}\mathbf{G}$ on its own forms a magnetic space group of type II in the notation of Bradley and Cracknell \cite{bradley2009mathematical}.

We use the notation $\sselement{B}{R}{\bs{t}}$ to denote the general symmetry element of those symmetry groups that allow for decoupled spin-space and real-space elements where $B$ acts on spin space and $\element{R}{\bs{t}}$ is the ordinary space group element that does not act on spin-space \cite{Litvin1974}. Such symmetry groups were named {\it spin-space groups} in the original papers of Brinkman and Elliott \cite{BrinkmanElliott1966,BrinkmanElliott1966b}. The action of a spin-space group element on the lattice moments $\hat{J}^\mu(\mathbf{r})$ is the {\it active} transformation
\beq
\sselement{B}{R}{\bs{t}} \hat{J}^\mu(\mathbf{r})  = \sum_{\nu} {\rm det }(B) B^{\mu\nu} \hat{J}^\nu(\element{R}{\bs{t}}\mathbf{r})
\eeq
where the determinant is present because magnetic moments are pseudovectors $-$ for example they are invariant under inversion.
It follows that
\begin{align}
&\sselement{B_1}{R_1}{\bs{t}_1}\sselement{B_2}{R_2}{\bs{t}_2} = \sselement{B_1 B_2}{R_1 R_2}{R_1 \bs{t}_2 + \bs{t}_1} \\
&\sselement{B}{R}{\bs{t}}^{-1}  = \sselement{B^{-1}}{R^{-1}}{-R^{-1} \bs{t}}.
\end{align}

In the case where the spin-orbit coupling vanishes, the Hamiltonian is Heisenberg-like and the group elements acting on spin space are completely decoupled from the real space elements $-$ the former being the 3D rotation group. When spin-orbit coupling is present and Hamiltonian is not fine-tuned, the moments are usually locked to the space group transformations. We then write
\beq
\sselement{R}{R}{\bs{t}} \hat{J}^\mu(\mathbf{r})  = \sum_{\nu} {\rm det} (R) R^{\mu\nu} \hat{J}^\nu(\element{R}{\bs{t}}\mathbf{r}).
\eeq
Since the magnetic Hamiltonian must be invariant under lattice symmetries such locking is always possible and the resulting group is one of the magnetic space groups. 
There are cases where the moments transform under lattice transformations but where there is a residual spin space invariance $-$ nontrivial elements that act purely on spin space.

The magnetic Hamiltonian in many insulating magnets is well approximated by a Heisenberg model in those cases where the spin-orbit coupling is weak \cite{Coldea2001}, where the orbital part of the moment is quenched or frozen out \cite{Ichikawa2005} or through the fortuitous cancellation of anisotropic terms \cite{Rau2018,Sala2019}. Such couplings are allowed by symmetry on all lattices as well as in amorphous solids. For example, the parent material La$_2$CuO$_4$ of one prominent high $T_c$ superconducting cuprate which has a Heisenberg exchange scale of about $100$ meV \cite{Coldea2001} while any magnetic anisotropies, for example inferred from the small spin wave gap, are at most a hundredth of this scale \cite{PhysRevB.37.5817}. 

In instances where there is spin-orbit coupling, the crystal field may lead to single ion anisotropies that break the spin rotation group from $SO(3)$ to the local site symmetry group. The degree to which the moment preserves its spin-only character or is mixed with the orbital moment is dependent on the magnetic ion and the material in which it appears but we now use $\hat{J}^\mu$ to denote the moment operators. In general, the single anisotropy takes the form
\beq
\hat{H}_{\rm SIA} = \sum_i \sum_{l,m} \Delta^l_m \hat{O}^l_{i,m}
\eeq 
where $\hat{O}^l_{i,m}$ is a Steven's operator which is the operator equivalent of spherical harmonic $Y^l_m(\theta,\phi)$ and is polynomial in the spin operators with degree $l \leq 6$ as fixed by the site symmetry. 

We now give some concrete examples of possible spin groups in lattices of moments arising from the single ion anisotropy. While the single ion anisotropy has the site symmetry of the magnetic ion, there may be a hierarchy of scales. For example, in tetragonal K$_2$CuF$_4$, the copper is almost isotropic with exchange scale $J\approx 1$ meV because the spin-orbit coupling is weak. Nevertheless, it does have a detectable easy plane anisotropy of about $10^{-2}J$ with single ion Hamiltonian $\hat{O}^2_0 = 3(\hat{J}^z)^2-J(J+1)$ and $\Delta^2_0 >0$. There is a even weaker but detectable four-fold anisotropy \cite{Yamazaki1981} that can be captured by a term in $H_{\rm SIA}$ of the form $\hat{O}^4_{4}=(1/2)((\hat{J}_i^+)^4 + (\hat{J}_i^-)^4)$. Thus, the 3D rotation group is broken by the easy plane anisotropy down to $U(1)\times \mathbb{Z}_2$. The weaker terms break this down to the site symmetry $D_4$ in principle allowing for five non-vanishing Steven's operator coefficients $\Delta^2_0$, $\Delta^4_0$, $\Delta^4_4$, $\Delta^6_0$, $\Delta^6_4$. There are materials where the easy plane anisotropy is much greater. 

Analogous symmetry considerations guide our understanding of interactions between magnetic moments. We mainly restrict our attention to couplings that are bilinear in the moments: 
\beq
\hat{H} = \sum_{ i,j }\mathsf{J}_{ij}^{\mu\nu} \hat{J}^\mu_i  \hat{J}^\nu_j.
\eeq
The Heisenberg coupling is $\mathsf{J}_{ij}^{\mu\nu}=\delta^{\mu\nu}\mathsf{J}_{ij}$ but, in general, the exchange may have anisotropies that respect the lattice symmetries. On a single bond, in the absence of symmetry constraints, there are nine allowed couplings: three diagonal, three off-diagonal and anti-symmetric and three off-diagonal and symmetric. Symmetry generally places constraints on these couplings. For example, consider a simple cubic lattice and a nearest neighbor $\hat{\boldsymbol{x}}$ bond. The $C_4$ about the axis through the bond takes $\hat{\boldsymbol{y}}\rightarrow \hat{\boldsymbol{z}}$ fixing the $\mathsf{J}^{yy}=\mathsf{J}^{zz}$. The mirrors in the planes of the cubic faces are equivalent to inversion and a $C_2$. Inversion leaves the magnetic moment invariant so only the $C_2$ acts nontrivially thus forcing the off-diagonal components to zero. The resulting $\mathsf{J}_{i i+\hat{\boldsymbol{x}}} = {\rm diag}(\mathsf{J}^{xx},\mathsf{J}^{yy},\mathsf{J}^{yy})$ and the other nearest neighbor bond coupling can be obtained from this using lattice symmetries. Since the most general exchange Hamiltonian has only the lattice symmetries, the spatial and spin transformations can be thought of as being locked to one another and this is the limit of strong spin-orbit coupling. In this case, the group $\mathbf{G}_H$ is just the group $\mathbf{G}\oplus \Ts\mathbf{G}$ as there are no residual spin-space transformations that are decoupled from the real-space transformations. A detailed worked example of this kind of argument is given in Appendix~\ref{app:exchange_constrain}.

However, this restriction, while strictly true in principle, does not allow for the existence of a hierarchy of exchange couplings. For example, as we noted above, there are materials in which the Heisenberg coupling is overwhelmingly the largest coupling. In other materials, there are various well-understood mechanisms (as well as cases with merely a degree of fine-tuning), that can lead to certain anisotropic couplings being significantly larger than others. We consider various examples. 

Cobalt(II) in an octahedral crystal field has a relatively small spin-orbit coupling that, in the presence of trigonal distortion, may lead to a single ion ground state doublet with easy plane anisotropy with a level splitting on the order of the exchange scale. An example is CoTiO$_3$ for which $g_\perp/g_\parallel \approx 1.7$ \cite{Elliott2020}. It is natural to write down an effective spin one-half model to understand the magnetism in this material and hence single ion anisotropy terms are trivial. The easy-plane anisotropy must therefore be included through the magnetic interactions. Indeed, in this material the leading order description of the magnetism is in terms of an XXZ model  
\beq
\hat{H} = \sum_{i,j }\frac{\mathsf{J}^{\perp}_{ij}}{2}\left( \hat{J}^+_i  \hat{J}^-_j + \hat{J}^-_i  \hat{J}^+_j \right) + \mathsf{J}^{zz}_{ij} \hat{J}^z_i  \hat{J}^z_j.
\eeq
and further anisotropies are sub-leading. The spin-space group is therefore $U(1)\times \mathbb{Z}_2$. The same group appears for Dzyaloshinskii-Moriya (or anti-symmetric) exchange with collinear $\boldsymbol{D}$ vector which may appear, for example, on second nearest neighbor bonds of the honeycomb lattice and which may be important in the CrX$_3$ magnets:
\beq
\hat{H} = D \sum_{\la\la i,j \ra\ra} \hat{J}^x_i  \hat{J}^y_j - \hat{J}^y_i  \hat{J}^x_j.
\eeq 

A further example is Kitaev exchange. Consider a honeycomb lattice of magnetic moments with Ising couplings along perpendicular directions on the three bonds originating from each lattice site. Thus:
\beq
\hat{H} = K \sum_{\la i,j \ra_{\gamma}} \hat{J}^\gamma_i  \hat{J}^\gamma_j
\eeq 
where $\gamma$ runs over ${x,y,z}$ and identically oriented bonds belong to the same Ising component. If we now perform a global spin-space rotation about any of the cubic spin-space axes $x$, $y$, $z$, the Hamiltonian will be left invariant. So the spin-space group is isomorphic to point group $D_2$. Kitaev-Heisenberg models can arise on several lattices with edge-sharing octahedra that supply a superexchange mechanism to generate such couplings. Such lattices include the honeycomb lattice, its three-dimensional generalizations including the hyperhoneycomb, the pyrochlore lattice, the kagome lattice and so on \cite{Kimchi2014}. The same group $D_2$ appears also for $90^\circ$ compass models such as the simple cubic lattice model with $\mathsf{J}^{\alpha\alpha}_{ii+\hat{\boldsymbol{\alpha}}}$ along the $\hat{\boldsymbol{\alpha}}=\hat{\boldsymbol{x}},\hat{\boldsymbol{y}},\hat{\boldsymbol{z}}$ bonds.

An immediate implication of the above remarks for magnon spectra is that given spin-space group of the Hamiltonian $\mathbf{G}_H$, the pure spin rotational part $\mathbf{R}$ of $\mathbf{G}_H$ can be used to find new spin orientations that give the same magnon spectrum. To take an almost trivial example: in the Heisenberg case, spin rotation invariance means that the magnon spectrum is completely invariant to changes in the moment orientation.

So far we have discussed purely magnetic models with the aim of studying magnon band topology. However, spin-space groups that we discuss extensively here may play a role in electronic systems too. For example, we may minimally couple any of the spin-space symmetric magnetic exchange models (with Hamiltonian $\hat{H}_{\rm mag}$) to electrons through a Kondo-like Hamiltonian 
\beq
\hat{H} = \sum_{\la i,j \ra, \alpha} t_{ij}c_{i\alpha}^\dagger c_{j\alpha} + {\rm h.c.} + \sum_{i, \alpha\beta} c^\dagger_{i\alpha} \hat{J}^\mu_i \cdot \bs{\sigma}_\mu^{\alpha\beta} c_{i\beta} + \hat{H}_{\rm mag}
\eeq
where the electronic band structure now inherits the spin-space symmetry of the magnetic subsystem. 

\subsection{Symmetries of the Magnetically Ordered Ground State}

In this section, we show how the considerations of symmetry in the previous section must be adjusted in the presence of magnetic order. Since we are ultimately interested in magnons, we suppose the magnetic ground state is characterized by local order parameter $\la  J^{\mu}_i \ra$. From the parent group $\mathbf{G}_H$ $-$ the group of operations that leave the magnetic Hamiltonian invariant $-$ we identify the subgroup of transformations that leave the magnetic structure invariant $\mathbf{G}_M$. Frequently, the onset of magnetic order enlarges the unit cell thus breaking down the group of primitive translations to a subgroup. The wavevector associated to the magnetic order may even be incommensurate significantly lowering the symmetry. In addition to the translation symmetries there will tend to be combinations of translations, point group operators on the lattice and spin transformations that leave the magnetic structure invariant. Unlike the parent group, $\mathbf{G}_M$ is generally not a simple product group. Instead the spin and space transformations tend to be coupled.

In the case where $\mathbf{G}_H =  \mathbf{G}\oplus \hat{\Ts}\mathbf{G}$ $-$ in other words, when the spin transformations are locked to the space group transformations $-$ the subgroup that leaves the magnetic structure invariant is another magnetic space group. Since the magnetic order breaks physical time reversal symmetry and $\mathbf{G}$ contains the identity $\mathbf{G}_M$ cannot be of the form $\mathbf{G}\oplus \hat{\Ts}\mathbf{G}$ but must instead be a type I group $-$  one with no anti-unitary elements, or a type III or IV magnetic space group of the form $\mathbf{G}\oplus \hat{A}\mathbf{H}$ where $\hat{A}=\hat{\Ts}\hat{U}$ is an antiunitary element and $\hat{U}$ is a nontrivial unitary and $\mathbf{H}$ is a (unitary) space group.

In the other extreme case of Heisenberg models, magnetic order breaks the decoupled spin space and real space transformations down to a discrete subgroup. For example, if the magnetic structure is collinear then the space group transformations acting only on real space leave the moments invariant  there are pure spin space rotations about the axis of the moments as well as  $\hat{\Ts}\sselement{C_{2\perp}}{E}{\bs{0}}$ where the $C_2$ rotation is about an axis perpendicular to the ordered moment. 

\subsection{Magnon Symmetries}
\label{sec:magnons}

Here we show in outline how to determine the single magnon excitation spectrum and the relationship between the symmetries of the magnetic Hamiltonian and those of the magnons. We consider the following general exchange Hamiltonian for localized moments defined on some lattice and including a Zeeman term
\beq
\mathsf{H} = \frac{1}{2} \sum_{ia,jb; \alpha,\beta} \mathsf{J}_{iajb}^{\alpha\beta} \hat{J}_{ia}^{\alpha} \hat{J}_{jb}^{\beta} - \sum_{ia,\alpha} \mathsf{B}^{\alpha}\hat{J}^{\alpha}_{ia}.
\eeq
The couplings have the symmetry property $\mathsf{J}_{ia;jb}^{\alpha\beta}=\mathsf{J}_{jb;ia}^{\beta\alpha}$. We are supposing that the moments have nonzero expectation values either through spontaneous or field-induced magnetic ordering. And we use the following notation, $\la \hat{J}_{ia}^{\alpha} \ra $, for the local order parameter with $i$ running over the $N$ magnetic primitive cells and $a$ running over the $m$ magnetic sublattices on a finite lattice. We further suppose that the moments are written in a local quantization frame defined with local $z$ component $\hat{\boldsymbol{z}}_a$ along the ordered moment direction as in $\la \hat{J}_{ia}^{\alpha} \ra \equiv S \delta^{\alpha z}$ and with uniform ordered moment from site to site. In principle, we could consider ordered structures with ferrimagnetic textures or models with different types of magnetic ion via a straightforward extension of the methods described here. Concurrently, we introduce local transverse directions, $\hat{\boldsymbol{x}}_a$ and $\hat{\boldsymbol{y}}_a$ that may be chosen arbitrarily - observable quantities should not depend on the choice of transverse axes - so there is a local phase invariance. 

The angular momentum operators are bosonized through the Holstein-Primakoff representation for spins of size $S$
\begin{align}
\hat{J}^{z} & = S - \hat{a}^\dagger \hat{a} \\
\hat{J}^{+} & = \sqrt{2S} \sqrt{1 - \frac{\hat{a}^\dagger \hat{a}}{2S}} \hat{a} = \sqrt{2S} \left( 1- \frac{\hat{a}^\dagger \hat{a}}{4S} \right) \hat{a} + \ldots  \\
\hat{J}^{-} & = \sqrt{2S} \hat{a}^\dagger \sqrt{1 - \frac{\hat{a}^\dagger \hat{a}}{2S}} = \sqrt{2S} \hat{a}^\dagger \left( 1- \frac{\hat{a}^\dagger \hat{a}}{4S} \right) + \ldots
\end{align}
where the bosons satisfy the usual commutation relations $[\hat{a},\hat{a}^\dagger]=1$. 

Expanding about the mean field ground state leads to the quadratic Hamiltonian
\begin{widetext}
\beq
\mathsf{H}_{\rm SW} = \frac{S}{2} \sum_{\boldsymbol{k}} \hat{\boldsymbol{\Upsilon}}^{\dagger}(\boldsymbol{k}) \left( \begin{array}{cc} \boldsymbol{\mathsf{A}}(\boldsymbol{k}) &  \boldsymbol{\mathsf{B}}(\boldsymbol{k}) \\  \boldsymbol{\mathsf{B}}^{\star}(-\boldsymbol{k}) &  \boldsymbol{\mathsf{A}}^{\star}(-\boldsymbol{k})  \end{array} \right) \hat{\boldsymbol{\Upsilon}}(\boldsymbol{k}) \equiv   \frac{S}{2} \sum_{\boldsymbol{k}}  \hat{\boldsymbol{\Upsilon}}^{\dagger}(\boldsymbol{k}) \boldsymbol{M}(\boldsymbol{k}) \hat{\boldsymbol{\Upsilon}}(\boldsymbol{k}) 
\label{eq:SWH}
\eeq
\end{widetext}
where 
\beq
\arraycolsep=0.8pt\def\arraystretch{1.0}
\hat{\boldsymbol{\Upsilon}}^{\dagger}(\boldsymbol{k}) = \left( \begin{array}{cccccc} \hat{a}^{\dagger}_{\boldsymbol{k}1} & \ldots & \hat{a}^{\dagger}_{\boldsymbol{k}m} & \hat{a}_{-\boldsymbol{k}1} & \ldots & \hat{a}_{-\boldsymbol{k}m} \end{array}   \right) \hspace{0.1cm}
\hat{\boldsymbol{\Upsilon}}(\boldsymbol{k}) = \left( \begin{array}{c} \hat{a}_{\boldsymbol{k}1} \\ \vdots \\ \hat{a}_{\boldsymbol{k}m} \\ \hat{a}^\dagger_{-\boldsymbol{k}1} \\ \vdots \\ \hat{a}^\dagger_{-\boldsymbol{k}m} \end{array}   \right)
\eeq
and the $\mathsf{A}_{ab}(\boldsymbol{k})$ and $\mathsf{B}_{ab}(\boldsymbol{k})$ depend on the exchange couplings in the local frame as follows:
\begin{align}
\mathsf{A}_{ab}(\boldsymbol{k}) & =\tilde{\mathsf{J}}_{ab}^{+-}(\boldsymbol{k}) - \delta_{ab} \sum_{c} \tilde{\mathsf{J}}_{ac}^{zz}(\boldsymbol{0}) \label{eq:Ablock} \\
\mathsf{B}_{ab}(\boldsymbol{k}) & = \frac{1}{2}\left( \tilde{\mathsf{J}}_{ab}^{xx}(\boldsymbol{k}) - \tilde{\mathsf{J}}_{ab}^{yy}(\boldsymbol{k}) -i \tilde{\mathsf{J}}_{ab}^{xy}(\boldsymbol{k}) - i\tilde{\mathsf{J}}_{ab}^{yx}(\boldsymbol{k})   \right) \nonumber \\
& =  \tilde{\mathsf{J}}_{ab}^{--}(\boldsymbol{k}).
\label{eq:Bblock}
\end{align}
Note that these expressions with the factor one-half define $\tilde{\mathsf{J}}_{ab}^{\alpha\beta}$ for $\alpha, \beta = \pm$.

The diagonalizing transformation on Eq.~\ref{eq:SWH} to find the spin wave spectrum,
\begin{align*}
\boldsymbol{U}^\dagger (\boldsymbol{k})  \boldsymbol{M}(\boldsymbol{k}) \boldsymbol{U}(\boldsymbol{k}) & = \boldsymbol{\Lambda}(\boldsymbol{k}) \\
\boldsymbol{U}(\boldsymbol{k}) \boldsymbol{\eta}\boldsymbol{U}^\dagger (\boldsymbol{k}) = \boldsymbol{\eta}
\end{align*}
where $\boldsymbol{\Lambda}(\boldsymbol{k})$ is diagonal, must preserve the commutation relations 
\beq
\left[ \Upsilon_a , \Upsilon_b^\dagger \right] = \eta_{ab}
\eeq
where $\epsilon_{ab}=1$ if $a=b\leq m$ and $\epsilon_{ab}=-1$ if $a=b\geq m+1$ and zero otherwise. 

It is straightforward to see that 
\beq
\boldsymbol{\eta} \boldsymbol{M}(\boldsymbol{k}) \boldsymbol{U}(\boldsymbol{k}) = \boldsymbol{U}(\boldsymbol{k}) \boldsymbol{\eta} \boldsymbol{\Lambda}(\boldsymbol{k})
\eeq
so the diagonalizing transformation can be found by solving this non-Hermitian eigenvalue problem. This diagonalizing transformation has important consequences for magnon band topology \cite{Xu2020}. In short, the ten-fold way that classifies single particle fermion problems in the absence of lattice symmetries, is reduced to a three-fold way with only time reversal symmetry that can be either absent, or present and squaring to $\pm 1$. This has the consequence of leaving only Chern insulators in 2D, or $\mathbb{Z}_2$ topological bands in 2D or 3D as gapped bands.

Symmetries of the spin wave Hamiltonian, Eq.~\ref{eq:SWH}, are inherited from the symmetries of the full magnetic Hamiltonian that leave the magnetic structure invariant. In other words, the appropriate symmetry group is $\mathbf{G}_M$ and a particular representation of the symmetry elements is determined from the transformation properties of the transverse spin components in the local quantization frame since the Holstein-Primakoff bosons are related to them through
\begin{align}
& \hat{J}^{+}_{\bs{k}a} \rightarrow \sqrt{2S} \hat{a}_{\bs{k}a} \\
& \hat{J}^{-}_{\bs{k}a} \rightarrow \sqrt{2S} \hat{a}^\dagger_{-\bs{k}a}.
\label{eq:transversetoHP}
\end{align}
Under a unitary element of $\mathbf{G}_M$, the transverse spin components will transform into one another via a permutation of magnetic sublattices, a rotation $-$ that, in the $\pm$ frame amounts to an overall phase $-$ a transformation in momentum space and a translation. More precisely, for spin-space transformation  $\sselement{B}{R}{\bs{t}}$ we may write
\beq
\hat{U}_{g_S}  \hat{\Upsilon}_{am}(\boldsymbol{k}) \hat{U}^\dagger_{g_S} = \sum_b e^{ -i  R\bs{k}\cdot\bs{t} }  \left[ U_B \right]_{ab} \hat{\Upsilon}_{bm}(R\boldsymbol{k})
\label{eq:tr_main}
\eeq
where $\left[ U_B \right]_{ab}$ is a matrix of phase factors $e^{ -i \phi^{ab}_S }$ where indices $a$ and $b$ are constrained by the real space transformation $\bs{i}+\bs{a}=R(\bs{j}+\bs{b})+\bs{t}$. The boson spinor $\hat{\Upsilon}_{am}$ is indexed by sublattice $a$ and particle-hole index $m=0,1$. In order to treat antiunitary elements we require the transformations under time reversal 
\begin{align}
& \hat{\Ts}i\hat{\Ts}^{-1} = -i \\
& \hat{\Ts}\hat{J}^{\pm}_{ia}\hat{\Ts}^{-1} = -\hat{J}^{\mp}_{ia} \\
& \hat{\Ts}\hat{J}^{\pm}_{\bs{k}a} \hat{\Ts}^{-1} = -\hat{J}^{\mp}_{-\bs{k}a}.
\label{eq:TR}
\end{align}
In ensuing sections, we describe in detail how to construct band representations of the full spin-space group. 

Before that, we make a couple of further important remarks. The first concerns the effect of higher order terms in the spin wave expansion. Such magnon interaction terms in general couple single magnon to multi-magnon states leading to damping and renormalization of the single magnon modes. In all of the following we assume that the effect of magnon damping is sufficiently weak that the single magnon states have a lineshape that is negligible compared to the magnon bandwidth. In other words, we shall rely on a robust single particle picture throughout. In practice, this assumption is often very reasonable. Indeed most magnetic materials are in such a regime and magnon interaction effects tend to be severe only in geometrically frustrated magnets with a high density of low-lying states and especially those with non-collinear moments or anisotropic exchange. Even where magnon interactions are important their effect can be reduced, at least in principle, by applying a large enough magnetic field.  Although we shall not consider magnon interactions further in this paper, it is worth mentioning that the symmetries of the magnetic Hamiltonian are symmetries of the magnon Hamiltonian order-by-order in the interactions. It follows that any symmetry protected degeneracies in the single magnon spectrum cannot be broken by higher order terms.

Although spin-space symmetries survive magnon interactions, it is possible for the symmetries of the linear spin wave Hamiltonian to be higher than one would expect on the basis of the foregoing discussion. In other words, the quadratic magnon Hamiltonian may have accidental symmetries. These are most commonly discussed in relation to order-by-disorder where an accidental symmetry appears at the mean field level and in linear spin wave theory \cite{Villain1980,shender1982antiferromagnetic,Henley1989,Elliott2020}. One example is the $U(1)$ symmetry of strong spin-orbit coupled pyrochlore moments \cite{savary2012,zhitomirsky2012}. This $U(1)$ is absent in the magnetic exchange but appears in linear spin wave theory and is broken down to the discrete lattice symmetries by magnon interactions. In the case of order-by-disorder both the ground state  and linear spin wave spectrum have an accidental symmetry. There are also cases where there is no order-by-disorder but still the linear spin wave spectrum is more symmetric \cite{mook2020interactionstabilized}. In this paper, we concentrate our attention on features that survive magnon interactions as they are protected by spin-space symmetries that appear to all orders of spin wave theory.

\section{Representation Theory and Magnon Band Topology}
\label{sec:BR}

\begin{figure}[!htb]
\hbox to \linewidth{ \hss
\includegraphics[width=\columnwidth]{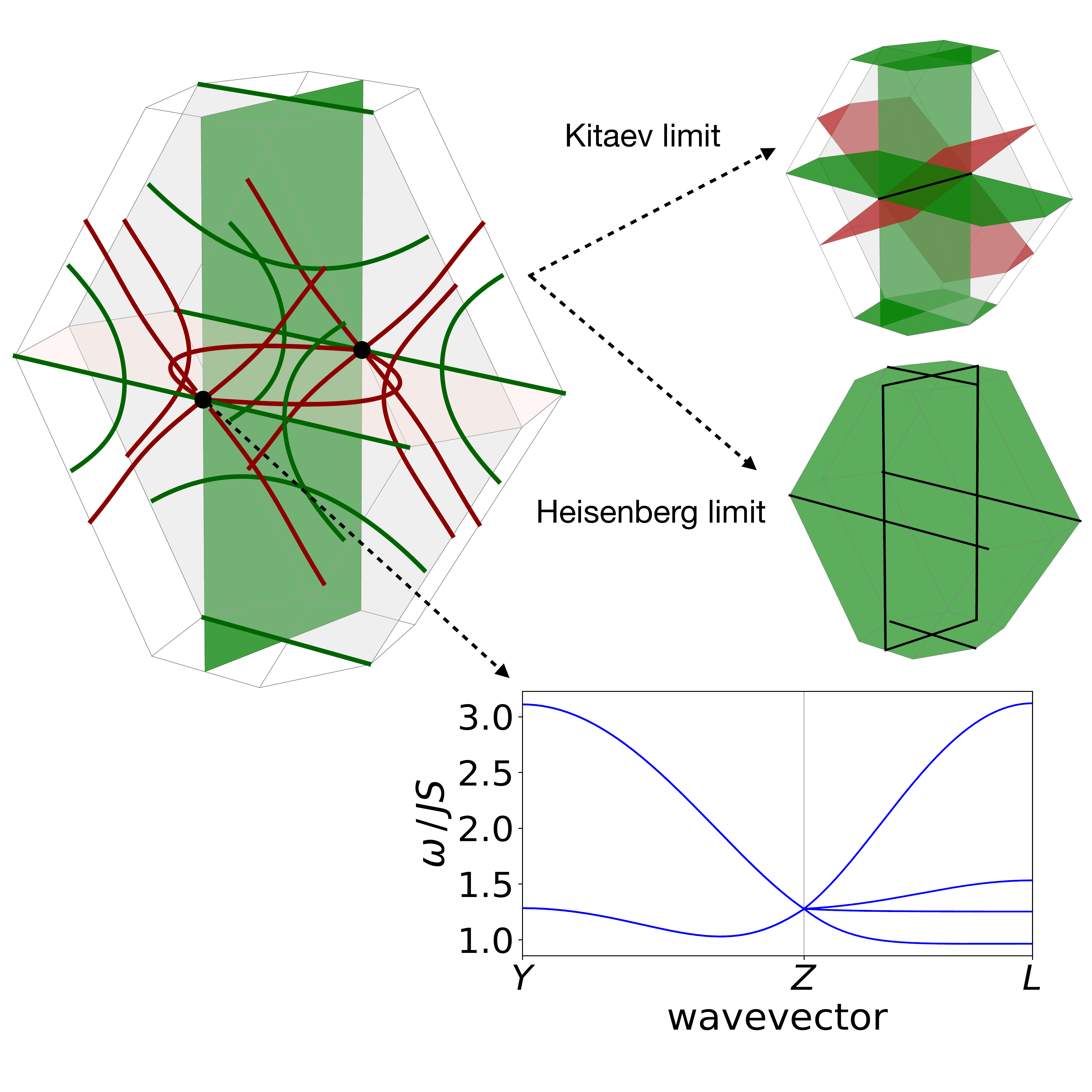}
\hss}
\caption{
(Main Panel, top left) Magnon spectrum for the N\'{e}el phase and $\varphi=0.35\pi$ (where $\varphi$ is defined following Eq.~\ref{eq:KHModel}) with the band degeneracies indicated. Nodal lines between bands $(1,2)$ and $(3,4)$ are shown as green lines and those between bands $(2,3)$ as red lines. There is a nodal plane (green surface) and 4-fold degenerate points (black points).
The inset at the bottom right shows the magnon dispersions along the high symmetry path $Y\rightarrow Z\rightarrow L$ in the vicinity of the 4-fold degenerate point showing the double degeneracy along $[Y Z]$ in the nodal plane. Section~\ref{sec:representationtheory} accounts for all these features through a symmetry analysis.
(Insets) The insets at the upper right show two coupling limits $-$ the Kitaev limit $\varphi=0.5\pi$ and Heisenberg AFM limit $\varphi=0$.
In the Kitaev limit the bands $(1,2)$ are degenerate everywhere in the zone, bands $(2,3)$ form two tilted nodal planes (red surfaces) and bands $(3,4)$ nodal planes on the high symmetry surface (green surfaces).
The two types of nodal planes meet along line $[\Gamma Z]$ (black line) which is indeed 4-fold degenerate, and can be seen as the continuous shrinking of the circular nodal loop at $\varphi=0.35\pi$.
In the Heisenberg limit the modes form a doubly degenerate nodal volume over the entire Brillouin zone between bands $(1,2)$ and $(3,4)$ (green volume). In addition the bands are four-fold degenerate on some high symmetry lines (black).
}
\label{fig:magnonsNeel}
\end{figure}

In the following sections, we work out spin-space symmetry constraints on magnon band structures using representation theory. The approach we take is to work from the atomic limit and build {\it band representations} introduced by Zak in Ref.~\onlinecite{Zak82}. In this section we outline how to construct band representations for magnons. The reader who is content to skip the details can see an outline of the method in the next paragraph. 

In a nutshell, the band representation ties together all symmetry information about the band structure at different momenta using group elements and local atomic orbitals as ingredients. From the band representation, one may extract the irreducible representations at different symmetry-distinct momenta as well as information about the connectivity of these irreps. The building blocks of the band representation for magnons are the on-site transverse spin components in the local quantization frame, $\hat{J}^{\pm}$. These components form a basis for a representation of a group that leaves the lattice site invariant. Since this site symmetry group is a subgroup of the full spin-space group one may carry out a well-defined induction procedure to obtain a representation of the full group that has dimension equal to the number of bands and that is also a function of momentum.

We now describe the process in more detail. Take a point $\bs{r}$ in the primitive cell and act on it with elements of the spin-space group $g\in \mathbf{G}_S$. Those elements that leave the point invariant form a group $G_{\bs{r}}$ called the {\it site symmetry group} or {\it stabilizer group} $-$ that may include translations. By construction, elements of the stabilizer group leave the magnetic moment invariant at the site. The stabilizer group will generally have both unitary and anti-unitary elements. 

Now act on the point $\bs{r}$ with elements $g\in \mathbf{G}$ that are not in the stabilizer group. The set of points thus defined
\beq
\left\{ \bs{r}_a = g_a  \bs{r} \vert g_a \notin G_{\bs{r}} \right\} \hspace{0.25cm} {\rm for} \hspace{0.1cm} a=1,\ldots,n
\eeq 
is associated to a {\it Wyckoff position} with multiplicity $n$ where $n$ is the number of points generated in this way that live in the primitive cell. The stabilizer groups associated to these points are isomorphic.

Starting from the atomic limit, we take a set of orbitals $-$ essentially maximally localized Wannier functions $W_{ia}(\bs{r})$ $-$ forming a representation $\rho$ of the stabilizer group $G_{\bs{r}}$. Suppose there are $n_r$ such functions. Then for each element $h$ of the stabilizer group the representation $\rho_{ij}(h)$ has indices that run from $1$ to $n_r$.

Now, given representation $\rho$ of the stabilizer group, one may induce to a representation of the full space group that we write as $\rho \uparrow \mathbf{G}$. In real space, the dimension of the representation is $(n_r \times n \times N)\times (n_r \times n \times N)$ where $N$ is the number of primitive cells. In momentum space, the translations are diagonalized and the momentum dependent representation is a matrix of dimension $(n_r \times n )\times (n_r \times n )$ that acts on the Fourier transformed Wannier functions:
\beq
a_{ia}(\bs{k},\bs{r}) = \sum_{\mu} e^{-i \bs{k}\cdot\bs{t}_\mu } W_{ia}(\bs{r}-\bs{t}_\mu).
\eeq
A formula for the induced representation for $g\in \mathbf{G}$ is
\begin{widetext}
\beq
\left( \rho_G(g) \right)_{ia;jb}(\bs{k}) a_{jb}(\bs{k},\bs{r}) = e^{-i(g\bs{k})\cdot \bs{t}_{ba}} \rho_{ji}\left( g_{b}^{-1} \element{E}{-\bs{t}_{ba}}g g_a \right)  a_{jb}(h\bs{k},\bs{r})
\eeq
\end{widetext}
where 
\beq
\bs{t}_{ba} = g \bs{r}_a - \bs{r}_b
\eeq
is a Bravais lattice vector. To each site $a$ there is exactly one $b$ and these are related through $g g_b = \element{E}{\bs{t}_{ba}} g_b h$ where $h \in G_{\bs{r}}$. This band representation contains complete symmetry information about the bands at the discrete momentum space points of distinct symmetry including constraints on the connectivity of the bands in the Brillouin zone.

At each point in the Brillouin zone, $\bs{k}$, the little group $\mathbf{G}_{\bs{k}}$ consists of elements $h \in \mathbf{G}$ such that $h\bs{k}=\bs{k}$ which is a momentum diagonal block in the band representation. We obtain a representation of the little group at $\bs{k}$, $\rho^{\bs{k}}_G= \rho_G \downarrow \mathbf{G}_{\bs{k}}$. By modding out translations we obtain the little co-group, $\bar{\mathbf{G}}_{\bs{k}}$ at this wavevector that is isomorphic to some spin point group. This is generally composite and can be decomposed into irreps using the orthogonality of irreps $\rho^{(\alpha)}_{\bs{k}}$
\beq
\rho^{\bs{k}}_G = \bigoplus_{\alpha}  n_\alpha \rho^{(\alpha)}_{\bs{k}}.
\eeq
 Thus the number of times the irrep $\alpha$ occurs is given by
 \beq
 n_\alpha = \frac{1}{h} \sum^{h}_{m=1} \chi^*_{\rho_{\bs{k}}^{(\alpha)}}(m) \chi_{\rho^{\bs{k}}_G}(m)
 \eeq
where the sum is over distinct relevant classes $m$, $\chi_\sigma$ is the character of the representation $\sigma$, and $h$ is the order of the group. In this way, we obtain the symmetry distinct modes at each wavevector. In practice, the decomposition requires the character table of the spin point group. Later on we give some examples of this decomposition. 

So far we have the representation theory for magnons in the case where the spin-space group is unitary. When there are anti-unitary elements there are some important new features. The $1651$ conventional magnetic space groups appear in four types: (I) the ordinary space groups ($230$ in all), (II) paramagnetic groups of the form ${\bf G}\oplus \Ts{\bf G}$ ($230$), (III) ${\bf H}\oplus \Ts\left( {\bf G}-{\bf H}\right)$ ($674$) and (IV) the black and white magnetic groups ${\bf G}\oplus \Ts \element{E}{{\bf t}} {\bf G}$ ($517$). It will be useful to bear these in mind as, in later sections, we show that certain spin-space groups are isomorphic to magnetic space groups. In general, a magnetic space group takes the form ${\bf G}\oplus \mathcal{A}{\bf G}$ where $\mathcal{A}$ is some anti-unitary element. 

In principle one can construct the full band representation for the spin-space group including magnetic elements. Then at a given momentum one can determine the decomposition into irreducible co-representations (or coreps) of the magnetic little co-group. However, since our principal focus is the symmetry-enforced degeneracies, it is possible to side-step this process and find the band representation for the unitary part of the group as described above. As above, we find the irreps from the subduced representation at a given momentum. We then determine irreducible co-representations (or coreps) of the magnetic group associated to each unitary irrep using the following criterion {\it requiring access only to characters of the unitary elements} that separates the coreps into three classes (a), (b) and (c)
\beq
\sum_{\alpha} \chi\left(  B_\alpha^2 \right) = \begin{cases}
                     				 +\vert {\bf G} \vert \,,        & (a) \\
                     				-\vert {\bf G} \vert \,,        & (b) \\
                     				 0\,,   & (c)
            		                     \end{cases}
\label{eq:AUTest}		                     
\eeq
where the sum runs over the anti-unitary elements $B_\alpha$. Each class is associated to a canonical form for the corep which, for class (a), has the same dimension as the unitary irrep from which it is derived while, for classes (b) and (c) the degeneracy is doubled in passing over to the magnetic group. 

Now consider two high symmetry points $\bs{k}_1$ and $\bs{k}_2$ joined by a high symmetry line $\bs{k}_1 + \lambda(\bs{k}_2-\bs{k}_1)$. The symmetry group along the line is a subgroup of the groups at the two endpoints. It follows that, at each high symmetry point, the symmetry group associated to that point corresponds to a set of irreducible representations $X_a$ that, in general, are reducible under the subgroup along the line connecting the endpoints. In terms of the characters, for each irrep $X_a$ at a high symmetry point, and irreps $Y_b$ along the high symmetry line, there is a compatibility relation 
\beq
\chi\left( X_a \right) = \sum_{b} \chi\left( Y_b \right)
\eeq
with a similar condition at the other endpoint of the line. Representation theory therefore supplies a discrete notion of band connectivity in momentum space. For, given the magnon group representation at each high symmetry point and line in the zone, there is a set of energy levels at each labelled by some irrep. Then the compatibility relations constrain the ways in which these levels connect to one another to form a continuous band structure through the zone. Depending on the ordering of the irreps in energy $-$ that is not fixed by symmetry $-$ the compatibility relations may enforce crossings between bands.

\section{Nodal Points, Lines and Planes from Spin-Space Symmetry: An Example}
\label{sec:hyperhoneycomb}

\subsection{Model and Overview of Results}
\label{sec:overview}

In this section, we illustrate the proliferation of nodal features in band structures caused by spin-space symmetries using the example of a Kitaev-Heisenberg model on a tri-coordinated lattice in three-dimensions: the hyperhoneycomb lattice (Fig.~\ref{fig:Lattice}). The Hamiltonian is
\beq
\hat{H} = J \sum_{<ij>} \bs{J}_i \cdot \bs{J}_j + K  \sum_{<ij>_{\gamma}} J_i^{\gamma} J_j^{\gamma} - \bs{h} \cdot \sum_i \bs{J}_i
\label{eq:KHModel}
\eeq
$-$ the Kitaev-Heisenberg interaction is parametrized using angle $\varphi$ so that $J=\cos{\varphi}$ and $K=\sin{\varphi}$.
As discussed in Section~\ref{sec:exchange}, the Heisenberg model alone has decoupled spin and space degrees of freedom and the inclusion of the Kitaev exchange coupling breaks the $SO(3)$ spin group of the Heisenberg model down to $D_2$. With the onset of magnetic order, the spin and space symmetries get broken down to a subgroup and become intertwined so that the group ceases to be a simple product of spin and space transformations. 
We consider the hyperhoneycomb lattice primarily because these couplings are allowed by symmetry. The lattice also has the attractive feature of having four sublattices thus allowing for up to four-fold degeneracies for $\bs{Q}=0$ magnetic order. The phases of the model in zero field as a function of $\varphi$ were studied in Ref.~\cite{Lee2014} and this analysis was extended to finite field in Ref.~\cite{Kruger2020}. There are four phases in zero field: the collinear ferromagnet, a N\'{e}el phase, and two further antiferromagnetic phases called skew-stripey and skew-zigzag. 

The hyperhoneycomb lattice is also the iridium $\text{Ir}^{4+}$ sublattice in $\beta \text{-}\text{Li}_2\text{IrO}_3$. The oxygen ions in this material form a lattice of edge-sharing octahedra such that the Ir-O-Ir bond angle is $90^\circ$ and this geometry provides the basis for a microscopic mechanism leading to Kitaev-Heisenberg couplings \cite{jackeli2009mott,chaloupka2010kitaev}. The spiral ground state of this magnet in zero field is suggestive of the presence of significant off-diagonal symmetric, or $\Gamma$, exchange in this system and, in general, we expect materials to deviate from the ideal Kitaev-Heisenberg model. However, when there is a family of magnetic materials with similar crystal structures a certain degree of fine-tuning is compatible with the existence of materials proximate to the Kitaev-Heisenberg limit. 

The particular example we take in this section is the hyperhoneycomb Kitaev-Heisenberg N\'{e}el antiferromagnet in zero applied magnetic field. We note that there is an order-by-disorder mechanism that fixes the moments to lie along one of the Cartesian axes and, without loss of generality, we choose this to be the $[001]$ direction. This example will turn out to have  enhanced magnetic symmetry described by a spin-space group that is {\it not isomorphic to a magnetic space group} and, therefore, that could not be inferred simply by requiring that the magnetic structure be left invariant by space group transformations. 

Fig.~\ref{fig:magnonsNeel} illustrates features of the magnon spectrum within the N\'{e}el phase for non-vanishing Kitaev coupling ($\varphi=0.35\pi$). The main panel shows the band degeneracies within the first Brillouin zone. There is one plane (shown in green) where the four magnon bands are symmetry-enforced to pair up into two two-fold degenerate bands. This plane is intersected by several nodal loops that can be inferred from compatibility relations. They are protected by glide symmetry on their respective mirror planes. One of these, between bands two and three, is confined to the plane through the $\Gamma$ point perpendicular to the nodal plane. The existence of this nodal loop implies that the nodal plane bands meet at a four-fold degenerate nodal point (shown in black) at the Z point. The band structure in the vicinity of the Z point, plotted with one axis in the $\Gamma-Z-Y$ plane and another perpendicular to this plane, is therefore a double cone emanating from a single point where the two cones touch along a plane. The magnon band dispersions through the Z point in a nodal plane direction ($Y\rightarrow Z$) and an out of plane direction ($Z\rightarrow L$) are shown in the lower panel of Fig.~\ref{fig:magnonsNeel}. The aforementioned nodal loop
shrinks as the Heisenberg coupling is reduced, forming a 4-fold degenerate line along $\Gamma \rightarrow Z$ in the Kitaev limit. One further symmetry-enforced nodal line runs along the line $Z\rightarrow A$ (and equivalently by reciprocal vector translation along $Y\rightarrow A1$).

\subsection{Representation Theory}
\label{sec:representationtheory}

We saw in Section~\ref{sec:BR} how to proceed from localized orbitals to band structures using symmetry considerations alone. Here we show that the representation approach allows us to account for the symmetry protected features of a magnon model and enumerate the kinds of magnon band topology that can arise for the given symmetry.

The full group is $\mathbf{G}_{\rm Neel}=\mathbf{H}_{\rm Neel} + \Ts C_2^{x}(\bs{s}) \mathbf{H}_{\rm Neel}$ with coset representatives
\begin{align}
& \mathbf{H}_{\rm Neel}= \nonumber \\ & E + d_1 + d_3 + C_2^a + C_2^{x,y}(\sv) \, (\Ps + d_2 + C_2^b + C_2^c) + C_2^{z}(\bs{s})
\label{eq:NeelUnitary}
\end{align}
giving $16$ elements in all and then there are translations in addition to these. Here we have used a notation for the symmetry elements where pure spin transformations are labelled with $\sv$ and the remaining elements are combined spin and space transformations $-$ they are locked to one another. All symmetry elements are defined in Appendix~\ref{app:lattice_data} and a short introduction to group notation is given in Appendix~\ref{app:group_notation}.

The hyperhoneycomb lattice belongs to Wyckoff positions $16g$. The magnon band representation (BR) $\rho_G^{\bs{k}}$ is induced from the representation  $\rho_{S_{\perp}}^{16g}$ of spin transverse components $(J^+, J^-)$ around the ordering vector.

If we take as representative of the orbit the position $q_1^{16g} = \bs{r}_1$, the spatial site-symmetry group is: 
\beq
\mathbf{G}_{q_1^{16g}} =  \element{E}{0} \, + \, \element{2_{001}}{-1/4,-1/4,0 } \cong C_2
\eeq
If we now consider also the possible additional spin rotations we get the enhanced magnetic (spin-space) site-symmetry group $\mathbf{G}_{16g}^{\rm SS}$ (noting that spin rotations leave the position invariant):
\begin{widetext}
\begin{align}
\mathbf{G}_{q_1^{16g}}^{\rm SS} = & ~~~~\, \sselement{E}{E}{0} \, + \, \sselement{2_{010}}{E}{0} \, + \, \sselement{4_{010}^{+}}{2_{001}}{-1/4,-1/4,0 }  \, + \, \sselement{4_{010}^{-}}{2_{001}}{-1/4,-1/4,0 } 
 \nonumber  \\
& + \, \sselement{2_{010}}{E}{0}'  \, + \, \sselement{2_{101}}{E}{0}' \, + \,  \sselement{2_{100}}{2_{001}}{-1/4,-1/4,0 }'  \, + \,  \sselement{2_{001}}{2_{001}}{-1/4,-1/4,0 }' \nonumber  
\end{align}
\end{widetext}
Note that the $\bs{a}$, $\bs{b}$, $\bs{c}$
coordinate system is used here and in all the group theory calculation (see Appendix~\ref{app:lattice_data}).
In addition we separate out the spin and space transformations using notation introduced in Section~\ref{sec:exchange}. The first line is the unitary part which is isomorphic to the $C_4$ point group. In the second line there are anti-unitary elements (prime sign) which always give $[E||E]$ when squared (the translation part is nonzero because of the choice of origin and cancels out). From the anti-unitary elements we obtain $(a)$ coreps (see Eq.~\ref{eq:AUTest}) and so no extra degeneracies in the irreps of $C_4$ are expected.

The transverse components  $(J^+, J^-)$ therefore transform as irreps of $C_4$. The symmetries act on the global conventional frame, while the transverse components are around the ordering directions, so in the local frames. The spin transformations in local frames $\widetilde{T_S}$ are then obtained as $\widetilde{T_S}=R_i^T T_S R_i$ where index $i$ appears in both rotation matrices because the symmetries considered here are site preserving.
In this case, the matrix representation of the spin rotations in the local $(J^+, J^-)$ basis is:
\begin{align}
D(2_{010}) = 
\begin{pmatrix}
-1 &0 \\ 0  &-1 
\end{pmatrix}_\pm
~~~~
D(4_{010}^{\pm}) = 
\begin{pmatrix}
\mp i &0 \\ 0  &\pm i 
\end{pmatrix}_\pm.
\end{align}
Comparing with the character table of $C_4$ we see that the representation decomposes to:
\beq
\rho_{S_{\perp}}^{16g} = \Gamma_3 + \Gamma_4
\eeq
and specifically $\rho_{S^{+}}^{16g} = \Gamma_3$ and $\rho_{S^{-}}^{16g} = \Gamma_4$. Here it is important to note that these two reps are complex conjugates of each other, as required by the relationship between the transverse spin components $S^{+}$ (that maps to the Holstein-Primakoff $a$ operator) and $S^{-}$ (that maps to $a^{\dagger}$).

To induce the local representation to the full group ($\rho_{S_{\perp}} = \rho_{S_{\perp}}^{16g} \uparrow \mathbf{G}^{\rm SS}$) we need to consider all the orbits of the Wyckoff position $\{q_{\alpha} = g_{\alpha} q_1 \, | \,g_{\alpha} \in \mathbf{G}^{\rm SS}\}$, $\alpha = 1, ..., n$ with multiplicity $n$  of the Wyckoff position. For $16g$ the multiplicity inside the primitive cell is $4$ (while $16$ in the conventional cell), therefore we can choose representative $g_{\alpha}$ as:
\begin{align}
\label{Eq:OrbitRep}
& g_1 = \sselement{E}{E}{\bs{0}} ~, ~  g_2 =  \sselement{2_{\text{-}101}}{2_{010}}{-1/4,-1/4,0} \nonumber \\   
& g_3 = \sselement{E}{m_{010}}{1/4, 0, 1/4}  ~, ~  g_4 = \sselement{2_{\text{-}101}}{-1}{\bs{0}} 
\end{align}
Now we have all the ingredients necessary to use the general formula for induction. Since the characters of the band representation include all the information we need, we can simply note that:
\begin{widetext}
\begin{align}
\label{eq:IndChar}
\chi_{\rho^{\bs{k}}_{G,S_{\perp}}} (h) = \begin{cases}
                                                \sum_{\alpha} \exp{-i \,(h \,\bs{k}) \cdot \bs{t}_{\alpha \alpha}}  \chi_{\rho_{S_{\perp}}^{16g}} (g_{\alpha}^{-1} \, \sselement{E}{E}{- \bs{t}_{\alpha \alpha}}\, h \, g_{\alpha}) & h \in \mathbf{G}_{q_1^{16g}}^{\rm SS} \\
                                                0 &h \notin \mathbf{G}_{q_1^{16g}}^{\rm SS}\\
                                                \end{cases}
\end{align}
\end{widetext}
where $ \bs{t}_{\alpha \alpha} = h \, \bs{q}_{\alpha} - \bs{q}_{\alpha}$. The second line is always zero since if $h \notin \mathbf{G}_{q_1^{16g}}^{\rm SS}$ then the symmetry will permute the sublattices giving an off-diagonal band representation matrix. Therefore for the N\'{e}el case we will have:
\begin{align}
\label{eq:IndCharNeel}
& \chi_{\rho_{G, S^{\pm}}^{\bs{k}}} ( \sselement{2_{010}}{E}{\bs{0}}) = -4  \\
& \chi_{\rho_{G, S^{\pm}}^{\bs{k}}} ( \sselement{4_{010}^{\pm}}{2_{001}}{-1/4, -1/4, 0} ) = 0
\end{align}
Now we have the full band representations and we can subduce it to different little groups $\rho^{\bs{k}}_{S_{\perp}} =\rho_{S_{\perp}} \downarrow \mathbf{G}^{\bs{k}}_{\rm SS}$.

The representations of the enhanced magnetic little groups  $\mathbf{G}^{\bs{k}}_{\rm SS}$ are straightforward to find for points $\bs{k}$ inside the Brillouin zone and, for symmorphic groups, also on boundary points. Indeed in these cases we need only find the point group isomorphic to the little co-group $\bar{\mathbf{G}}^{\bs{k}}_{\rm SS}$ (little group without primitive and non-symmorphic translations) and decompose into the  irreducible representations of that group.

{\it The Nodal Surface} $-$ For example we can explicitly calculate the enforcement of nodal surface $\bold{E} = (0,u,w)$ (plane {\bf $\Gamma - Z - T$} in conventional basis) for the N\'{e}el antiferromagnet with $[00\pm1]$ moments.
The little group (factored out primitive translations) on this high symmetry surface is:
\begin{widetext}
\begin{align}
\mathbf{G}^{\bold{E}}_{\rm SS} / \bold{T} =  &~~~~\, \sselement{E}{E}{\bs{0}} \, + \, \sselement{2_{010}}{E}{\bs{0}} \, + \, \sselement{4_{010}^{+}}{m_{100}}{0,1/4,1/4}  \, + \, \sselement{4_{010}^{-}}{m_{100}}{0,1/4,1/4} 
 \nonumber  \\
& + \, \sselement{E}{-1}{\bs{0}}'  \, + \, \sselement{2_{010}}{-1}{\bs{0}}' \, + \,  \sselement{4_{010}^{+}}{2_{100}}{0, -1/4, -1/4}'  \, + \,  \sselement{4_{010}^{-}}{2_{100}}{0, -1/4,-1/4}' \nonumber  
\end{align}
where we note that the spin and space transformations are coupled but distinct, highlighting the importance of the enhanced symmetry coming from the internal spin symmetry. The unitary part is isomorphic to $C_4$ with character table \ref{tab:ReptableE}.
The coreps will therefore be given by the test:
\begin{align}
\label{eq:ETest}
\sum_{h_{\bs{k}'}} \chi_{p}^{\bold{E}}(h_{\bs{k}'}^2)  
= \, 2 \, ( \chi_{p}^{\bold{E}}(\sselement{E}{E}{\bs{0}})  +  \chi_{p}^{\bold{E}}(\sselement{2_{010}}{E}{\bs{0}}) ) =
\begin{cases}
                                                4 =  |\bar{\mathbf{G}}^{\bold{E}}_{\rm SS}|  ~~~ &\text{Type (a)} ~~ \text{if} ~~ p = E_{1}^{+}, \, E_{2}^{+}  \\
                                                0     ~~~ &\text{Type (c)} ~~ \text{if} ~~ p = E_{1}^{-}, \, E_{2}^{-}  \\
\end{cases}
\end{align}
\end{widetext}
where $h_{\kv'}$ are all the anti-unitary elements of the little co-group (such that $h_{\bs{k}'} \bs{k} = - \bs{k} + \bs{g}_i$).
We obtain therefore a doubly-degenerate corep $DE^{-}(2) = (E_{1}^{-}, E_{2}^{-})(2)$ with $\chi(\sselement{2_{010}}{E}{\bs{0}}) = -2$ and $\chi(\sselement{4_{010}^{\pm}}{m_{100}}{0,1/4,1/4}) = 0$.

Since we know the band representation $\rho^{\bs{k}}_{G,S_{\perp}}$ for every $\bs{k}$ we can now subduce it to the surface $\bold{E}$ and from its characters  in Eqs.~\ref{eq:IndChar} and \ref{eq:IndCharNeel} we get (the number in parenthesis indicates the dimension of corep):
\begin{align}
\label{eq:ERep}
\rho^{\bold{E}}_{S_{\pm}} = 2 \, DE^-(2)
\end{align}
So we conclude, on the basis of spin-space symmetry, that the magnons on plane $\bold{E}$ are two-fold degenerate where the anti-unitary symmetry and the resulting binding of irreps is responsible for the degeneracy.
%
\begin{table*}[!htb]
\centering
\hbox to \linewidth{ \hss
\begin{tabular}{ |c|c|c|c|c|c|  }
 \hline
 $\mathcal{G}^{\bold{E}}_{\rm SS}$~ & ~$\sselement{E}{E}{\bs{0}}$~ & ~$\sselement{2_{010}}{E}{\bs{0}}$~ & ~$ \sselement{4_{010}^+}{m_{100}}{0,1/4,1/4}$
 ~ & ~$ \sselement{4_{010}^-}{m_{100}}{0,1/4,1/4}$ ~  & ~Type Coreps~ \\ 
 \hline
  $E_{1}^{+}$   & 1   & 1    &   $\xi$   &  $\xi$  & (a)  \\
 \hline
 $E_{2}^{+}$   & 1   & 1     &  -\,$\xi$  &  -\,$\xi$  & (a)  \\
 \hline
 $E_{1}^{-}$    & 1   & -1    &  $\xi$   &  -\,$\xi$   & (c) \\
 \hline
 $E_{2}^{-}$    & 1    & -1   &  -\,$\xi$  &   $\xi$   & (c)  \\
 \hline
\end{tabular}
\hss}
\caption{
Character table of the unitary part of $\mathbf{G}^{\bold{E}}_{\rm SS}$. The phase factor is $\xi = \exp(i \, \bold{E} \cdot (\textstyle 0 \frac{1}{4} \frac{1}{4}) ) = \exp(i \, \frac{\pi}{2} (u + w) ) $.
}
\label{tab:ReptableE}
\end{table*}
%

{\it Four-fold Degenerate Nodal Point} $-$ To analyse the degeneracies of the $\bold{E}$ plane, we were able to rely on the little co-group being isomorphic to a point group and use standard tables to decompose the representation into irreps. But, when $\bs{k}$ is a  boundary point and there are non-symmorphic elements, the situation is, in principle, more complicated and we may need to consider projective representations.
A representation is said to be projective when $\Delta(h_i) \Delta(h_j) = \mu(h_i, h_j) \Delta(h_k)$, where $\Delta$ are matrix representations of group elements $h_i \in \mathbf{G}^{\bs{k}}_{\rm SS}$ and $\mu(h_i, h_j) = \exp( - i\bs{g}_i \cdot \bs{w}_j)$ is an element of the factor system, with $\bs{g}_i = h_i^{-1} \bs{k} - \bs{k}$ and $\bs{w}_j$ the translation associated to $h_j$.
If $\mu(h_i, h_j) = 1$ for all cases then we reduce to ordinary (non-projective) representations. 
If this is not the case we proceed by studying the representations of the central extension of the little co-group $\bar{\mathbf{G}}^{\bs{k}^*}_{\rm SS} = \bar{\mathbf{G}}^{\bs{k}}_{\rm SS} \otimes \bs{Z}_g$ with kernel $\bs{Z}_g$, the cyclic group of integers $0, 1, ..., (g-1)$.  The number $g$ comes from the parametrization of the factor system as $\mu(h_i, h_j) = \exp( 2 \pi i a(h_i, h_j) / g)$, where $a(h_i, h_j) = 0,1,..., (g-1)$ and the group elements are of the kind $(h_i, \alpha)$ with product rule $(h_i, \alpha) (h_j, \beta) = (h_i h_j, \alpha + \beta + a(h_i, h_j))$.
Of all the irreps of $\bar{\mathbf{G}}^{\bs{k}^*}_{\rm SS}$ we are interested only in the ones giving the right factor system, that is the ones with $\Delta(E,\alpha) =  \exp( 2 \pi i \alpha / g) \, \mathbb{I}$. Since the set of elements $(h_i, 0)$ is isomorphic to $\bar{\mathbf{G}}^{\bs{k}}_{\rm SS}$, we can now extrapolate the character tables of those irreps and build the table of projective irreducible representations of $\bar{\mathbf{G}}^{\bs{k}}_{\rm SS}$ (and therefore the one of $\mathbf{G}^{\bs{k}}_{\rm SS}$, adding the right phase factors coming from translations).

 These considerations are relevant to the point $Z = (0,0,-1)$ (in the conventional basis), which in the N\'{e}el $[00\pm1]$ case is four-fold degenerate. This point is highly symmetric $-$ the little group $\mathbf{G}^{\bold{Z}}_{\rm SS}$ is the full spin-space group. The details of the representation theory for this point can be found in Appendix~\ref{app:central_extension}. Here we summarize the chain of reasoning. First of all one can show that the factor system at the $Z$ point is nontrivial.  In particular, $\mu( \sselement{2_{-101}}{-1}{\bs{0}}) =-1$. 
 
 We then find the central extension group for $\bar{\mathbf{G}}^{\bold{Z}^*}_{\rm SS}$ with $g =2$. This is a group with $32$ elements that is isomorphic to $\bar{\mathbf{G}}^{\bold{Z}^*}_{\rm SS} \cong D_{4h} + D_{4h} \times ( \sselement{4_{010}^{+}}{m_{100}}{\bs{0}})$.
The irreps of  $\bar{\mathbf{G}}^{\bold{Z}^*}_{\rm SS}$ can be obtained by conjugating the ones of the subgroup $D_{4h}$ by the symmetry $\sselement{4_{010}^{+}}{m_{100}}{\bs{0}}$. Of these irreps we are only interested in the ones with $\Delta( \sselement{E}{E}{\bs{0}},1) =  -\, \mathbb{I}$. The table of relevant irreps is given in Table~\ref{tab:ProjReptableZ} and these are all two-dimensional. We then return to the anti-unitary elements and look for additional degeneracy in the corepresentations. The standard test reveals that the magnon bands belong to corep with class (c) binding two two-dimensional irreps. The overall degeneracy is therefore four-fold as was to be shown.

{\it Other symmetry constraints on the band structure} $-$ All the degeneracies of the Neel case can be seen in Fig.~\ref{fig:magnonsNeel}. Here the group theory enforced degeneracies are the nodal plane, the 4-fold degenerate points and the straight lines $A=[Z A],[Y A_1]$. All the other curved lines are given by compatibility relations and are protected by glide symmetries on mirror planes. The bands $(2,3)$ (red) are degenerate on a chain of loops on mirror planes $d_1$ and $d_3$. The band $(1,2)$ (green) has a closed nodal line on mirror plane $d_3$ intersecting the nodal plane.

\section{Spin-Space Groups and Nodal Volumes}
\label{sec:heisenberg}

Heisenberg models have the property that the spin space part of the symmetry group is completely decoupled from the spatial part. In the paramagnetic phase, this symmetry group is the three dimensional rotation group. The effect of the spin group on the magnon band structure for Heisenberg models has some general features that we discuss in this section. While this section largely reviews known results \cite{BrinkmanElliott1966,BrinkmanElliott1966b}, it is useful to re-visit them and cement their spin-space origin before breaking new ground.

The simplest case is that of the Heisenberg ferromagnet on an arbitrary lattice. The ground state is collinear, the magnetic structure preserves the translational symmetry of the lattice and, because spin space and real space are decoupled, the other lattice symmetries are also preserved. It follows that the symmetry group of the magnetic Hamiltonian that preserves the magnetic structure is the space group of the underlying lattice times a spin-space group $\mathbf{G}_M =  \mathbf{G}\otimes \mathbf{G}_S$ where the elements of $\mathbf{G}_S$ are: axial rotations through angle $\phi$ about the moment direction, $\hat{\Ts} C_2$ where the $C_2$ is about an axis perpendicular to the moment direction $-$ the choice of axis is unimportant as a change in the axis may be absorbed into an axial rotation. Because spin and real space have decoupled, we label the magnon eigenstates with irreps coming from the space group and an irrep from the spin group. This means that there is no interplay between the spin and space parts of the symmetry group.

The unitary part of the spin group is the continuous group $C_{\infty}$. This is the only nontrivial element of the little cogroup at a general position in the zone for a general ferromagnet. This symmetry element leads to an infinite number of irreps labelled by integers but the invariance of the magnetic moment under the group operations implies that the rotation is fixed to $J^{\pm}_{ia}\rightarrow e^{\mp i\phi} J^{\pm}_{ia}$. In other words, the rotation of the transverse spin components picks out irreps with $n = \pm 1$ and also these rotation operators are diagonal operators in the basis of transverse components. Invariance of the magnon Hamiltonian under such axial rotations (or otherwise, inspection of Eq.~\ref{eq:Bblock}) shows that $\mathsf{B}_{ab}(\boldsymbol{k})=0$ for collinear Heisenberg ferromagnets. This ensures that magnons are eigenstates of the global transverse spin rotations. Diagonalizing the Hamiltonian unitarily then reveals that the upper $\mathsf{A}$ block components and the lower $\mathsf{A}$ components are eigenstates of the Hamiltonian but with different irrep label $n$. It follows that the magnons are each labelled by a common 1D irrep of $C_{\infty}$. They also have an irrep label originating from the space group symmetry. For collinear Heisenberg ferromagnets, any degeneracy in the magnon spectrum that is enforced by symmetry must come from the space group symmetries.

For collinear ferromagnets with inversion symmetry that maps magnetic sublattices into themselves, the nontrivial part of the group at a general position is larger 
\beq
\mathbf{G}^{\rm GP}_{\rm FM_I} = C_n^{\parallel}(\bs{s}) \, (E + \Ts C_2^{\perp}(\bs{s}) \Ps)
\eeq
and, in particular, it contains an anti-unitary part.
The symmetries act on the moments like: $\Ps:  J^{\pm}_{a} \rightarrow J^{\pm}_{a}$, $\Ts: J^{\pm}_{a} \rightarrow -J^{\mp}_{a}$ and $C_2^{\perp}(\bs{s}): J^{\pm}_{a} \rightarrow -J^{\mp}_{a}$ (where we choose $C_2^{\perp}(\bs{s}) = C_2^{y}(\bs{s})$).
Therefore the additional element $\Ts C_2^{\perp}(\bs{s}) \Ps$ acts like an identity and does not mix the irreps coming from the axial rotations.

The next simplest type of Heisenberg model is one with a collinear antiferromagnetic ground state. The two moment directions may be related by
\beq
J^{x}_{a} \rightarrow -J^{x}_{b} \,\, J^{y}_{a} \rightarrow J^{y}_{b} \,\, J^{z}_{a} \rightarrow -J^{z}_{b}
\eeq
where $z$ is the quantization direction. As in the ferromagnetic case, $C_{\infty}$ is an element at a general position and the oppositely oriented moments transform under irreps of this group with opposite signs $+1$ and $-1$.

Let us first consider the classic example of a collinear antiferromagnet on a black-and-white lattice that includes the cubic lattice with a simple N\'{e}el ground state and the rutile lattice \cite{BrinkmanElliott1966}. Rather than enumerating all the symmetry elements we observe that there is a simple translation $\bs{\tau}$ that maps the magnetic sublattices into one another. Then the general position has symmetry elements
\beq
\mathbf{G}^{\rm GP}_{\rm AFM_{\rm bw}} = ( C_n^{\parallel}(\bs{s}) + \sselement{C^{\perp}_2}{E}{\bs{\tau}} ) \cong D_n^{\parallel}(\bs{s}) 
\eeq
The $\sselement{C^{\perp}_2}{E}{\bs{\tau}}$ mixes the two sublattices thus binding the 1D irreps of $C_\infty$ into 2D irreps of $D_{\infty}$. This accounts for the double degeneracy of the modes. For certain lattices of this sort including the cubic lattice, there is an additional symmetry at the general position coming from the lattice inversion symmetry that swaps the magnetic sublattices
\begin{align}
\mathbf{G}^{\rm GP}_{\rm AFM_{ bw+I}} &= ( C_n^{\parallel}(\bs{s}) + \sselement{C^{\perp}_2}{E}{\bs{\tau}}) \, (E + \Ts \Ps) \nonumber \\
&\cong D_n^{\parallel}(\bs{s}) \, (E + \Ts \Ps)
\end{align}
The additional $\mathcal{PT}$ symmetry, taking the group into one isomorphic to a type II magnetic dihedral point group $D_n1'$, does not further bind irreps at the general position. 
In passing, we mention a case where a volume degeneracy can arise on a black-and-white lattice in the presence of a single ion anisotropy of the form $\hat{O}^{2}_{\pm 2}$ thanks to inversion symmetry \cite{BrinkmanElliott1966}. Here the general position has
\begin{align}
\mathbf{G}^{\rm GP}_{\rm AFM_{bw+I}} &= ( C_2^{z}(\bs{s}) + \sselement{C^{x/y}_2}{E}{\bs{\tau}}) \, (E + \Ts C_2^{xy}(\bs{s}) \Ps) \nonumber \\
&\cong D_2^{\parallel}(\bs{s}) \, (E + \Ts C_2^{xy}(\bs{s}) \Ps) \cong D_4^{' \parallel}(\bs{s})
\end{align}
The unitary part $D_2^{\parallel}(\bs{s})$ has 1D irreps, but anti-unitary $C_4$ are created by tilted $C^{x}_2$ elements and $\Ts C^{xy}_2$. The final isomorphism is $D_4^{'\parallel}(\bs{s})$, a type III magnetic point group (explicitly $4'2'2$, i.e. with anti-unitary $C_4$ rotations) which has 2D irreps at the general position.

For collinear antiferromagnets that do not lie on black-and-white lattices $-$ lattices composed of a pair of sublattices separated by a translation $-$ such as the honeycomb and hyperhoneycomb lattices, the magnon bands are doubly degenerate originating from inversion symmetry. At the general position
\beq
\mathbf{G}^{\rm GP}_{\rm AFM_I} = C_n^{\parallel}(\bs{s}) \, (E + \Ts \Ps). 
\eeq
The key here is that the inversion symmetry swaps the magnetic sublattices. The inversion acts as $\Ps: J^{\pm}_{a} = -J^{\mp}_{b}$, therefore $\Ps \Ts: J^{\pm}_{a} = J^{\pm}_{b}$ mixes the two magnetic sublattices, pairing the conjugate irreps into 2D irreps. Indeed, in this case, $\mathbf{G}^{\rm GP}_{\rm AFM_I}$ is isomorphic to a type II magnetic cyclic point group $C_n1'$ which gives 2D irreps for $n > 2$ (for $C_2$ there are no conjugate irreps, only a single real one). When there is inversion symmetry that does not swap the magnetic sublattices, the analysis is similar to the ferromagnetic case considered above. 

We have established that the hyperhoneycomb antiferromagnet has a double degeneracy at the general position. In fact, it has more degeneracy still enforced by lattice symmetries at more symmetric wavevectors as we show in Appendix \ref{app:HeisenNeel}.

\begin{figure}[htbp!]%
\centering
\subfloat[Kitaev]{{\includegraphics[width=0.8\columnwidth]{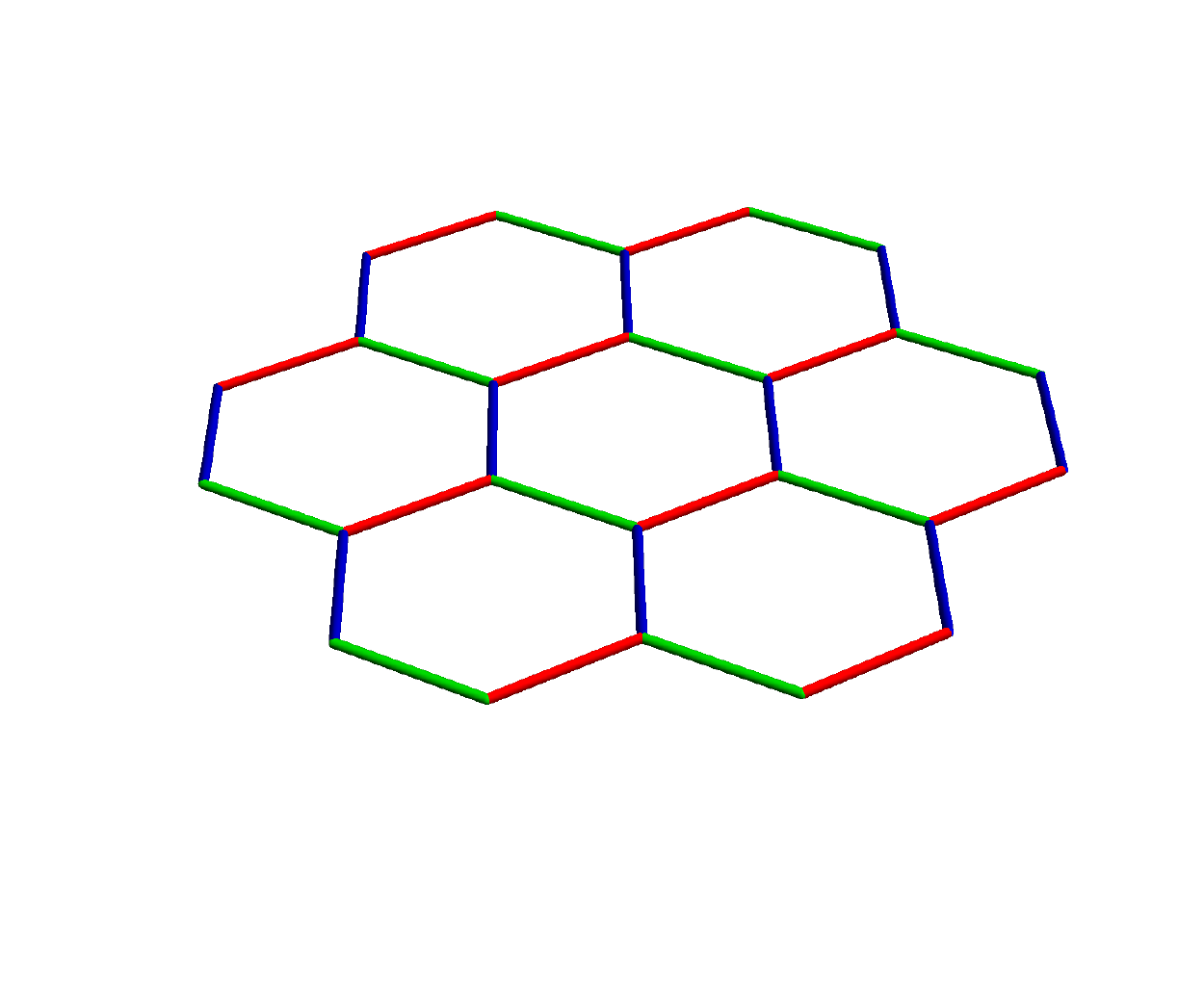} }}%
\qquad
\subfloat[Second neighbor Dzyaloshinskii-Moriya exchange]{{\includegraphics[width=0.8\columnwidth]{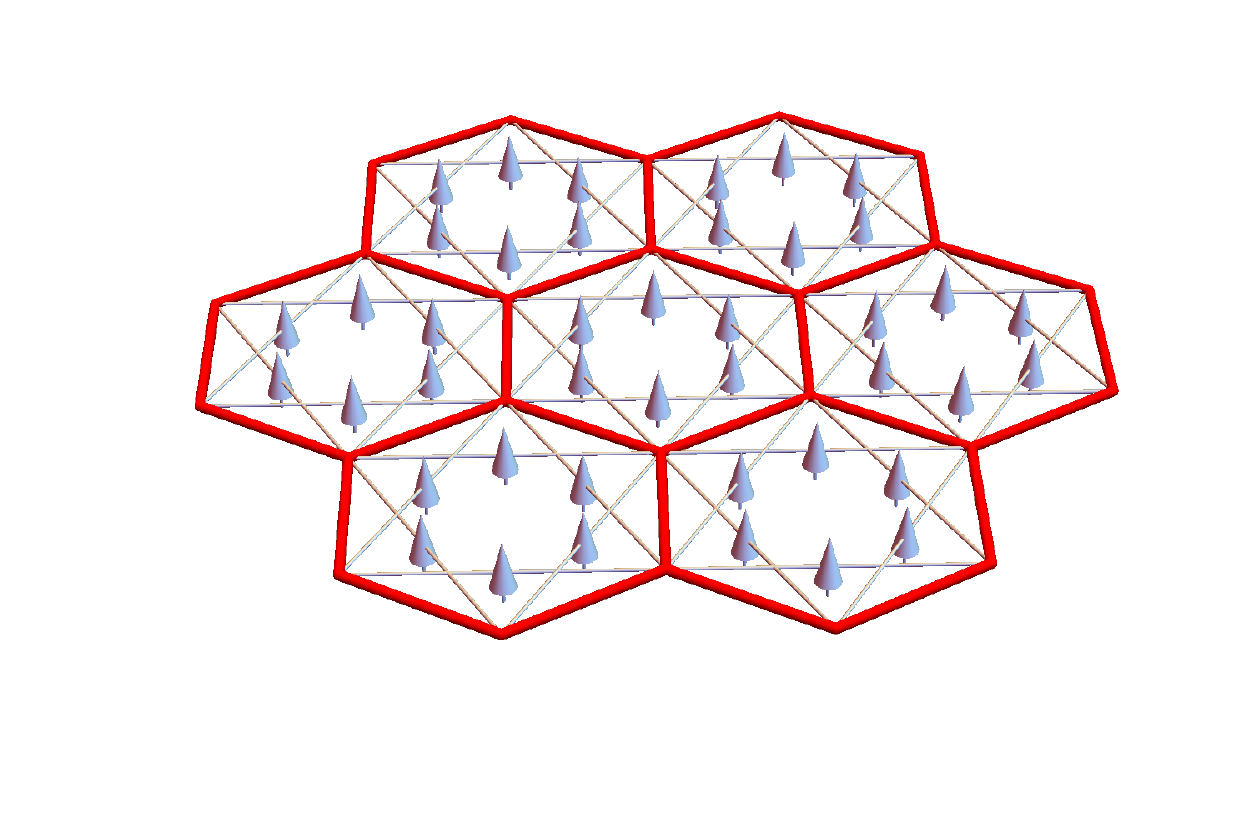} }}%
\qquad
\subfloat[Nearest neighbor Dzyaloshinskii-Moriya exchange]{{\includegraphics[width=0.8\columnwidth]{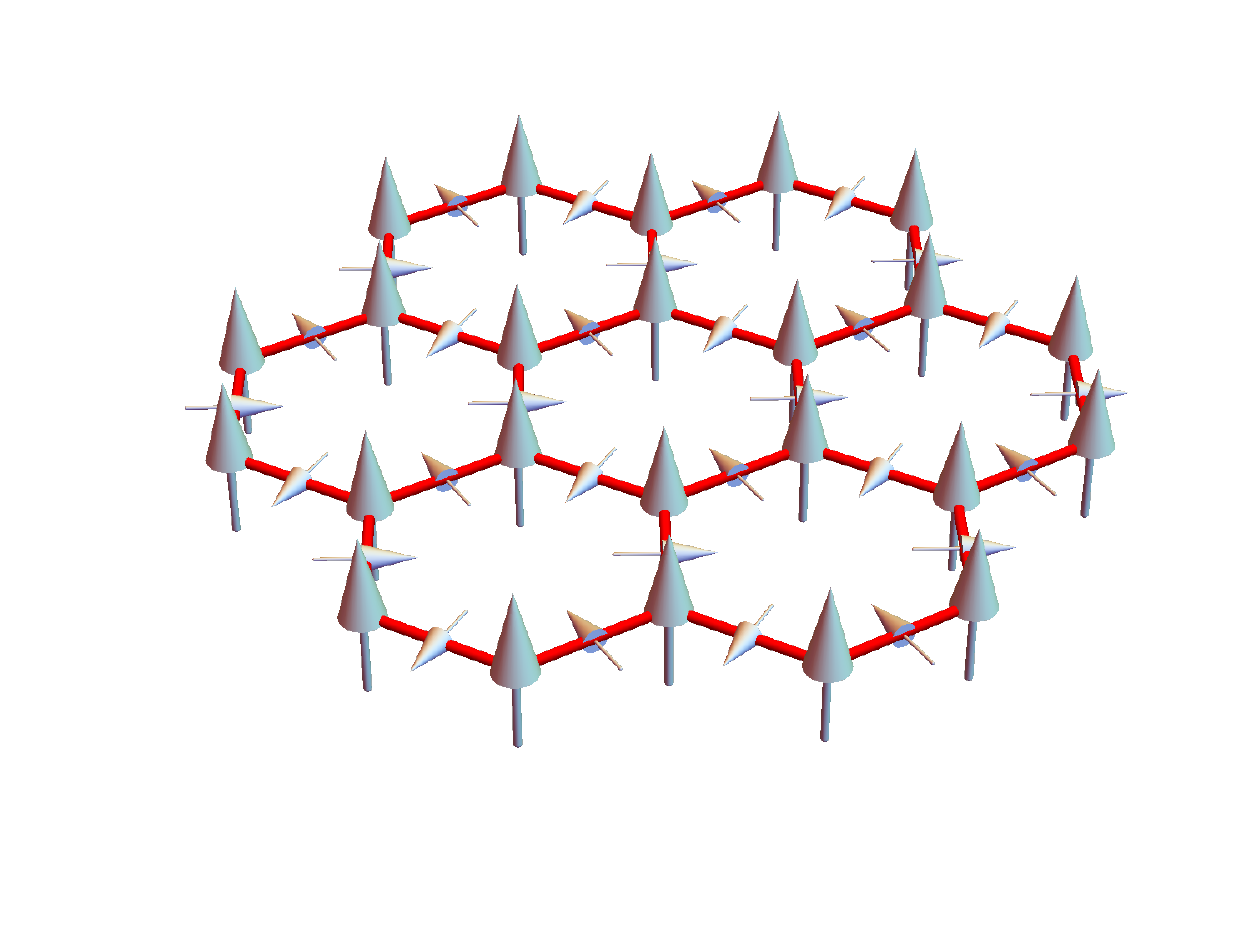} }}%
\caption{
Honeycomb tiling: (a) with different coloured bonds indicating the pattern of distinct Ising couplings of the Kitaev exchange, (b) showing the second neighbor bonds and arrows indicating the direction of the DMI vector where the bond orientation is always anti-clockwise on the triangles within each hexagon, (c) showing the DMI vectors for the nearest neighbor DMI for a single honeycomb layer on a substrate where, here, the bond orientation is from the $A$ to the $B$ sites.
}
\label{fig:honeycomb}
\end{figure}

\section{Anisotropic Honeycomb Lattice Models}
\label{sec:honeycomb}

We turn now to the honeycomb lattice in 2D with anisotropic exchange couplings of different types as in Fig.~\ref{fig:honeycomb} and use $k\cdot p$ arguments to understand the role of spin space symmetries in the protection of Dirac points and how these can be gapped out leading to topological bands. 

The first model we consider is the much-studied nearest neighbor Kitaev-Heisenberg model. The relevant point group under which the exchange is left invariant is $D_{3d}$ with nontrivial symmetries including three-fold rotations about the hexagonal centers, three two-fold rotational symmetries about axes through opposite hexagonal vertices and inversion symmetry again about the hexagonal centers. In addition there is the pure spin space $D_2$ symmetry discussed above and time reversal symmetry. 

The Kitaev-Heisenberg model has a rich semi-classical phase diagram in an applied magnetic field \cite{janssen2016honeycomb}. In the polarized phase $-$ both the zero field ferromagnet and the continuously connected field-polarized regime $-$ the two magnon bands are connected at Dirac points when the field direction is along meridians $[xy0]$, $[x0z]$ and $[0yz]$ relative to the Cartesian frame that defines the Kitaev Ising directions. For other field directions, there is an energy gap between the magnon bands and they carry Chern number $\pm 1$ \cite{mcclarty2018,joshi2018}. The sign of the Chern number swaps between the bands when crossing one of the meridians. These facts can be understood on the basis of the spin-space symmetries of the magnetically ordered configurations. 

As a reminder, the stability of Dirac points in honeycomb tight-binding models follows from the presence of time-reversal symmetry $\Ts$ and inversion $\Ps$. Inversion swaps sublattices and takes $\bs{k}\rightarrow -\bs{k}$. A possible time reversal operation is complex conjugation and $\bs{k}\rightarrow -\bs{k}$. The effective model in the vicinity of the Dirac point is:
\beq
H_{\bs{k}} = d_x(\bs{k}) \sigma_x + d_y(\bs{k}) \sigma_y + d_z(\bs{k}) \sigma_z 
\eeq
that maps to
\beq
H_{\bs{k}} = d_x(\bs{k}) \sigma_x + d_y(\bs{k}) \sigma_y - d_z(\bs{k}) \sigma_z 
\eeq
under $\Ps\Ts$ so the mass term must vanish if $\Ps\Ts$ is a symmetry. Furthermore, the Dirac points are fixed at the $K$ and $K'$ points if $C_{3z}$ is also a symmetry.

Consider the highly symmetric situation where the moment points perpendicular to the honeycomb plane $-$ the $[111]$ direction. The symmetries are then $E$, $C_3$, $C'_2\Ts$, and $\Ps$. $C'_2$ is a lattice symmetry but it reverses the moment direction which can be restored under the action of $\Ts$. Note that there are no spin space symmetries. The only anti-unitary symmetries are combined with lattice transformations and, in particular, the combination of one of these elements with inversion does not map $\bs{k}$ to itself. The result is there is no constraint that the gap close between the magnon bands and since pure time reversal is broken, although inversion is present, the Berry curvature may be non-vanishing. The model therefore lies in the bosonic Altland-Zirnbauer class $A$ \cite{xuthreefold} and the bands may therefore be topologically nontrivial. 

Now consider the case where moments are polarized along the meridian $[xy0]$. The symmetry elements are $E$, $\Ts C_2^z(\sv)$, $\Ps$ and products of these. As in the case of $[111]$ moments, the time reversal operator is a composite with a $C_2$ symmetry. However, in this case, the $C_2$ is a pure spin space transformation that, therefore, does not transform the momentum. The presence of inversion symmetry and the $\Ts C_2^z(\sv)$ are sufficient to forbid the mass term so Dirac points are present albeit not at the $K$, $K'$ points owing to the absence of $C_3$. It follows that there are Dirac points whenever the field is aligned along any one of the cubic meridians with one spin coordinate vanishing.

As an aside, the $[100]$ moment direction has Dirac points but also more symmetry than a general point along the cubic meridians. These elements include $E$, $C_2^x(\sv)$, $\Ts C_2^y(\sv)$, $\Ts C_2^z(\sv)$, $C^{'x}_2\Ts$ and $\Ps$.

For a general moment direction $[xyz]$ with none of these vanishing, the only symmetries are $E$, $\Ps$ so the symmetry constraints are not sufficient to close the gap between the magnon bands.

Spin-space symmetry is also important to understand magnetic models with Dzyaloshinskii-Moriya exchange where the $\bs{D}$ vector in $\bs{D}\cdot\left(\hat{\bs{J}}_i\times\hat{\bs{J}}_j\right)$ is collinear. In this case, there is a continuous spin space rotation symmetry about an axis parallel to $\bs{D}$. We now consider the honeycomb ferromagnet with second nearest neighbor Dzyaloshinskii-Moriya exchange with $\bs{D}$ perpendicular to the plane. When the moments are polarized along $\bs{D}$, the only manifestation of the spin-space symmetry group is the $C^z_{\infty}$ that simply adds a quantum number $\pm 1$ to the magnons as discussed for the case of Heisenberg exchange. The remaining symmetries are identical to those of the Heisenberg-Kitaev model as discussed above so there is a gap between the magnon bands and nonzero Berry curvature. In fact, once again the magnon bands have nonzero Chern number. 

If, instead, the moments are aligned in the honeycomb plane, all the honeycomb lattice symmetries are present once we allow for rotations in spin space around the $\bs{D}$ axis. There is, in addition, a $\Ts C_2^z(\sv)$ symmetry that, taken together with the inversion symmetry, ensures the presence of Dirac points that now, owing to the $C_3^z$, are located at the $K$, $K'$ points.

We contrast the model with second neighbor Dzyaloshinskii-Moriya exchange and parallel $\bs{D}$ vector with the model with nearest neighbor Dzyaloshinskii-Moriya with in-plane $\bs{D}$ that may be present for single layer honeycomb magnets grown or mounted on a substrate. In this case, there are no spin space symmetries and, with moments perpendicular to the plane, the only nontrivial symmetry is $C^z_3$. The two bands are therefore gapped and topological. As noted in Ref.~\cite{mook2020interactionstabilized}, linear spin wave theory fails to capture this feature because there are no $O(S)$ quadratic terms coming from the antisymmetric $S^{\pm}_i S^z_j$ couplings. In this case, symmetry breaking due to the presence of the $\bs{D}$ coupling comes from higher order corrections.  

\section{Nodal Lines, Weyl Points, Spin-Space Groups and Magnetic Space Groups}
\label{sec:hyperhoneycombfm}

We now return to the case of the Kitaev-Heisenberg model on the hyperhoneycomb lattice. In Section~\ref{sec:hyperhoneycomb} we identified the N\'{e}el phase of this model as having nontrivial spin-space group symmetry with dramatic consequences for the magnon band structure including a nodal plane, four-fold degenerate point and an abundance of nodal lines. We now consider the collinear ferromagnetic phase. This offers some important lessons about spin-space symmetry groups for magnons and their effect on band topology. For, as we shall see, this phase makes accessible a tower of different spin-space symmetry groups, by tuning the direction of the moments, with a variety of properties. For the several spin-space groups relevant to this model, we present the features of the band structures that are imposed by symmetry, including Weyl points and nodal lines. For comparison we also report the features expected on the grounds of invariance of the magnetic structure thereby giving several examples of how spin-space symmetry leads to richer band structures. We shall also see that spin-space groups may be isomorphic to ordinary magnetic space groups including type II groups $-$ of the form $\bs{G}+\Ts\bs{G}$ $-$ that ordinarily would not describe magnetically ordered systems. 

\begin{figure}[!htb]
\hbox to \linewidth{ \hss
\includegraphics[width=0.99\linewidth]{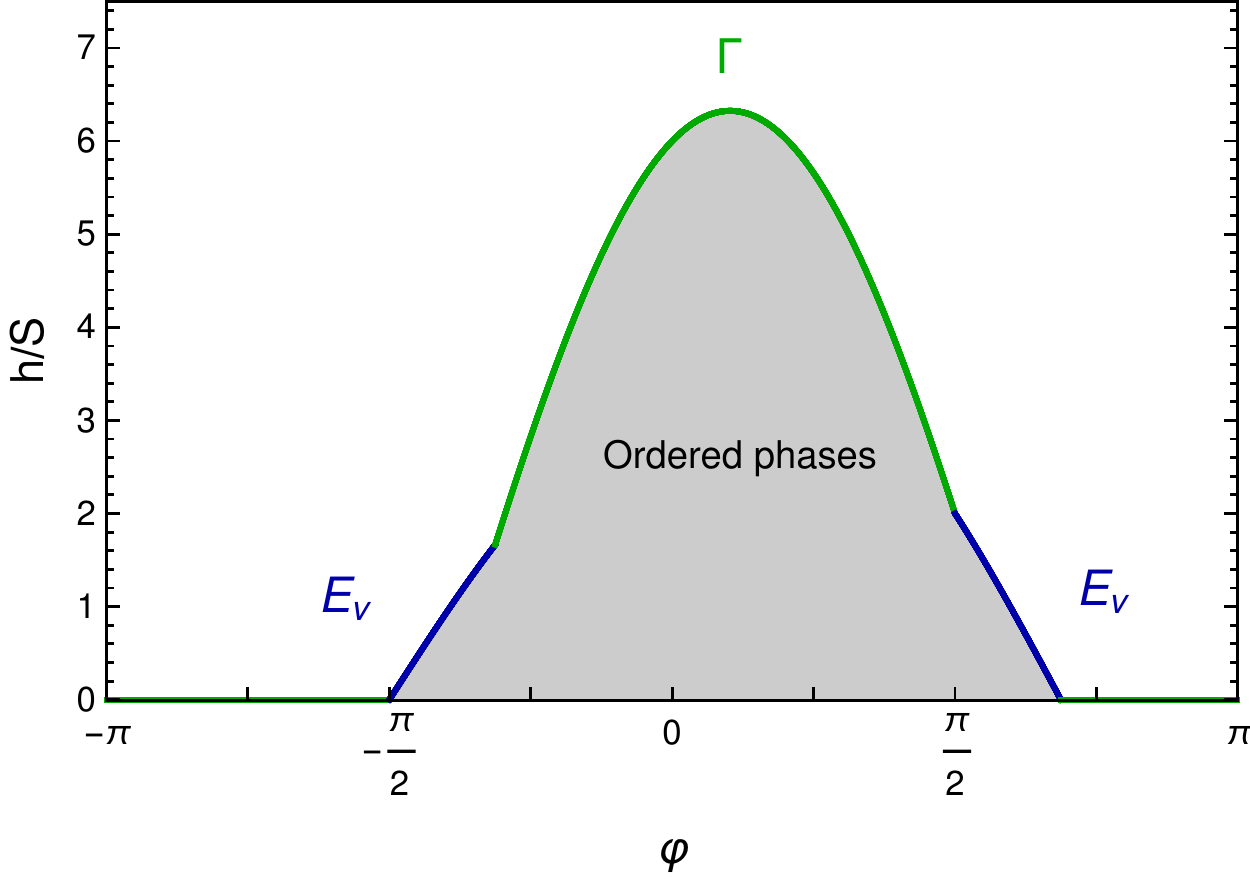}
\hss}
\caption{
Phase diagram of the Kitaev-Heisenberg model in a $[111]$ field. The phase boundary is obtained from the condensation of magnons in linear spin wave theory about the collinear field-polarized ferromagnetic state. The condensation wavevector is indicated. The wavevector $E_v = (-\frac{2}{3}, 0, 0)$ is the one characteristic of the vortex and AF vortex phase, as in Ref.~\cite{Kruger2020}, which is indeed associated with a second order phase transition. 
}
\label{fig:PD111}
\end{figure}

To begin, we add a Zeeman term to the Hamiltonian, Eq.~\ref{eq:KHModel}. At sufficiently large fields, the moments form a collinear ferromagnetic state and both the symmetry group associated with the magnetic order and the extent of the polarized phase depend on the field direction. For the $[111]$ direction, Fig.~\ref{fig:PD111} shows the stability region of the field-polarized ferromagnet as obtained from the condensation of magnons within linear spin wave theory that also gives the ordering wavevectors indicated. One observes that the polarized phase is continuously connected to the zero field ferromagnetic phase over a broad swathe of $\varphi$. The phase diagram is qualitatively similar for all field directions. 

\begin{figure}[htbp!]%
\centering
\subfloat{{\includegraphics[width=0.99\columnwidth]{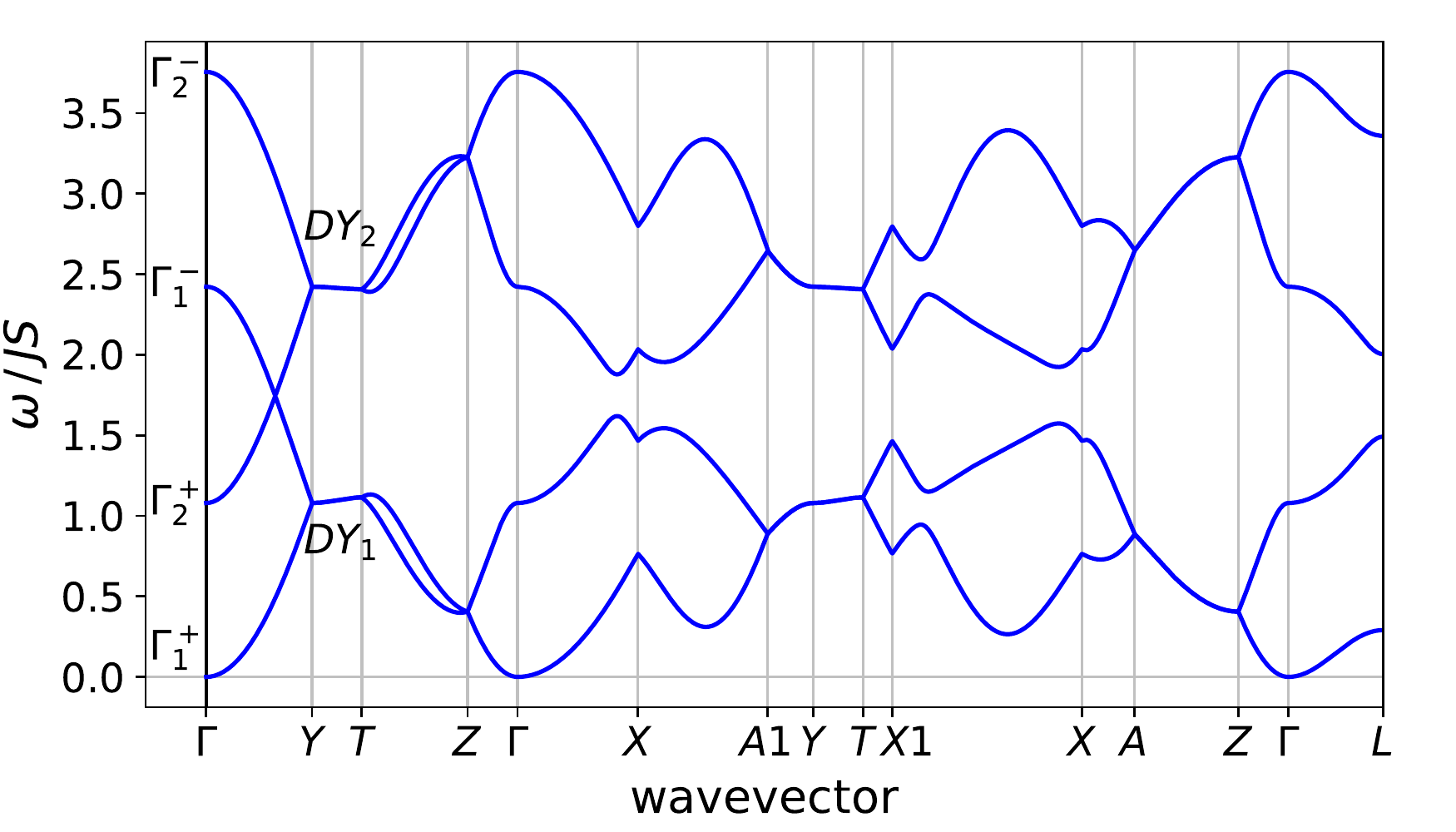} }}%
\qquad
\subfloat[$\phi = -0.6\pi$]{{\includegraphics[width=0.5\columnwidth]{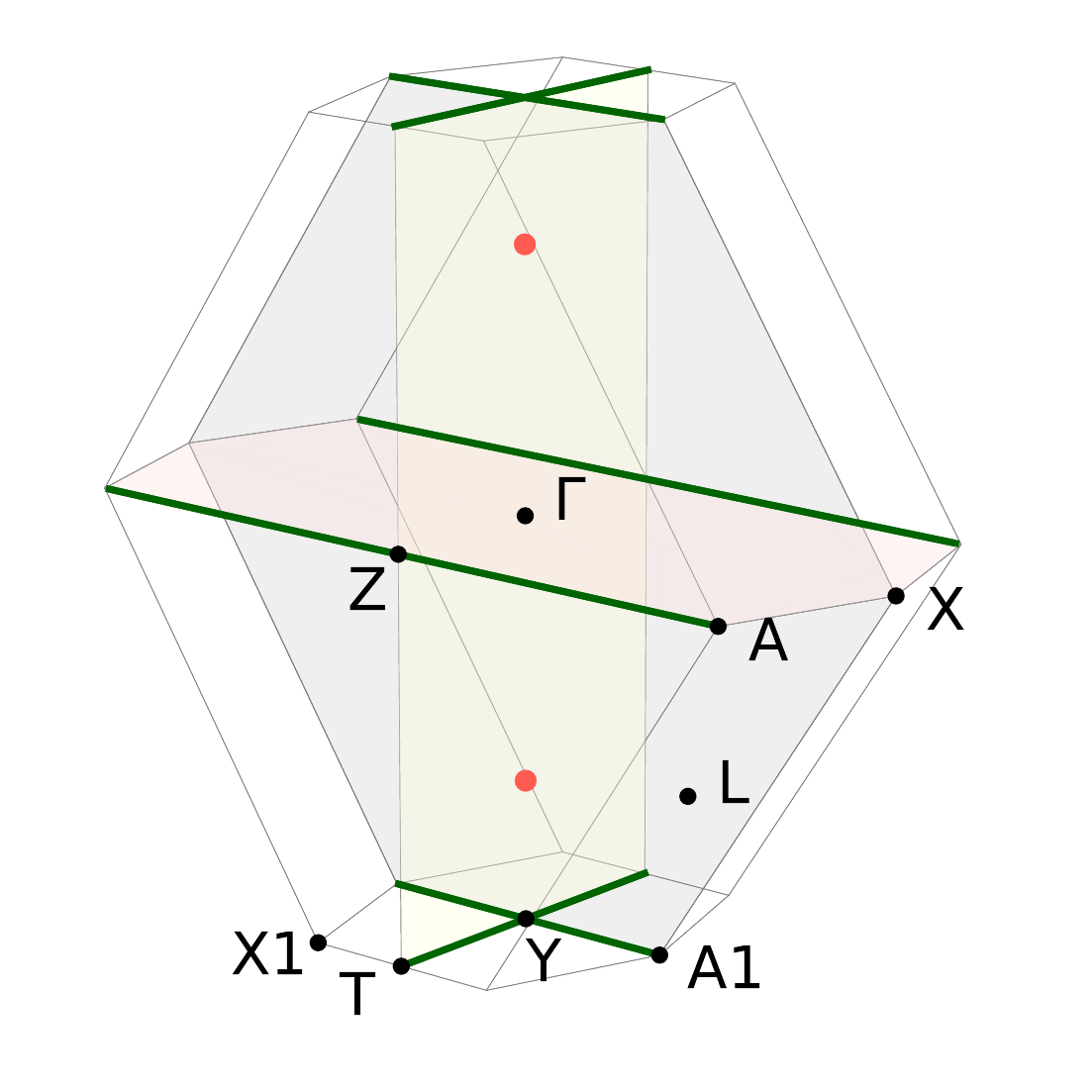} }}%
\subfloat[$\phi = -0.4\pi$ ]{{\includegraphics[width=0.5\columnwidth]{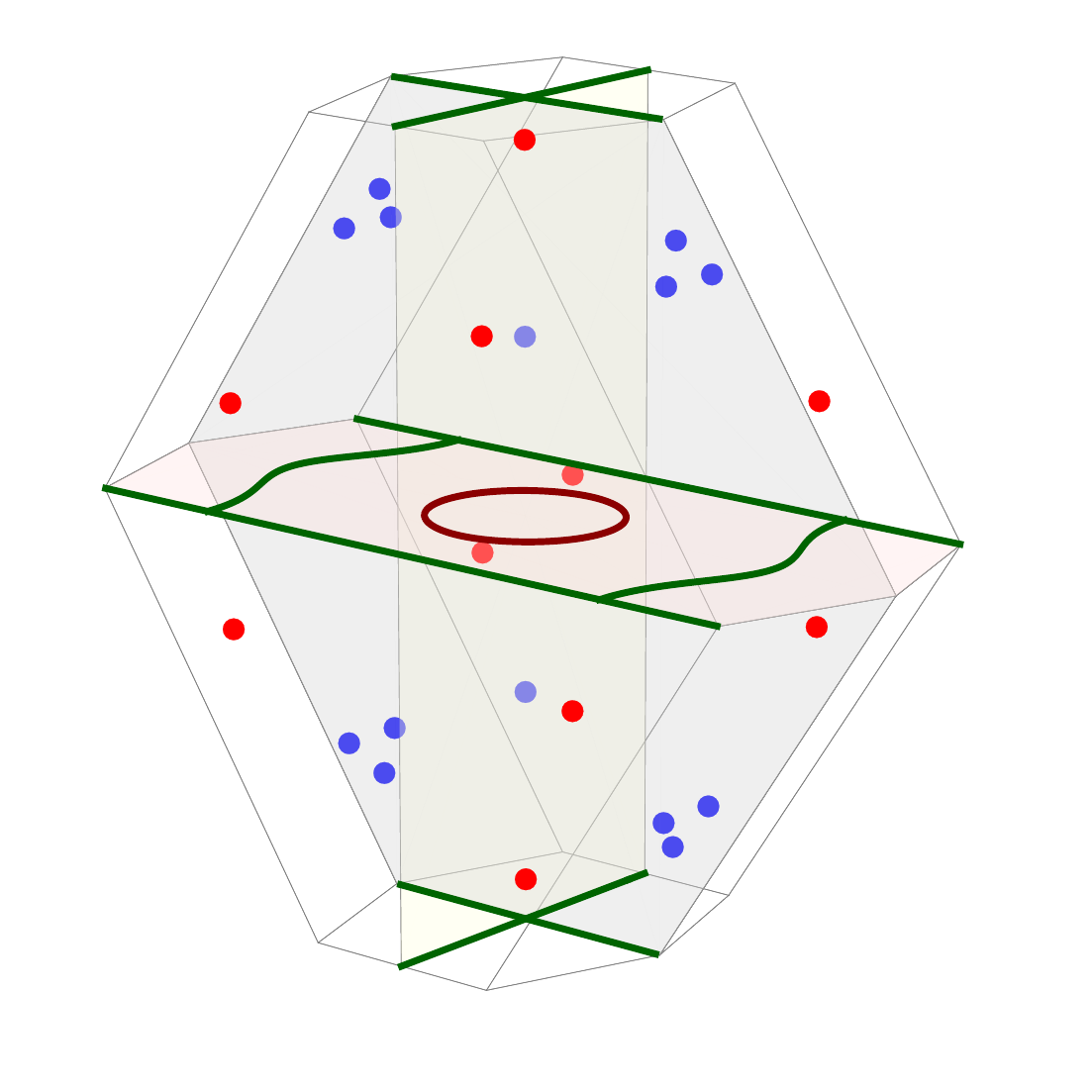} }}%
\caption{
Example of a field direction without $\Ps \Ts$ in the magnetic group isomorphism, therefore allowing Weyl points.
Magnon band structure in $[111]$ field and $\varphi=-3\pi/5$ along high symmetry paths in the Brillouin zone (top). Here the irrep labels are marked for points $\Gamma$ and $Y$, showing how the Weyl node along $\Gamma Y$ arises from compatibility relations.
On the bottom the degeneracies of the band structure are shown for case $\varphi=-3\pi/5$ (left) and $\varphi=-2\pi/5$ (right).
The green lines are nodal lines between bands $(1,2)$ and/or $(3,4)$, while red lines are for nodal lines between $(2,3)$. The colored point are Weyl points, between $(1,2)$ (blue) and $(2,3)$ (red). 
The case $\varphi=-3\pi/5$ has a stable Weyl point along $\Gamma Y$. Indeed moving to $\varphi=-2\pi/5$ this point is still present (closer to $Y$). At this fine-tuned point in parameter space, there is a proliferation of other  Weyl points. There are also new nodal lines that are protected by the glide $d_1$ (not by $\Ps \Ts$) and which are therefore constrained to the mirror plane.
}
\label{fig:magnons111}
\end{figure}

For this case, Fig.~\ref{fig:magnons111} shows the magnon band structure along high symmetry directions in the Brillouin zone for $\varphi=-3\pi/5$ and zero applied field showing Goldstone modes with quadratic dispersion at $\Gamma$. At finite field, the spectrum is completely gapped and details of the band structure change. However, some features are robust to changes of parameter $\varphi$ and the magnitude of the field: these are the Weyl point along the line $\Gamma Y$ and the double degeneracy of the lower and upper magnon bands along zone boundary lines: $AZ$, $YT$ and $YA_1$. These nodal lines cross at point $Y$.

We now examine the magnon band structure from the point of view of symmetry. Once again, the spin space part of the group $\mathbf{G}_H$ consists of $SO(3)$ for the Heisenberg coupling alone. For the Kitaev coupling, the exchange on the $x$ bond is
\beq
\begin{pmatrix} 
K     & 0     & 0     \\
0     & 0     & 0     \\
0     & 0     & 0 
\end{pmatrix} 
\eeq
and the transformations that leave this invariant are the axial rotations $C_{\infty}^x$ and perpendicular rotation $C_2^{[0yz]} $ that together make up $D_{\infty}$. Taking the three inequivalent bonds together leaves the symmetry elements $\{ C_2^x, \, C_2^y,  \, C_2^z \} = D_2$. The paramagnetic parent group is $\mathbf{G}_H= (\mathbf{G}+\Ts \mathbf{G})\otimes D_2$. 

We now find the subgroup $\mathbf{G}_M \leq \mathbf{G}_H$ that leaves the collinear ferromagnetic structure invariant using separate allowed real space and spin space transformations for the $[11z]$ direction with $z\neq 0$. This turns out to be
\begin{align}
\mathbf{G}_{M} = & \, E + \Ps +C_2^z(\sv) (d_1 + C_2^{\bs{b}}) \nonumber  \\
                             &+ \Ts \, (d_2 + C_2^{\bs{a}}+ C_2^z(\sv) (d_3 +  C_2^{\bs{c}}))
\end{align}
The group contains distinct spin space and real space transformations as well as anti-unitary elements. However, the group is isomorphic to the magnetic space group $Fd'd'd$ as we shall discuss in more detail below.

Now, in contrast, if we lock the spin and space transformations and determine the symmetries of the magnetic structure $-$ in other words those elements of the group $\mathbf{G}+\Ts \mathbf{G}$ that leave the magnetic structure invariant $-$ we find
\beq
\mathbf{G}_{M} = \, E + \Ps + \Ts d_2 +  \Ts C_2^{\bs{a}} = (E + \Ps) +  \Ts d_2 (E + \Ps)
\eeq
which is the magnetic space group $C2'/c'$. 

The magnetic symmetry of the magnons is therefore higher than that of the underlying magnetic structure owing to the freedom to perform $D_2$ transformations in spin space. This enhanced symmetry has important consequences for the magnon band structure and band topology. In particular, one can show that the lower symmetry magnetic group $C2'/c'$ enforces degeneracy along line $ZA$ but not along other directions. But when the full symmetry group is taken into account, as we show in Appendix~\ref{app:111FM}, all the aforementioned observed magnon band degeneracies are enforced by symmetry $-$ the symmetry guarantees the presence of nodal lines. In addition to arguments based on representation theory we give some more direct justification of zone boundary nodal line degeneracies for the $[111]$ hyperhoneycomb ferromagnet protected by magnetic glides in Appendix~\ref{app:111FMInformal}. 

In a similar way, the spin-space group may be found for the remaining symmetry-distinct field directions. Table~\ref{tab:HHEnhancedMagnGroupHK} summarizes the results for the collinear ferromagnet and for three zero field antiferromagnetic phases $-$ simple N\'{e}el order, the skew-stripy phase and the skew-zigzag phase. For the ferromagnetic phase, there are $10$ different groups. One of these is the Heisenberg ferromagnet for which the group is independent of the field direction. The remainder are for the Kitaev-Heisenberg model for different high symmetry directions and the general direction $[xyz]$. 

\begin{figure}[htbp!]%
\centering
\subfloat[Heisenberg ]{{\includegraphics[width=0.5\columnwidth]{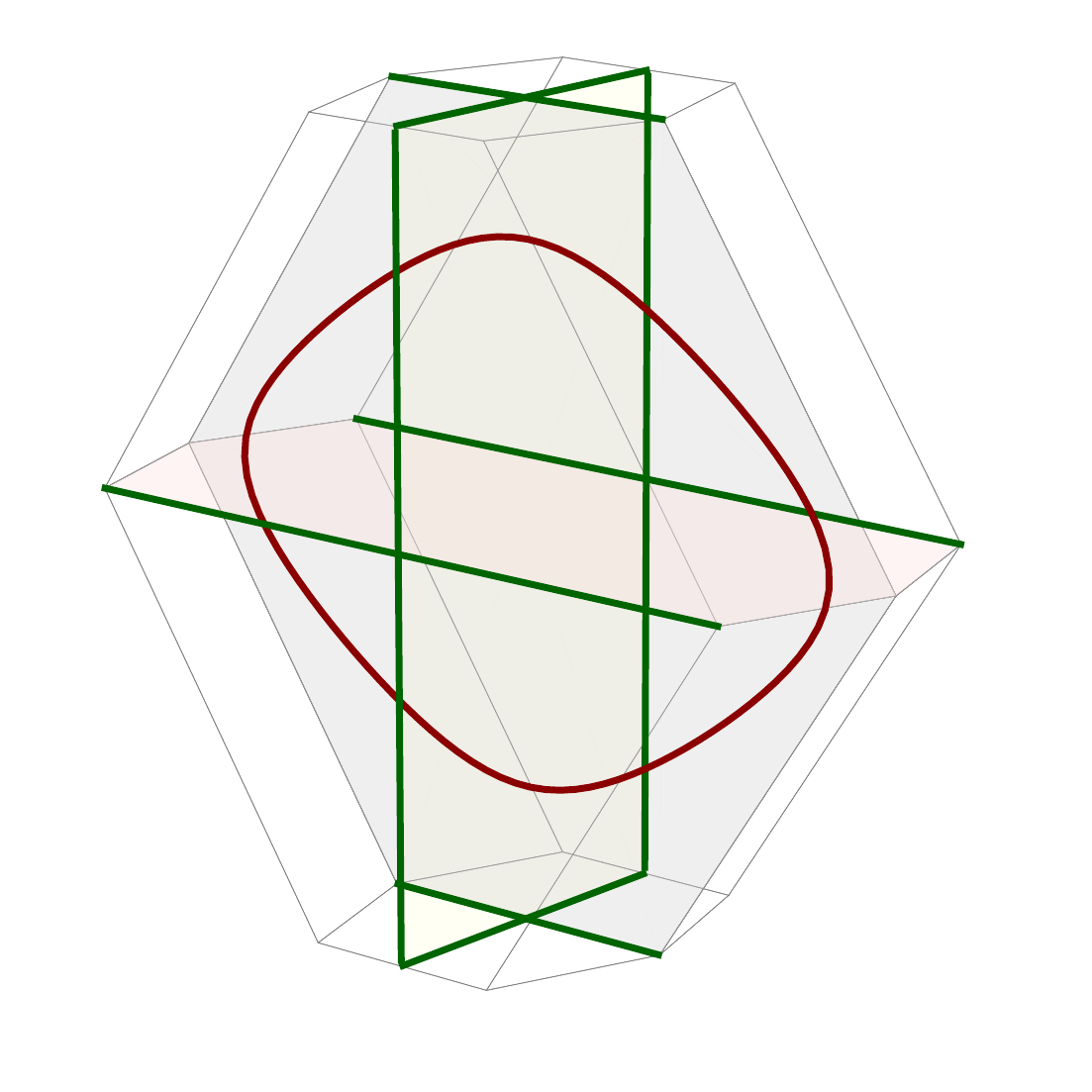} }}%
\subfloat[FM $\text{[}x0z\text{]}$]{{\includegraphics[width=0.5\columnwidth]{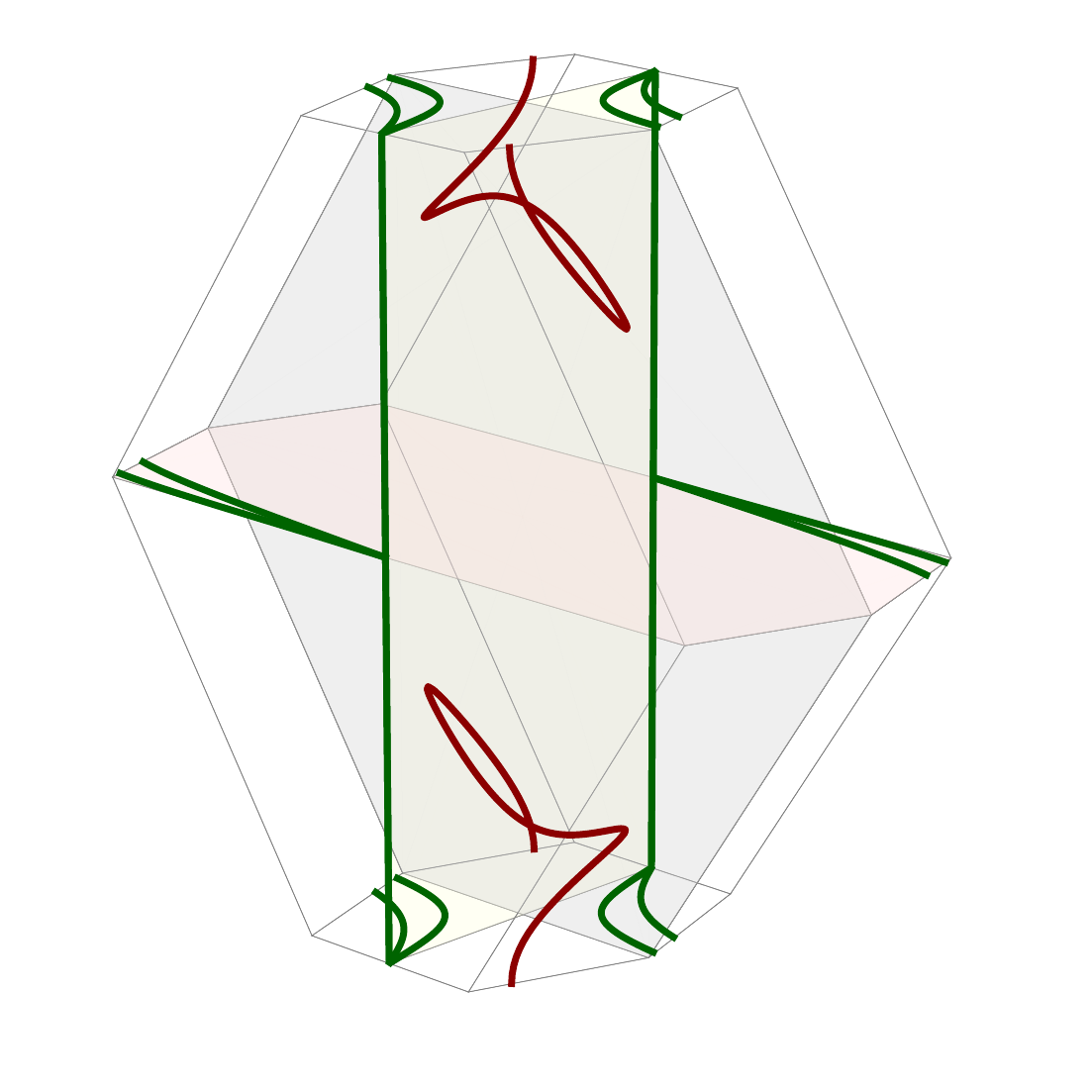} }}%
\caption{
Example of field directions with $\Ps \Ts$ in the isomorphic magnetic space group with, therefore, nodal lines allowed everywhere in the zone.
The degeneracies of the band structure are shown for the Heisenberg case (left) and for field direction $[x0z]$ with $\varphi=-3\pi/5$ (right).
The green lines are nodal lines between bands $(1,2)$ and/or $(3,4)$, while red lines are nodal lines between $(2,3)$. 
The highly symmetric Heisenberg case has all possible enforced degeneracy boundary zone lines (green) and the characteristic nodal line (red).
The $[x0z]$ has a much lower symmetry, but still exhibits a nodal line away from symmetric positions in the BZ, since it has pure time reversal symmetry in the isomorphic magnetic group, exactly as in the Heisenberg case, see Table \ref{tab:HHEnhancedMagnGroupHK}.
}
\label{fig:DiracLinesFM}
\end{figure}

There are several observations to make about the groups listed in the table. First of all, because the lattice has inversion symmetry and the moments are invariant under this operation, all the ferromagnetic spin-space groups have inversion symmetry. They also all have spin-space elements and non-symmorphic elements. Inspection of the anti-unitary part of the group reveals that, for a random magnetic field direction, the group has no magnetic elements. Then, of the more symmetric field directions, two ($[\pm1 \pm 1 z]$ and $[\pm1 \mp 1 z]$) have time reversal multiplying a glide and the rest have time reversal multiplying an element that acts on spin space. As we saw when discussing honeycomb lattice models, the $\Ts C_2(\bs{s})$ elements behave like pure time reversal symmetry because they simply reverse the sign of $\bs{k}$. Since inversion is also present, these cases have $\mathcal{PT}$ symmetry. Fig.~\ref{fig:DiracLinesFM} shows the degeneracies in magnon band structures in instances where there is effective $\mathcal{PT}$. Both have nodal lines in the spectra protected by $\mathcal{PT}$ with the Heisenberg case (in the left panel) having various zone boundary nodal lines and one pinned to a bisecting plane in the Brillouin zone. The less symmetric $[x0z]$ case (right panel) has interior nodal lines that are not pinned.

\begin{figure}[!htb]
\hbox to \linewidth{ \hss
\includegraphics[width=0.99\linewidth]{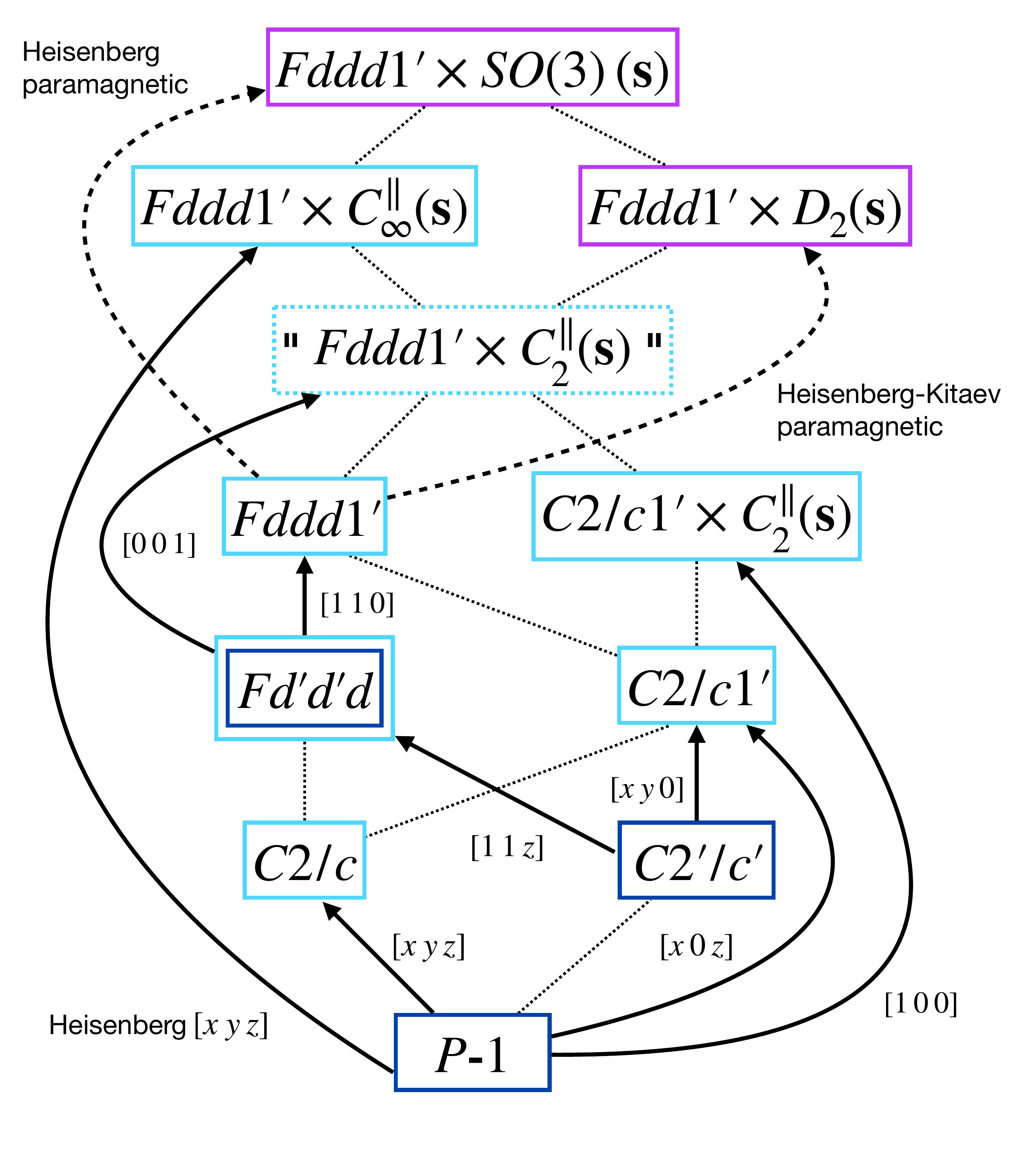}
\hss}
\caption{
The hierarchy of spin-space groups for the Heisenberg-Kitaev ferromagnet on the hyperhoneycomb lattice. At the top, there are the paramagnetic parent groups of Heisenberg and Heisenberg-Kitaev coupling (magenta) and, below, all the relevant subgroups for different ferromagnetic directions. The paramagnetic spin-space group is enhanced from the normal grey group $Fddd1'$ (dashed arrow).  For each ferromagnetic case, the magnetic group of the ground state is given (dark blue) and its enhancement (arrow) by the Hamiltonian spin-space symmetries to a group relevant for the magnon spectra (cyan). The ground state symmetries are enhanced in all cases as a result of the parent group pure spin-space symmetry. For all groups we have used magnetic space group notation. For most cases, the spin-space group is isomorphic to a magnetic space group. The only exception is the $[001]$ case. However, for $[001]$, the group $Fddd1' \times C_2^{\parallel}(\sv)$ (dotted cyan) nevertheless correctly captures the magnon band degeneracies. 
}
\label{fig:MagnGroupHierarchy}
\end{figure}

A summary of the groups that appear by rotating the field direction is given in Fig.~\ref{fig:MagnGroupHierarchy}. This plot shows the groups  arranged by increasing symmetry with the most symmetric group $-$ that of the Heisenberg Hamiltonian $-$ at the top. The decoupled spin group $SO(3)$ for the Heisenberg exchange is broken down to $D_2$ by the Kitaev coupling. The figure includes the groups that leave the magnetic structure invariant without spin-space transformations allowed by the Hamiltonian (dark blue). The figure also includes those groups that leave the magnetic structure invariant up to spin-space transformations (cyan) and these are all indexed using isomorphic magnetic space group notation. The arrows connecting different groups reveal the symmetry enhancement in going from dark blue to cyan groups.

Table~\ref{tab:HHEnhancedMagnGroupHK} illustrates that, for the ferromagnet, the spin-space groups are isomorphic to magnetic space groups except for the $[100]$ moment direction. In contrast, the spin-space groups for the antiferromagnetic phases do not coincide with any of the magnetic space groups. To understand these coincidences better, we first note that for collinear ferromagnets the only possible net spin transformations are axial rotations. These commute with one another and, when they match the real space point group transformation, the group multiplication table matches that of a magnetic space group simply because the spin space transformation is not distinct.  

For example the FM $[\pm 1 \,\pm 1\,\, 0]$ has unitary elements:
\beq
\mathbf{H}_{\rm SS} = E + \Ps + d_3 + C_2^{\cv} + C_2^z(\sv) \, (d_1 + d_2 + C_2^{\av} + C_2^{\bv}) .
\eeq
Elements like $C_2^z(\sv)C_2^{\bs{c}} = \sselement{2_{100}}{2_{001}}{-1/4,-1/4,0 }$ can be mapped to real space symmetry only elements $\sselement{E}{2_{001}}{-1/4,-1/4,0 }$ preserving the
multiplication table of the group. The isomorphic group is therefore an usual magnetic group with unitary elements:
\beq
\mathbf{H}_{ISO} = E + \Ps + d_3 + C_2^{\cv} + d_1 + d_2 + C_2^{\av} + C_2^{\bv},
\eeq
which corresponds to the unitary part of $Fddd1'$. The $[\pm 1 \,\pm 1\,\, 0]$ spin-space group, despite being isomorphic to a normal magnetic space group, has an enhancement with respect to the ground state magnetic group $Fd'd'd$ with fewer unitary elements:
\beq
\mathbf{H}_{GS} = E + \Ps + d_3 + C_2^{\cv}
\eeq
We further note that even where there is an isomorphism to a magnetic space group it is important to keep track of the spin space transformations in representation computations.  Secondly, not all ferromagnetic spin space groups are isomorphic to magnetic space groups. For example, for the direction $[0 \,0 \, \pm 1]$, if we look for a magnetic space group that is closest in form to the spin-space group, we would find $Fddd1'\times C_2^{\parallel}(\sv)$. But if we inspect the elements of the spin-space group we find $C_2^x(\sv)C_2^{\bs{c}} = \sselement{4_{010}^+}{2_{001}}{-1/4,-1/4,0 }$ which does not respect the multiplication rule of the $Fddd1'$ group (trivially the square of the element is not identity).
Even so, $Fddd1'\times C_2^{\parallel}(\sv)$ {\it does} correctly capture the band structure degeneracies. This is because in this specific case, the  degeneracies are restricted to the Brillouin zone boundary, and there the central extension groups of the two groups are isomorphic because only spatial elements are relevant.
For AFM cases, spin space elements may include rotations about axes  perpendicular to the moment. Since these generally do not  commute with other spin space operators one tends to find different multiplication tables to the usual magnetic space groups. 

So far we have commented on the types of spin-space groups that can arise in the Kitaev-Heisenberg ferromagnet. We now examine the consequences of these groups for the magnon bands structures. Table~\ref{tab:DegeneracyEnhancedHK} summarizes the symmetry enforced degeneracies for all high symmetry moment directions for the hyperhoneycomb Kitaev-Heisenberg model deduced from the representation theory of the groups given in Table~\ref{tab:HHEnhancedMagnGroupHK} and, in particular, both the groups expected for the ground state alone and the full spin-space group. The spin-space group with its enhanced symmetry leads always to a more degenerate spectrum except for the $[xy0]$ case.

As we have already observed, some of the spin-space groups have a time reversal element multiplying a pure spin space element. This is isomorphic to a pure time reversal element and the system then has $\mathcal{PT}$ symmetry that imposes a reality condition on the Hamiltonian that protects nodal lines and forbids the presence of Weyl points as in Fig.~\ref{fig:DiracLinesFM}. In the other cases ($[xyz]$ and $[11z]$) we have the opposite scenario, with presence of Weyl nodes and no closed nodal lines (apart from the nodal line protected by glide symmetry constrained to its mirror plane) as in Fig.~\ref{fig:magnons111}. In the AFM cases, since the symmetry is highly enhanced, the presence of closed nodal lines is the norm, and, as we discussed, nodal planes may also arise.

\begin{table*}[!htb]
\centering
\hbox to \linewidth{ \hss
\begin{tabular}{ |c|p{10.0cm}|p{3.5cm}|p{2.9cm}|   }
 \hline
 \multicolumn{1}{|c|}{Spin direction}  & Representative of the unitary subgroup  $\mathbf{H}_{\rm SS}$ & Spin-Space Group $\mathbf{G}_{\rm SS}$  & Isomorphism \\ 
 \hline
 $ \text{Heisen. FM } [x \,y\, z]  $        &  $(E + \Ps + d_1 + C_2^{\bv}  + d_2 + C_2^{\av} + d_3 + C_2^{\cv}) (\rv) + C_{\infty}^{\parallel}(\sv)$         & $\mathbf{H}_{\rm SS} + \Ts C_2^{\perp}(\sv)  ~  \mathbf{H}_{\rm SS}$  & ~~$ \cong Fddd1'  \times C_{\infty}^{\parallel}(\sv)$   \\
 $ \text{Heisen. N\'{e}el } [x \,y\, z]  $        & $(E + d_1 + d_2 + C_2^{\cv})(\rv) + C_2^{\perp}(\sv) \, [(\Ps + d_3 +  C_2^{\bv} +  C_2^{\cv})(\rv)] + C_{\infty}^{\parallel}(\sv)$          & $\mathbf{H}_{\rm SS} + \Ts C_2^{\perp}(\sv)  ~  \mathbf{H}_{\rm SS}$ & ~~ -    \\
 $[x \,y\, z] $         &  $E + \Ps + C_2^z(\sv) \, ( d_1 + C_2^{\bv}) $         & $\mathbf{H}_{\rm SS}$ & ~ $\cong C2/c$   \\
 $[x \,0\, z], [0 \,y\, z]  $         &  $E + \Ps + C_2^z(\sv) \, ( d_1 + C_2^{\bv}) $         & $\mathbf{H}_{\rm SS} +  \Ts C_2^{y/x}(\sv) ~ \mathbf{H}_{\rm SS}$ & ~ $ \cong C2/c1'$     \\
 $ [x \,y\, 0]  $        &  $E + \Ps + C_2^z(\sv) \, ( d_1 + C_2^{\bv}) $         & $\mathbf{H}_{\rm SS} +  \Ts C_2^z(\sv)  ~\mathbf{H}_{\rm SS}$ & ~ $ \cong C2/c1'$     \\
 $[\pm 1 \,0\, 0], [0 \, \pm 1\, 0]  $         &  $E + \Ps + C_2^z(\sv) \, ( d_1 + C_2^{\bv}) + C_2^{x/y}(\sv) $         & $\mathbf{H}_{\rm SS} +  \Ts C_2^{z}(\sv) ~ \mathbf{H}_{\rm SS}$ & ~ $ \cong C2/c1'  \times C_2^{\parallel}(\sv)$     \\
 $ [\pm 1 \,\pm 1\,\, z]  $        &  $E + \Ps + C_2^z(\sv) \, ( d_1 + C_2^{\bv}) $         & $\mathbf{H}_{\rm SS} +   \Ts d_2 ~ \mathbf{H}_{\rm SS}$ & ~ $ \cong Fd'd'd$    \\
 $ [\pm 1 \,\mp 1\,\, z]  $        &  $E + \Ps + C_2^z(\sv) \, ( d_1 + C_2^{\bv}) $         & $\mathbf{H}_{\rm SS} +   \Ts d_3 ~ \mathbf{H}_{\rm SS}$ & ~ $ \cong Fd'd'd$    \\
 $ [\pm 1 \,\pm 1\,\, 0]  $         &  $E + \Ps + d_3 + C_2^{\cv} + C_2^z(\sv) \, (d_1 + d_2 + C_2^{\av} + C_2^{\bv}) $       & $\mathbf{H}_{\rm SS} + \Ts C_2^z(\sv)  ~  \mathbf{H}_{\rm SS}$ & ~ $ \cong Fddd1'$    \\
 $ [\pm 1 \,\mp 1\,\, 0]  $         &  $E + \Ps + d_2 + C_2^{\av} + C_2^z(\sv) \, (d_1 + d_3 + C_2^{\bv} + C_2^{\cv}) $         & $\mathbf{H}_{\rm SS} +  \Ts C_2^z(\sv) ~ \mathbf{H}_{\rm SS}$ & ~ $ \cong Fddd1'$    \\
 $ [0 \,0 \, \pm 1]  $        &  $E + \Ps + d_1 + C_2^{\bv} + C_2^{x,y}(\sv) \, (d_2 + d_3 + C_2^{\av} + C_2^{\cv}) + C_2^{z}(\sv) $         & $\mathbf{H}_{\rm SS} + \Ts C_2^{x,y}(\sv)  ~  \mathbf{H}_{\rm SS}$ & $\cong "Fddd1' \times C_2^{\parallel}(\sv)"$    \\
 $ \text{N\'{e}el } [0 \,0 \, \pm 1]  $        &  $E + d_1 + d_3 + C_2^{\av} + C_2^{x,y}(\sv) \, (\Ps + d_2 + C_2^{\bv} + C_2^{\cv}) + C_2^{z}(\sv) $         & $\mathbf{H}_{\rm SS} + \Ts C_2^{x,y}(\sv)  ~  \mathbf{H}_{\rm SS}$ & ~~  -   \\
 $ \text{Skew-Stripy } [0 \,0 \, \pm 1]  $        &  $E + d_2 + d_3 + C_2^{\bv} + C_2^{x,y}(\sv) \, (\Ps + d_1  + C_2^{\av} + C_2^{\cv}) + C_2^{z}(\sv) $         & $\mathbf{H}_{\rm SS} + \Ts C_2^{x,y}(\sv)  ~  \mathbf{H}_{\rm SS}$ & ~~ -   \\
 $ \text{Skew-Zigzag } [0 \,0 \, \pm 1]  $        &  $E + \Ps + d_2  + C_2^{\av} + C_2^{x,y}(\sv) \, (d_1 + d_3 + C_2^{\bv} + C_2^{\cv}) + C_2^{z}(\sv) $         & $\mathbf{H}_{\rm SS} + \Ts C_2^{x,y}(\sv)  ~  \mathbf{H}_{\rm SS}$ & ~~ -    \\
 \hline
\end{tabular}
\hss}
\caption{
Spin-space groups $\mathbf{G}_{\rm SS}$ (and their relative unitary subgroups $\mathbf{H}_{\rm SS}$) for various phases of the Heisenberg-Kitaev model on a hyperhoneycomb lattice. For the collinear ferromagnet, the symmetry groups for different moment directions are listed. The right-most column indicates whether there is an isomorphism between the spin-space group and a magnetic space group.
The short-hand notation used for the symmetry elements is: pure spin transformations are labelled with $\sv$, pure real space with $\rv$ and the combined spin-space (locked) symmetry without extra label. Appendix~\ref{app:group_notation} has further information about the group theory notation.
}
\label{tab:HHEnhancedMagnGroupHK}
\end{table*}

\begin{table*}[!htb]
\centering
\hbox to \linewidth{ \hss
\begin{tabular}{ |c|*{11}{c c|}}
 \hline
 & \multicolumn{2}{|c|}{H. N\'{e}el $[x \,y \, z]$}
 & \multicolumn{2}{|c|}{N\'{e}el $[0 \,0 \, \pm 1]$}
 & \multicolumn{2}{|c|}{SZ $[0 \,0 \, \pm 1]$}
 & \multicolumn{2}{|c|}{SS $[0 \,0 \, \pm 1]$} 
 & \multicolumn{2}{|c|}{H. FM $[x \,y \, z]$}
 & \multicolumn{2}{|c|}{$[x \,y \, z]$}
 & \multicolumn{2}{|c|}{$[x \,0 \, z]$}
 & \multicolumn{2}{|c|}{$[x \,y \, 0]$}
 & \multicolumn{2}{|c|}{$[\pm 1 \,\pm 1\,\, z]$}
 & \multicolumn{2}{|c|}{$[\pm 1 \,\pm 1\,\, 0]$}
 & \multicolumn{2}{|c|}{$[0 \,0 \, \pm 1]$} \\
 \hline
  IBZ & GS & SW & GS & SW & GS & SW & GS & SW & GS & SW & GS & SW & GS & SW & GS & SW & GS & SW & GS & SW & GS & SW \\ 
 
\hline 
{\bf $\Gamma$}                               &                                          &  \checkmark                 &                                          &  \checkmark                 &                              & \checkmark                &                              & \checkmark           &                                          &                                          &                              &                                        &                              &                            &                                          &                                          &                              &                               &                                          &                                          &                                          &                                         \\
 
\hline
{\bf $Y$}                                             &                             & $(4)$                             & \checkmark                  &  \checkmark                 &                              & \checkmark                &                              &                                   &                                          &  \checkmark                 &                              &                                        &                              &                              &                                          &                                          &                              &   \checkmark     &                                          &  \checkmark                 &  \checkmark                 &  \checkmark                 \\
 
\hline
{\bf $T$}                                             &                             & $(4)$                             &                                          &  \checkmark                 & \checkmark      & $(4)$                           & \checkmark      & \checkmark          &                                          &   \checkmark                &                              &  \checkmark                 &                             &  \checkmark                         &   \checkmark               &   \checkmark                 &                              &    \checkmark     &  \checkmark                 &  \checkmark                 &  \checkmark                 &  \checkmark                 \\

 \hline
{\bf $Z$}                                             &                                          & $(4)$                             & \checkmark                  & $(4)$                             & \checkmark      & \checkmark                &  \checkmark     & \checkmark          &                                          &   \checkmark                &                              &   \checkmark                &                              &  \checkmark      &    \checkmark               &   \checkmark                &   \checkmark     &   \checkmark     &  \checkmark                 &  \checkmark                  &  \checkmark                 &  \checkmark                 \\

 \hline
{\bf $L$}                                             &                                          &  \checkmark                 &                                          &                                          &                              &                                        &                              &                                 &                                          &                                          &                              &                                        &                              &                              &                                          &                                          &                              &                               &                                          &                                          &                                          &                                            \\
 \hline
{\bf $\Delta = [\Gamma Y]$}        &                                          &  \checkmark                 &                                          &  \checkmark                &                               & \checkmark                &                              &                                 &                                          &                                          &                              &                                        &                              &                              &                                          &                                          &                              &                               &                                          &                                          &                                         &                                             \\

\hline
{\bf $\Lambda = [\Gamma Z]$}   &                                          &  \checkmark                 &                                          &  \checkmark                 &                              & \checkmark                &                              & \checkmark          &                                          &                                          &                              &                                        &                              &                              &                                          &                                          &                              &                               &                                          &                                          &                                         &                                             \\

\hline
{\bf $\Sigma = [\Gamma X]$}      &                                          &  \checkmark                 &                                          &                                          &                              & \checkmark                &                              & \checkmark          &                                          &                                          &                              &                                        &                              &                                 &                                          &                                          &                              &                              &                                          &                                          &                                          &                                             \\

\hline
{\bf $H =[Y T]$}                                &                             & $(4)$                             &                                          &  \checkmark                 & \checkmark      & \checkmark                &                              &                                   &                                          &   \checkmark               &                              &                                        &                              &                                &                                          &                                          &                              &    \checkmark     &                                          &  \checkmark                 &  \checkmark                &  \checkmark                   \\

\hline
{\bf $B =[Z T]$}                                &                                          & $(4)$                             &                                          &  \checkmark                 & \checkmark       & \checkmark                &  \checkmark     & \checkmark          &                                          &   \checkmark               &                              &                                        &                               &  \checkmark      &  \checkmark                &   \checkmark                 &                              &                               &  \checkmark                 &  \checkmark                 &                                          &  \checkmark                  \\

\hline
{\bf $A =[Z A]$}                                &                                          & $(4)$                             & \checkmark                  &  \checkmark                 &                              &                                        &                              &                                 &                                          &   \checkmark                &                              &                                        &                              &                                &                                          &                                          &   \checkmark     &   \checkmark    &  \checkmark                 &  \checkmark                 &  \checkmark                  &  \checkmark                 \\

\hline
{\bf $E =[\Gamma Z T]$}               &                                          &  \checkmark                 &                                          &  \checkmark                 &                              & \checkmark                &                              &                                  &                                          &                                          &                              &                                        &                              &                              &                                          &                                          &                              &                              &                                          &                                          &                                          &                                            \\

\hline
{\bf $J =[\Gamma X Z]$}               &                                          &  \checkmark                 &                                          &                                          &                              &                                        &                              &                                  &                                          &                                          &                              &                                        &                              &                              &                                          &                                          &                              &                              &                                          &                                          &                                          &                                             \\

\hline
{\bf $M =[\Gamma X Y]$}             &                                          &  \checkmark                 &                                          &                                          &                              &                                        &                              &                                 &                                          &                                          &                              &                                        &                              &                                &                                          &                                          &                              &                              &                                          &                                          &                                         &                                               \\

\hline
{\bf $GP$}                                         &                                          &  \checkmark                 &                                          &                                          &                              &                                        &                              &                                 &                                          &                                          &                              &                                        &                              &                               &                                          &                                          &                              &                              &                                          &                                          &                                        &                                                \\
 \hline
\end{tabular}
\hss}
\caption{
Degeneracies in ordered Heisenberg-Kitaev models for the hyperhoneycomb lattice at all high symmetry points, lines and surfaces. The first 4 cases are antiferromagnets (where SZ $=$ skew-zigzag and SS $=$ skew-stripy) while the others are ferromagnets. For Heisenberg N\'{e}el and FM ground states the symmetries are given for a generic spin direction $[x \,y \, z]$.
The column $GS$ lists the degeneracies expected on the basis of the magnetic space group that leaves the magnetic structure invariant (and are therefore independent of the coupling). The column $SW$ lists those degeneracies coming from the spin-space group. 
The excitations have not only all the degeneracies coming from the ground state invariance but also a significant enhancement due to the presence of spin space symmetries (especially in the AFM cases thanks to the mixing of spin rotations with the perpendicular axis). A checkmark represents a double degeneracy, while a number $(4)$ a 4-fold degeneracy. Nodal planes are present along {\bf $E =[\Gamma Z T]$} for two AFM cases, while the nodal volume only in the Heisenberg N\'{e}el case (see general position GP).
}
\label{tab:DegeneracyEnhancedHK}
\end{table*}

\section{Summary and Conclusions}

The study of band topology has been a gigantic enterprise in condensed matter physics over roughly the last fifteen years. During that time, important insights have arisen as more symmetries have been considered. The first known topological band insulator was a Chern insulator that was the inspiration for time reversal symmetric gapped band topology. Later people devised topological band structures with particle-hole and chiral symmetries and, later still, lattice symmetries and their interplay with time reversal symmetry. In this paper, we have extended this programme further to include yet more symmetric cases by including spin rotation symmetry and analysing the resulting spin-space groups. 

We have given a number of examples of magnetic couplings where spin-space groups are the appropriate symmetry groups in the magnetically ordered phase. These include Heisenberg models, Kitaev Heisenberg models, collinear Dzyaloshinskii-Moriya couplings and various kinds of single ion anisotropy. 

For various cases, we have worked out the representation theory of the relevant spin-space group thus providing the underlying symmetry reason for topological features in the magnetic excitations on top of magnetically ordered states. These calculations show in unprecedented detail how to work out band degeneracies from the relevant group $-$ methods that are applicable to any other spin-space group. In many of these cases we have contrasted our findings with the expected band structures one would obtain with foreknowledge only of the magnetic structure and the corresponding magnetic space group. 

We have found that spin-space groups of various sorts can lead to nodal points (Dirac points in 2D, Weyl points and four-fold degenerate points in 3D), nodal lines ($2$-fold and $4$-fold degenerate) protected by non-symmorphic spin-space groups, nodal planes including those with intersecting nodal lines and, in some instances, degenerate volumes. The rich band structures presented in this work provide a mere glimpse of the types of gapless band topology that can arise from spin-space groups in the magnons of magnetic insulators and the electronic band structures of itinerant magnetic materials. As we should expect of groups of higher symmetry, magnetic groups including spin rotations are particularly efficient at generating degeneracies in band structures.

The magnetic space groups are a subset of spin-space groups where the spin and real space point group elements are locked. However, we have also shown, for various ferromagnetic Kitaev-Heisenberg models, that certain spin-space groups with nontrivial spin rotation elements can also be isomorphic to magnetic space groups, albeit groups of higher symmetry than one would expect on the basis of the magnetic structure alone.

All of the models we have considered are in some sense fine-tuned. That is to say that spin-orbit coupling is almost omnipresent in condensed matter systems so that the magnetic Hamiltonian will tend to include all terms allowed by symmetry and the moments will then be locked to real space. Nevertheless, all the couplings we have included are physically allowed couplings and a degree of fine-tuning is often feasible in condensed matter systems because the richness of chemistry admits an exploration of possible couplings. So, for example, in many first row transition metal magnets and elsewhere, spin-orbit is weak compared to the principal exchange scale and then Heisenberg models may be an excellent approximation to the magnetism. Indeed Heisenberg exchange and XY exchange both of which admit nontrivial spin-space symmetries have been the canonical models of magnetism for decades and satisfactorily account for the properties of a great many magnetic materials.

Going beyond Heisenberg couplings, antisymmetric exchange that often appears as the leading exchange contribution in the spin-orbit coupling can be associated to spin-space group symmetries. Also, Kitaev-Heisenberg exchange has been argued to be the dominant set of couplings in various magnets with magnetic ions in edge-shared octahedral cages.

Even where the couplings depart from the spin-space symmetric surfaces in parameter space, one can imagine that there are materials where the magnon spectra contain near degeneracies that arise from a nearby parent Hamiltonian with spin-space symmetries and that these degeneracies would be otherwise mysterious. In fact, such near degeneracies could be used {\it to diagnose} the presence of dominant Kitaev terms for example or at least the absence of certain symmetry breaking terms. Similarly, even if symmetries are violated that would otherwise protect topological surface states, the boundary states can remain when the couplings are proximate to symmetric surfaces in parameter space. 

Our work also sheds some light on the phenomenon of order-by-disorder where accidental mean field ground state degeneracies are broken down by fluctuations. One symptom of order-by-disorder is the presence of spurious Goldstone modes in linear spin wave theory that cannot be present in the full interacting model by symmetry. One may ask whether linear spin wave theory may have other incongruous features in the spectrum that may be lifted by fluctuations. For the cases we have considered in this paper where order-by-disorder is present $-$ such as the hyperhoneycomb N\'{e}el state $-$ the answer is that linear spin wave theory faithfully reflects the spin-space symmetries of the magnetically ordered state up to the appearance of Goldstone modes.  

This paper points to a number of interesting future directions most notably the considerable task of carrying out a complete classification of physically relevant spin-space groups, their representations and associated band topology. More immediately, it would be interesting to investigate in detail the implications of spin-space symmetries for electronic systems such as spin-orbit coupled magnetic semi-metals and to magnetic excitons that are not magnons. More speculatively, one may ask whether there are physically natural generalizations, say for $SU(N)$ magnets or exotic order parameters, of the spin-space symmetries considered here to higher symmetries still such as the polychromatic groups. 

\sectitle{Acknowledgments} $-$ This work was in part supported by the Deutsche Forschungsgemeinschaft  under grants SFB 1143 (project-id 247310070) and the cluster of excellence ct.qmat (EXC 2147, project-id 390858490).

\clearpage

\appendix

\section{Hyperhoneycomb Lattice Conventions and Symmetries}
\label{app:lattice_data}

The hyperhoneycomb lattice is a three-dimensional lattice (\ref{fig:Lattice}) based around a face-centered orthorhombic cell with a four site basis belonging to the space group 
Fddd (\#70). The primitive lattice vectors in $x$, $y$, $z$ Kitaev coordinate system are
\beq
\bs{R}_1 = (2,4,0),~~ \bs{R}_2 = (3,3,2), ~~\bs{R}_3 = (-1,1,2)
\eeq
and the basis is
\beq
\bs{r}_1 = (0,0,0),~ \bs{r}_2 = (1,1,0), ~\bs{r}_3 = (1,2,1),~\bs{r}_4= (2,3,1).
\eeq
The orthorhombic conventional unit cell vectors are
\beq
\bs{a} = (-2,2,0),~~ \bs{b} = (0,0,4), ~~\bs{c}= (6,6,0)
\eeq
The reciprocal lattice vectors are 
\begin{align}
\bs{G}_1  &= 2\pi \Big(\frac{1}{6},-\frac{1}{3},\frac{1}{4}\Big), \nonumber \\ 
\bs{G}_2  &= 2\pi \Big(\!\!-\!\frac{1}{3},\frac{1}{6},-\frac{1}{4}\Big), \nonumber \\
 \bs{G}_3 &= 2\pi \Big(\frac{1}{3},-\frac{1}{6},-\frac{1}{4}\Big).
\end{align}
This coordinate system is useful to highlight the effect of global spin rotations compatible with the Heisenberg-Kitaev Hamiltonian.

Throughout the paper also another coordinate system has been used, the one used by the Bilbao Crystallographic Server $-$ the coordinate system of the conventional unit cell $\bs{a}$, $\bs{b}$, $\bs{c}$ with origin centered at the inversion point. The latter is simpler for dealing with lattice symmetries and has been used for all the group theory calculations.
In this coordinate system the primitive lattice vectors are:
\beq
\bs{R}_1 = \Big(\frac{1}{2},0,\frac{1}{2}\Big),~~ \bs{R}_2 = \Big(0,\frac{1}{2},\frac{1}{2}\Big), ~~\bs{R}_3 = \Big(\frac{1}{2},\frac{1}{2},0\Big)
\eeq
and the reciprocal lattice vectors in the dual basis $\bs{a}^*$, $\bs{b}^*$, $\bs{c}^*$ (such that for example $\bs{a} \cdot \bs{a}^* = 2 \pi$) are 
\begin{align}
\bs{G}_1  &= (-1,1,-1), \nonumber \\ 
\bs{G}_2  &= (1,-1,-1), \nonumber \\
 \bs{G}_3 &= (-1,-1,1).
\end{align}

The high symmetry points in the first Brillouin Zone in both coordinate systems reads:
\begin{align}
&\Gamma =(0,0,0)_{xyz} = (0,0,0)_{abc} \nonumber \\ 
&Y =\Big(0,0,-\frac{\pi}{2}\Big)_{xyz} = (0,-1,0)_{abc} \nonumber \\ 
&T =\Big(\!\!-\!\frac{\pi}{6},-\frac{\pi}{6},-\frac{\pi}{2}\Big)_{xyz} = (0,-1,-1)_{abc} \nonumber \\ 
&Z=\Big(\!\!-\!\frac{\pi}{6},-\frac{\pi}{6},0\Big)_{xyz} = (0,0,-1)_{abc} \nonumber \\ 
&X =\Big(\frac{29\pi}{72},-\frac{29\pi}{72},0\Big)_{xyz} = (-29/36,0,0)_{abc} \nonumber \\ 
&A_1 =\Big(\frac{11\pi}{72},-\frac{11\pi}{72},-\frac{\pi}{2}\Big)_{xyz} = (-11/36,-1,0)_{abc} \nonumber \\ 
&X_1 =\Big(\!\!-\!\frac{19\pi}{72},-\frac{5\pi}{72},-\frac{\pi}{2}\Big)_{xyz} = (7/36,-1,-1)_{abc} \nonumber \\ 
&A =\Big(\frac{13\pi}{72},-\frac{37\pi}{72},0\Big)_{xyz} = (-25/36,0,-1)_{abc} \nonumber \\ 
&L =\Big(\frac{\pi}{6},-\frac{\pi}{3},-\frac{\pi}{4}\Big)_{xyz} = (-1/2,-1/2,-1/2)_{abc} 
\end{align}
The first Brillouin zone together with the high symmetry directions and points are shown in Fig.~\ref{fig:Lattice}.

The lattice symmetries that constitute the space group $\mathbf{G}$ of the hyperhoneycomb include:
\begin{itemize}
\item Primitive translations $\bs{R}_i$
	\item Inversion $\Ps$ at the bond center of sublattices $2,3$ and $1,4$  (green and red bonds in
 Fig.~\ref{fig:Lattice});
	\item Three orthogonal $\text{C}_2$ axes at the bond center of bonds connecting sublattices $1,2$ and $3,4$
 (blue bonds). These axes are parallel to the face-centered-orthorhombic lattice vectors ${\bs a}$, ${\bs b}$, and
 ${\bs c}$. Bonds $2,3$ and $1,4$ are interchanged via these $\text{C}_2$ axes;
	\item Glide planes, $d_1$, $d_2$, $d_3$, with translation $\bs{R}_i/2$ interchanges bonds $1,2$ and $3,4$.
\end{itemize}

Therefore we have for the $7$ nontrival symmetry elements $\element{R}{\bs{t}}$, in $x$, $y$, $z$ Kitaev coordinates with $l$ sublattice index and in $\bs{a}$, $\bs{b}$, $\bs{c}$ system:

\begin{widetext}
\beq
\label{eq:d1xyz}
  d_1 : (x, y, z, l) = \begin{cases}
                     			(x, y, -z, 3)\,,            & l=1\\
                     			(x, y, -z, 4)\,,            & l=2\\
                     			(x+2, y+4, -z, 1)\,,  & l=3 \\
                     			(x+2, y+4, -z, 2)\,,  & l=4\,.
            		      \end{cases}
~~ = ~ \{m_{010}|\textstyle \frac{1}{4}, 0, \frac{1}{4} \}_{abc}
\eeq
\beq
\label{eq:d2xyz}
  d_2 : (x, y, z, l) = \begin{cases}
                     			(y, x, z, 3)\,,                 & l=1\\
                     			(y, x, z, 4)\,,                 & l=2\\
                     			(y+3, x+3, z+2, 1)\,,  & l=3 \\
                     			(y+3, x+3, z+2, 2)\,,  & l=4\,.
            		      \end{cases}
~~ = ~ \{m_{100}|\textstyle 0, \frac{1}{4}, \frac{1}{4} \}_{abc}
\eeq
\beq
\label{eq:d3xyz}
  d_3 : (x, y, z, l) = \begin{cases}
                     			(-y, -x, z, 4)\,,                & l=1\\
                     			(-y, -x, z, 3)\,,                & l=2\\
                     			(-y-1, -x+1, z+2, 2)\,,  & l=3 \\
                     			(-y-1, -x+1, z+2, 1)\,,  & l=4\,.
            		      \end{cases}
~~ = \{m_{001}|\textstyle \frac{1}{4}, \frac{1}{4}, 0 \}_{abc}
\eeq
\beq
\label{eq:Pxyz}
  \Ps=d_1d_2^{-1}d_3 : (x_1 ,x_2, x_3, l) = \begin{cases}
                     				                            (-x, -y, -z, 4)\,,    & l=1\\
                     				                            (-x, -y, -z,  3)\,,   & l=2\\
                     				                            (-x, -y, -z, 2)\,,    & l=3 \\
                     				                            (-x, -y, -z,  1)\,,   & l=4\,.
            		                                                \end{cases}
~~ = ~ \{-1| 0, 0, 0\}_{abc}
\eeq
\beq
\label{eq:C2axyz}
  C_2^{\bs{a}}=d_3^{-1}d_1 : (x, y, z, l) = \begin{cases}
                     			                                     (-y, -x, -z, 2)\,,               & l=1\\
                     			                                     (-y, -x, -z, 1)\,,               & l=2\\
                     			                                     (-y-3, -x-3, -z-2, 4)\,,    & l=3 \\
                     			                                     (-y-3, -x-3, -z-2, 3)\,,    & l=4\,.
            		                                                 \end{cases}
~~ = ~ \{2_{100}|\textstyle 0, -\frac{1}{4}, -\frac{1}{4} \}_{abc}
\eeq
\beq
\label{eq:C2bxyz}
  C_2^{\bs{b}}=d_3^{-1}d_2 : (x, y, z, l) = \begin{cases}
                     			                                     (-x, -y, z, 2)\,,            & l=1\\
                     			                                     (-x, -y, z, 1)\,,            & l=2\\
                     			                                     (-x-2, -y-4, z, 4)\,,    & l=3 \\
                     			                                     (-x-2, -y-4, z, 3)\,,    & l=4\,.
            		                                                 \end{cases}
~~ = ~ \{2_{010}|\textstyle -\frac{1}{4}, 0, -\frac{1}{4} \}_{abc}
\eeq
\beq
\label{eq:C2cxyz}
  C_2^{\bs{c}}=d_2^{-1}d_1 : (x, y, z, l) = \begin{cases}
                     			                                     (y, x, -z, 1)\,,                 & l=1\\
                     			                                     (y, x, -z,  2)\,,                & l=2\\
                     			                                     (y+1, x-1, -z-2, 3)\,,    & l=3 \\
                     			                                     (y+1, x-1, -z-2, 4)\,,    & l=4\,.
            		                                                 \end{cases}
~~ = ~ \{2_{001}|\textstyle -\frac{1}{4}, -\frac{1}{4}, 0 \}_{abc}
\eeq
\end{widetext}

\section{Hyperhoneycomb symmetry-allowed exchange Hamiltonian}
\label{app:exchange_constrain}

Here, taking the example of the hyperhoneycomb lattice, we give an explicit example of how spin-orbit constrains the exchange Hamiltonian of a spin system and how additional global spin symmetries can arise beyond those allowed by spin-orbit coupling.

The hyperhoneycomb unit cell has $i,j = \{1,2,3,4\}$ spins and it is tri-coordinated, so there are in total $6$ bonds per unit-cell, $(i,j)$.
We consider for this system a generic bilinear exchange Hamiltonian
\beq
\hat{H} = \sum_{ i,j }\mathsf{J}_{ij}^{\mu\nu} \hat{J}^\mu_i  \hat{J}^\nu_j,
\eeq
which accounting for the primitive translations can be specified by $6$ exchange $3 \times 3$ matrices, one for each bond.
Applying the space group symmetries in Appendix \ref{app:lattice_data} we obtain two sets of equivalent bonds. A set of equivalent bonds contains bonds that can be transformed into each other by symmetry operations, and therefore bonds belonging to different equivalence sets are said to be inequivalent and have, in principle, independent exchange Hamiltonians.
We call set $a$ the one with ${(1,2)_z,(3,4)_z}$ while set $b$ the one ${(2,3)_x,(2,3)_y,(4,1)_x,(4,1)_y}$ where the subscript {\it a posteriori} provides the Kitaev bond label.

In the {\it absence} of spin-orbit the application of the space group symmetries does not act on the spins but only on their positions.
This would result in a generic bilinear exchange Hamiltonian to that is completely specified by only two generic symmetric matrices, one for each set of equivalent bonds. However free spins are $SU(2)$ symmetric and the only possible isotropic coupling is Heisenberg.

In case spin-orbit is {\it present}, the symmetries will act also on spin-space as pseudo-vector transformations.
After a proper parametrization the constraints for set $a$ are:
\begin{widetext}
\begin{align}
\mathsf{J}_{(1,2)_z} = & 
\begin{pmatrix} 
\hphantom{-}J_a                   & \hphantom{-}\Gamma_a      & -D     \\
\hphantom{-}\Gamma_a     & \hphantom{-}J_a                    &  \hphantom{-}D     \\
\hphantom{-}D                       & -D                                                & \hphantom{-}(J_a+K_a) 
\end{pmatrix} 
\hspace{40px}&
\mathsf{J}_{(3,4)_z} = &
\begin{pmatrix} 
\hphantom{-}J_a                   & \hphantom{-}\Gamma_a      &   \hphantom{-}D     \\
\hphantom{-}\Gamma_a     & \hphantom{-}J_a                    & -D     \\
-D                                                & \hphantom{-}D                       & \hphantom{-}(J_a+K_a) 
\end{pmatrix}
\end{align}
and for set $b$:
\begin{align}
\mathsf{J}_{(2,3)_x} = & 
\begin{pmatrix} 
(J_{b(xy)}+K_b)    & \Gamma'        &  \Gamma''      \\
\Gamma'                & J_{b(xy)}         &  \Gamma_b      \\
\Gamma''               & \Gamma_b     & J_{b(z)}
\end{pmatrix} 
&
\mathsf{J}_{(4,1)_x} = &
\begin{pmatrix} 
(J_{b(xy)}+K_b)                     & \hphantom{-}\Gamma'         &  -\Gamma''      \\
\hphantom{-}\Gamma'       & \hphantom{-}J_{b(xy)}         &  -\Gamma_b      \\
-\Gamma''                              & -\Gamma_b                              & \hphantom{-}J_{b(z)}
\end{pmatrix}
\label{Eq:WmatSymm:xbond}  \\[6px]
\mathsf{J}_{(2,3)_y} = & 
\begin{pmatrix} 
\hphantom{-}J_{b(xy)}        & \hphantom{-}\Gamma'     &  -\Gamma_b      \\
\hphantom{-}\Gamma'       & (J_{b(xy)}+K_b)                    &  -\Gamma''       \\
-\Gamma_b                            & -\Gamma''                             & \hphantom{-}J_{b(z)}  
\end{pmatrix}
&
\mathsf{J}_{(4,1)_y} = &
\begin{pmatrix} 
J_{b(xy)}         & \Gamma'                 &  \Gamma_b      \\
\Gamma'        & (J_{b(xy)}+K_b)     &  \Gamma''      \\
\Gamma_b    & \Gamma''                & J_{b(z)}
\end{pmatrix} 
\end{align}
\end{widetext}

In total we count $10$ possible exchange couplings.
We see that the lattice supports Heisenberg-Kitaev couplings, antisymmetric or Dzyaloshinski-Moriya (DM) coupling $D$ for $z$ bonds and different off-diagonal symmetric exchange $\Gamma_{\alpha}$.
For a given bond, the couplings are invariant under higher symmetries than the lattice as a whole. For example,
$\Gamma_{\alpha} \rightarrow D_2(\sv)$ and $D \rightarrow C_{\infty}(\sv)$. However, only for a subset of the couplings, is there a global residual symmetry: for the Heisenberg case (with global spin rotation symmetry) and Heisenberg-Kitaev with $D_2$ symmetry. 

In the main text, we make the standard approximation that the Kitaev couplings on inequivalent bonds are the same.

\section{Magnetic Group Notation}
\label{app:group_notation}

In the main text, unless otherwise indicated, group elements such as two-fold rotation $C_2$ are assumed to act both on spin and real space degrees of freedom. This substantially abbreviates the full spin-space notation $\sselement{C_2}{C_2}{\bs{0}}$. Sometimes we also use an argument: $C_2(\bs{s})$ to denote an operation acting only on spins and $C_2(\bs{r})$ acting only on real space.

The magnetic group notation used through the paper is the one adopted in the International Tables For Crystallography, i.e. Hermann–Mauguin notation. We do not explain this notation in detail here but only give a general overview so that the meaning of the symbols used in this paper can be appreciated without a detailed knowledge of these tables.

As an example we take the hyperhoneycomb space group $Fddd$ and its subgroups $C2/c$ and $P\text{-}1$. 
The symbols always start with an uppercase letter describing the Bravais lattice type, for example $F$ stands for face-centered lattice, $C$ for single-face centered $-$ on C faces only $-$ and $P$ for primitive. Right after the letter we have the symmetry elements of the corresponding point group (the group that remains if one removes all translational components from the space group). The order of the symmetry elements follows the hierarchy of axes if present, from the most symmetric to the lowest.
The possible symbols are: number $n$ for rotations $C_n$, $m$ for mirror planes, $\text{-}n$ for improper rotations $S_n$. Moreover when two elements refer to the same axis they are written as $n/m$.
In case the group is non-symmorphic some of the point group elements will be replaced by the symbol for the screw axis (a number with a number subscript, i.e. $2_1$) or glide planes (letter $a, b, c, n, d$). 
Taking again the examples above we have: $Fddd$ having 3 glide planes with translation along a quarter of the 3 different face diagonals ("diamond" glide plane), $C2/c$ having a $C_2$ rotation and a perpendicular glide plane with a translation along half the lattice vector of face $C$, $P\text{-}1$ having only inversion.
The standard form uses a short notation which shows only the generators in the shortest unambiguous way, i.e. the symmetry elements which, when composed, gives all the others. As an example we observe that the glide planes $d_1, d_2, d_3$ generate all the other operations in appendix \ref{app:lattice_data}.

For magnetic groups, the notation is extended by indicating with a prime an anti-unitary operation. For type $II$ groups one has the symbol of the space group from which it is derived plus a $1'$ at the end, indicating the pure time reversal operation. For type $III$ groups the space group symbols are modified adding a $'$ to the symmetry elements which become anti-unitary. Finally for type $IV$ groups, the black and white Bravais lattice is represented by the original uppercase letter lattice symbol with an additional subscript.
For example, the group $Fddd1'$ is type $II$ with pure time reversal and $3$ normal glides, while the group $Fd'd'd$ is type $III$ with one normal glide and two magnetic glides.

\section{Central Extension Method for hyperhoneycomb N\'{e}el Kitaev-Heisenberg $Z$ Point}
\label{app:central_extension}

The representation theory of spin-space groups mirrors that of (magnetic) space groups. One lesson coming from space groups is that, for most wavevectors, the decomposition of the little co-group of that wavevector into irreps is simply a matter of identifying the associated point group at that point and using standard tables. An important exception is for certain boundary points in non-symmorphic groups where the group composition rule is projective. Here we give an example of such a case: the $Z$ point in the N\'{e}el phase of the Kitaev-Heisenberg model on a hyperhoneycomb lattice.

First we must determine the factor system as explained in Section~\ref{sec:representationtheory}. The extra reciprocal lattice translation for each spatial symmetry (noting that the spin part does not act in reciprocal space) is:
\begin{align}
\label{eq:ExtraReciprocalTranslation}
&\bs{g}_{-1} = \bs{g}_{2_{100}} = \bs{g}_{2_{010}} = \bs{g}_{m_{001}} = (0, 0, 2) \nonumber  \\
&\bs{g}_{2_{001}} = \bs{g}_{m_{100}} = \bs{g}_{m_{010}} = (0, 0, 0) 
\end{align}

The factor system will then have $\mu(h_i, h_j) = \pm 1$ (as expected for glide symmetry with non-symmorphic translation equal to half of primitive vectors). For example,
\begin{align*}
\label{eq:FactorSystemNeel}
\mu( &\sselement{2_{-101}}{-1}{\bs{0}} ,  \sselement{2_{001}}{m_{001}}{1/4,1/4,0} ) \nonumber \\
&=  \exp \left( - 2 \pi i \,  (0, 0, 2) \cdot ( 1/4, 1/4, 0)\right) = +1 
\end{align*}

\begin{align}
\mu( &\sselement{2_{-101}}{-1}{\bs{0}} ,  \sselement{2_{010}}{m_{010}}{1/4,0,1/4} ) \nonumber \\
&=  \exp\left( - 2 \pi i \,  (0, 0, 2) \cdot ( 1/4, 0, 1/4)\right) = -1.
\end{align}
Since the factor system is not trivial we need to find the projective irreducible representations. One way to do this is by first finding the irreps  of the unitary central extension of the little co-group $\bar{\mathbf{G}}^{\bold{Z}^*}_{\rm SS}$ with $g =2$ which is a group with $32$ elements. 
The element of this group are of the kind $(h_i, \alpha)$, where $h_i \in \bar{\mathbf{G}}^{\bold{Z}}_{\rm SS}$ and $\alpha \in \{0,1,..., (g-1)\}$.
The group has the isomorphism $\bar{\mathbf{G}}^{\bold{Z}^*}_{\rm SS} \cong D_{4h} + D_{4h} \times (\sselement{4_{010}^{+}}{m_{100}}{\bs{0}}, 0)$ where we can identify the elements of $D_{4h}$ as:
\begin{align}
\label{eq:IsoZNeel1}
&2_{001} \rightarrow (\sselement{E}{E}{\bs{0}} , 1)  \nonumber  \\
&4_{001}^+ \rightarrow ( \sselement{2_{100}}{2_{100}}{\bs{0}} , 0)  \nonumber  \\
& -1 \rightarrow ( \sselement{2_{010}}{E}{\bs{0}} , 0)  \nonumber \\
& 2_{100} \rightarrow ( \sselement{E}{m_{010}}{\bs{0}} , 0)  \nonumber  \\
& 2_{1\text{-}10} \rightarrow ( \sselement{2_{100}}{m_{001}}{\bs{0}}, 0)
\end{align}
The irreps of $\bar{\mathbf{G}}^{\bold{Z}^*}_{\rm SS}$ can be obtained by conjugating the ones of the subgroup $D_{4h}$ by the symmetry $(\sselement{4_{010}^{+}}{m_{100}}{\bs{0}}, 0)$. Of these irres we are only interested in the one with $\Delta( \sselement{E}{E}{\bs{0}},1) =  -\, \mathbb{I}$ (constrain to get the proper phase factors), therefore $Z_{5a(b)}^{\pm}$ where $(a,b)$ label comes from the fact that $Z_5^{\pm}$ of $D_{4h}$ are self-conjugated under $(\sselement{4_{010}^{+}}{m_{100}}{\bs{0}}, 0)$. Considering only the elements $(h_i, 0)$ and adding the right phase factors for translation we can build the character table of the unitary part of $\mathbf{G}^{\bold{Z}}_{\rm SS}$ as in Table~\ref{tab:ProjReptableZ}.

Finally we can add now the anti-unitary elements and compose the coreps. The test gives:
\begin{widetext}
\begin{align}
\label{eq:ZTest}
\sum_{h_{\bs{k}'}} \chi_{p}^{\bold{Z}}(h_{\bs{k}'}^2)  
= \, 4 \, ( \chi_{p}^{\bold{Z}}(\sselement{E}{E}{\bs{0}})  +  \chi_{p}^{\bold{Z}}(\sselement{2_{010}}{E}{\bs{0}}) ) =
\begin{cases}
                                                16 =  |\bar{\mathbf{G}}^{\bold{Z}^*}_{\rm SS}|  ~~~ &\text{Type (a)} ~~ \text{if} ~~ p = Z_{5a}^{+}, \, Z_{5b}^{+}  \\
                                                0     ~~~ &\text{Type (c)} ~~ \text{if} ~~ p = Z_{5a}^{-}, \, Z_{5b}^{-}  \\
\end{cases}
\end{align}
\end{widetext}
where $h_{\kv'}$ are all the anti-unitary elements for which $h_{\bs{k}'} \bs{k} = - \bs{k} + \bs{g}_i$, so for $\bold{Z}$ are all the anti-unitary elements in the full spin space group.
We get therefore a $4$-degenerate coreps $DZ_{5}^{-}(4) = (Z_{5a}^{-}, Z_{5b}^{-})$ with $\chi(\sselement{2_{010}}{E}{\bs{0}}) = -4$ and $\chi(\sselement{4_{010}^{\pm}}{m_{100}}{0,1/4,1/4}) = 0$.

Since we know the characters of the band representations $\rho^{\bs{k}}_{G,S_{\perp}}$ for every $\bs{k}$ we can now subduce it to the point $\bold{Z}$, therefore from its characters  in \ref{eq:IndChar} and \ref{eq:IndCharNeel} we get (the number in parenthesis indicates the dimension of the corep):
\begin{align}
\label{eq:ZRep}
\rho^{\bold{Z}}_{S_{\pm}} = DZ_5^-(4)
\end{align}
\begin{table*}[!htb]
\centering
\hbox to \linewidth{ \hss
\begin{tabular}{ |c|c|c|c|c|c|c|  }
 \hline
 $\mathcal{G}^{\bold{Z}}_{\rm SS}$~ & ~$\sselement{E}{E}{\bs{0}}$~ & ~$\sselement{2_{010}}{E}{\bs{0}}$~ & ~$ \sselement{4_{010}^+}{m_{100}}{0,1/4,1/4}$
 ~ & ~$ \sselement{4_{010}^-}{m_{100}}{0,1/4,1/4}$ ~ & ~$...$~ & ~Type Coreps~ \\ 
 \hline
  $Z_{5a}^{+}$   & 2   & 2    &   2\,$\xi$   &  2\,$\xi$    &  0 & (a)  \\
 \hline
 $Z_{5b}^{+}$   & 2   & 2     &  -2\,$\xi$  &  -2\,$\xi$   &  0 & (a)  \\
 \hline
 $Z_{5a}^{-}$    & 2   & -2    &  2\,$\xi$   &  -2\,$\xi$   &  0  & (c)\\
 \hline
 $Z_{5b}^{-}$   & 2    & -2   &  -2\,$\xi$  &   2\,$\xi$    &  0  & (c)    \\
 \hline
\end{tabular}
\hss}
\caption{
Character table giving the relevant irreps of the unitary part of $\mathbf{G}^{\bold{Z}}_{\rm SS}$. The phase factor is $\xi = \exp(i \, \bold{Z} \cdot (\textstyle 0 \frac{1}{4} \frac{1}{4}) ) = - i$ and the dot $(...)$ indicates irrelevant symmetries $-$ those with trivial characters.
}
\label{tab:ProjReptableZ}
\end{table*}
%

\section{Representation theory for Heisenberg N\'{e}el antiferromagnet}
\label{app:HeisenNeel}

For the Heisenberg coupling, space group symmetry elements act only on real space (since we can always cancel out their effect on the spins with a $SO(3)$ rotation).
In the antiferromagnetic case we can then divide the space group symmetries into two groups, those that do not swap the magnetic sublattice $G_{\uparrow \uparrow}$ and those that do. The latter have to be coupled with an additional spin rotation to preserve the magnetic order $C^{\perp}_2(\bs{s}) \, G_{\uparrow \downarrow}$.
In addition, we have to consider pure spin rotations about axes parallel to moment directions $C^{\parallel}_{\infty}(\bs{s})$ (rotations $\vartheta_z$).
The full group is therefore $\mathbf{G}_{\rm H-Neel}=\mathbf{H}_{\rm H-Neel} + \Ts C_2^{\perp}(\bs{s}) \mathbf{H}_{\rm H-Neel}$ with elements:
\begin{align}
& \mathbf{H}_{\rm H-Neel}= C^{\parallel}_{\infty}(\bs{s}) \times G_{\uparrow \uparrow}(\rv) \times C^{\perp}_2(\bs{s}) \, G_{\uparrow \downarrow}(\rv) 
=C_{\infty}^{\parallel}(\sv) + \nonumber \\ & (E + d_1 + d_2 + C_2^{\cv})(\rv) + C_2^{\perp}(\sv) \, [(\Ps + d_3 +  C_2^{\bv} +  C_2^{\cv})(\rv)]
\label{eq:HeisenNeelUnitary}
\end{align}
excluding primitive translations. Here we have used a short-hand notation for the symmetry elements where pure spin transformations are labelled with $\sv$ and pure real space with $\rv$.
All symmetry elements are defined in Appendix~\ref{app:lattice_data}.

The representations of the enhanced magnetic little groups  $\mathbf{G}^{\bs{k}}_{\rm SS}$ are straightforward to find for points $\bs{k}$ inside the Brillouin zone.

{\rm GP $= (u, \, v, \, w)$ } $-$ the least symmetric (general) position has little group:
\beq
\mathbf{G}^{\rm{GP}}_{\rm SS} / \bold{T} = C_{\infty}^{\parallel}(\bs{s}) \, (E + \Ts \Ps) \, \cong \, C_{\infty}1'
\eeq
and therefore doubly degenerate modes as explained in Section~\ref{sec:heisenberg}.

{$\Gamma = (0, \, 0, \, 0)$} $-$ the most symmetric point has little group:
\begin{align}
\mathbf{G}^{\Gamma}_{\rm SS} / \bold{T} = &[(G_{\uparrow \uparrow}(\rv) + C^{\parallel}_{\infty}(\bs{s})) \times (E + C_2^{\perp}(\sv)\Ps)] \nonumber \\
&\times (E + \Ts C_2^{\perp}(\sv))
\end{align}
where we have chosen $C_2^{\perp}(\sv)\Ps$ as a means of swapping the magnetic sublattices. We can build the final coreps in 4 steps.
First, we consider only the real space group $G_{\uparrow \uparrow}(\rv) = Fdd2$ which has only 1D irreps $\Gamma_n$. Second, multiply $C_{\infty}(\bs{s})$ into the group, which self-conjugates each irrep of $G_{\uparrow \uparrow}(\rv)$ to $\Gamma^{\pm}_n$ (since transverse spin components transform as $m=\pm 1$ reps of $C_{\infty}$).
Third, we add the representative $C_2^{\perp}(\sv)\Ps$ which mixes spin and space transformations.
In deducing the reps of this mixed group we need to check the conjugation of elements with $C_2^{\perp}(\sv)\Ps$:
\begin{align}
&(C_2^{\perp}(\sv)\Ps) ~ S(\rv) ~ (C_2^{\perp}(\sv)\Ps)^{-1} = S(\rv)  ~~~ \forall \, S(\rv) \in G_{\uparrow \uparrow}(\rv) \nonumber \\ 
&(C_2^{\perp}(\sv)\Ps) ~ \vartheta_z(\sv) ~ (C_2^{\perp}(\sv)\Ps)^{-1} = \vartheta_z^{-1}(\sv)  ~~~ \forall \, \vartheta_z \in C^{\parallel}_{\infty}(\sv).
\end{align}
Therefore the irreps will pair into two dimensional irreps $(\Gamma_n^+, \Gamma_n^-)(2)$.
The last step is to consider anti-unitary elements $\Ts C_2^{\perp}(\sv)$. Coreps types are assessed by computing:
\begin{align}
\label{eq:GTestHeisen}
\sum_{h_{\bs{k}'}} \chi_{p}^{\Gamma}(h_{\bs{k}'}^2)  
&= \, 4 \, \sum_{\vartheta} \chi_{p}^{\Gamma}(2\vartheta(\sv))  +  4 \, \sum_{2_{\perp}} \chi_{p}^{\Gamma}(E) \nonumber \\
&= 4 \int_0^{2\pi} 2 \cos{2\vartheta} + 4 \, n \, \chi_{p}^{\Gamma}(E)  \nonumber \\ &= 8 \, n = |\mathbf{H}^{\Gamma}_{\rm SS}| ~~~ \text{Type (a)}
\end{align}
where $h_{\kv'}$ are all the anti-unitary elements, $n \rightarrow \infty$ is the multiplicity of the class $C_2^{\perp}(\sv)$ and $\chi_{p}^{\Gamma}(E) = 2$ for 2D irreps.
Since the coreps are type $(a)$ there is no further degeneracy.

Since the least symmetric and most symmetric points inside the Brillouin Zone have the same degeneracy, all the intermediate cases must have the same degeneracy.
Nevertheless the presence of non-symmorphic symmetries can lead to extra degeneracies corresponding to projective representations on the zone boundaries.

{\rm B $= (0, \, u, \, -1)$} $-$ Using spin-space group notation the $[Z T]$ line has little group elements:
\begin{widetext}
\begin{align}
\mathbf{G}^{B}_{\rm SS} / \bold{T} = &~~~~ (\sselement{\vartheta_{z}}{E}{\bs{0}} \, + \, \sselement{\vartheta_{z}}{m_{100}}{\textstyle 0,\frac{1}{4},\frac{1}{4}} \, + \, \sselement{2_{\perp}}{m_{001}}{\textstyle \frac{1}{4},\frac{1}{4},0}  \, + \, \sselement{2_{\perp}}{2_{010}}{\textstyle -\frac{1}{4},0,-\frac{1}{4}}) \nonumber  \\
& + (\sselement{\vartheta_{z}}{-1}{\bs{0}}' \, + \, \sselement{\vartheta_{z}}{2_{100}}{\textstyle 0,-\frac{1}{4},-\frac{1}{4}}' \, + \, \sselement{2_{\perp}}{2_{001}}{\textstyle -\frac{1}{4},-\frac{1}{4},0}'  \, + \, \sselement{2_{\perp}}{m_{010}}{\textstyle \frac{1}{4},0,\frac{1}{4}}') \nonumber  \\
\end{align}
\end{widetext}
where the elements of the kind $\sselement{\vartheta_{z}}{E}{\bs{0}}$ (or  $\sselement{2_{\perp}}{E}{\bs{0}})$) represent all the rotations around the collinear axis (or $\pi$ rotation around the infinite perpendicular axes). Also the anti-unitary elements are denoted by a prime.
Here the $2_{\perp}$ rotations intertwine spin and space transformations, making the little group not a simple direct product and therefore making it impossible to use the tabulated space group projective representations.

To build the projective representations from scratch we must determine first the factor system as explained in Section~\ref{sec:representationtheory}. The extra reciprocal lattice translation for each spatial symmetry (noting that the spin part does not act on reciprocal space) is:
\begin{align}
\label{eq:ExtraReciprocalTranslationHeisen}
&\bs{g}_{2_{010}} = \bs{g}_{m_{001}} = (0, 0, 2) \nonumber  \\
&\bs{g}_{m_{100}} = (0, 0, 0).
\end{align}

The factor system will then have $\mu(h_i, h_j) = \pm 1$ and therefore gives nontrivial projective irreducible representations. 

The unitary central extension of the little co-group $\bar{\mathbf{G}}^{B^*}_{\rm SS}$ with $g =2$ is a group with $8 \,n$ elements ($n$ is, again, the multiplicity of the perpendicular axis).
The group has the isomorphism $\bar{\mathbf{G}}^{B^*}_{\rm SS} \cong D_{\infty h} + D_{\infty h} \times (\sselement{E}{m_{100}}{\bs{0}}, 0)$ where we can identify the elements of $D_{\infty h}$ as:
\begin{align*}
&\vartheta_z \rightarrow (\sselement{\vartheta_z}{E}{\bs{0}} , 0)  \nonumber  \\
&2_{\perp} \rightarrow ( \sselement{2_{\perp}}{2_{010}}{\bs{0}} , 0)  \nonumber  \\
& \text{-} 1 \rightarrow ( \sselement{E}{E}{\bs{0}} , 1).
\end{align*}
The irreps of $\bar{\mathbf{G}}^{B^*}_{\rm SS}$ can be obtained by conjugating those of the subgroup $D_{\infty h}$ with the symmetry $(\sselement{E}{m_{100}}{\bs{0}}, 0)$. Of these representations we are only interested in the ones with $\Delta(\sselement{E}{E}{\bs{0}},1) =  -\, \mathbb{I}$ (this constraint ensures the correct projective phase factors) and with $\Delta(\sselement{\vartheta_z}{E}{\bs{0}},0) =  2\cos{\vartheta}$ (coming from the transverse spin component band representation). This points to $B_{1u}^{a(b)}$. The $(a,b)$ label comes from the fact that $B_{1u}$ of $D_{\infty h}$ are self-conjugate under $(\sselement{E}{m_{100}}{\bs{0}}, 0)$.
Considering only the elements $(h_i, 0)$ and adding the right phase factors for translation we can build the character table of the unitary part of $\mathbf{G}^{B}_{\rm SS}$ as in Table~\ref{tab:ProjReptableBHeisen}.

Finally we consider the anti-unitary elements and compose the coreps. The test gives:
\begin{widetext}
\begin{align}
\label{eq:BTestHeisen}
\sum_{h_{\bs{k}'}} \chi_{p}^{\Gamma}(h_{\bs{k}'}^2)  
&= \, 2 \, \sum_{\vartheta} \chi_{p}^{B}(\sselement{2\vartheta_z^{\pm}}{E}{\bs{0}})  + (1 +  \exp(i \, B \cdot ({\textstyle \frac{1}{2} 0 \frac{1}{2}}) ) ) \sum_{2_{\perp}} \chi_{p}^{B}(\sselement{E}{E}{\bs{0}})
\nonumber \\
&= 2 \int_0^{2\pi} 2 \cos{2\vartheta} = 0 ~~~ \text{Type (c)}
\end{align}
\end{widetext}
where $h_{\kv'}$ are all the anti-unitary elements for which $h_{\bs{k}'} \bs{k} = - \bs{k} + \bs{g}_i$, so for $B$ are the anti-unitary elements with real space transformation $\{m_{010}, 2_{001}, 2_{100}, -1\}$.
We get therefore a $4$-fold degenerate co-representation $DB_{1u}(4) = (B_{1u}^a, B_{1u}^b)(4)$ with $\chi(\sselement{\vartheta_z^{\pm}}{E}{\bs{0}}) = 4 \cos{\vartheta}$ and $\chi(\sselement{E}{m_{100}}{0,1/4,1/4}) = 0$.

For band representation $\rho^{\bs{k}}_{G,S_{\perp}}$ we obtain:
\begin{align}
\label{eq:BRepHeisen}
\rho^{B}_{S_{\pm}} = DB_{1u}(4).
\end{align}

The same argument carries through for the high symmetry line {\rm A $= (u, \, 0, \, -1)$}, which again is 4-degenerate (line $[Z A]$).

{\rm H $= (u, \, 0, \, -1)$} $-$ Lastly the line $[Y T]$ line has little group:
\begin{widetext}
\begin{align}
\mathbf{G}^{H}_{\rm SS} / \bold{T} = &~~~~ (\sselement{\vartheta_{z}}{E}{\bs{0}} \, + \, \sselement{\vartheta_{z}}{m_{100}}{\textstyle 0,\frac{1}{4},\frac{1}{4}} \, + \, \sselement{\vartheta_{z}}{2_{001}}{\textstyle -\frac{1}{4},-\frac{1}{4},0}  \, + \, \sselement{\vartheta_{z}}{m_{010}}{\textstyle \frac{1}{4},0,\frac{1}{4}}) \nonumber  \\
& + (\sselement{\vartheta_{z}}{-1}{\bs{0}}' \, + \, \sselement{\vartheta_{z}}{2_{100}}{\textstyle 0,-\frac{1}{4},-\frac{1}{4}}' \, + \, \sselement{\vartheta_{z}}{m_{001}}{\textstyle \frac{1}{4},\frac{1}{4},0}'  \, + \, \sselement{\vartheta_{z}}{2_{010}}{\textstyle -\frac{1}{4},0,-\frac{1}{4}}') \nonumber  \\
= &~~~~  (C^{\parallel}_{\infty}(\bs{s}) \times G_{\uparrow \uparrow}(\rv)) \times (E + \Ps\Ts).
\end{align}
\end{widetext}
Here the little group is a simple direct product of spin and space.
The final coreps are therefore easier to build, exploiting tabulated projective irreps of group $G_{\uparrow \uparrow}(\rv) = Fdd2$ which are the 2D $H_1(2)$. Adding the direct product with $C_{\infty}(\bs{s})$, the irreps $H_1(2)$ will self-conjugate under these new elements, producing two new irreps $H^{\pm}_1(2)$ (since again the transverse spin component transform as $m=\pm 1$ irreps of $C_{\infty}$).
Finally we can consider the anti-unitary $\Ps\Ts$ symmetry, which produces no extra degeneracy on the space part of the direct product (indeed the irreps of $Fdd2 \times (E + \Ps\Ts) = Fddd'$ are still 2D), while pairing up the irreps $m=\pm 1$ of the spin part, therefore giving a 4-fold degenerate line $DH_1(4) = (H^{+}_1,H^{-}_1)(4)$ as well.

\begin{table*}[!htb]
\centering
\hbox to \linewidth{ \hss
\begin{tabular}{ |c|c|c|c|c|c|c|  }
 \hline
 $\mathcal{G}^{B}_{\rm SS}$~ & ~$\sselement{E}{E}{\bs{0}}$~ & ~$\sselement{\vartheta_z^{\pm}}{E}{\bs{0}}$~ & ~$...$~ & ~$ \infty ~ \sselement{2_{\perp}}{2_{010}}{-1/4,0,-1/4}$
 ~ & ~$ \sselement{E}{m_{100}}{0,1/4,1/4}$ ~ & ~Type Coreps~ \\ 
 \hline
  $B_{1u}^{a}$   & 2   & 2 $\cos{\vartheta}$ & ~$...$~   & 0  & 2\,$\xi$ & (c)  \\
 \hline
 $B_{1u}^{b}$   & 2   & 2 $\cos{\vartheta}$  & ~$...$~   & 0  & -2\,$\xi$ & (c)  \\
 \hline
\end{tabular}
\hss}
\caption{
Character table giving the relevant irreps of the unitary part of $\mathbf{G}^{B}_{\rm SS}$. The phase factor is $\xi = \exp(i \, B \cdot (\textstyle 0 \frac{1}{4} \frac{1}{4}) ) = \exp(i \frac{\pi}{2} (v - 1))$ and the dot $(...)$ indicates here the infinite class of axial rotations.
}
\label{tab:ProjReptableBHeisen}
\end{table*}
%

\section{Symmetry Analysis for $[111]$ $JK$ Ferromagnet}
\label{app:111FM}

To see this, we exploit the fact that the enhanced magnetic symmetry group is isomorphic to an ordinary magnetic space group $-$ which need not be the case as we shall see later $-$ and use the tabulated character tables at high symmetry points and lines to determine the enforced degeneracies. The enhanced magnetic group isomorphism for $[111]$ is $Fd'd'd$:
\begin{align}
\mathbf{G}_{M} = \, E + \Ps + d_1 + C_2^{\bs{b}} + \Ts \, (d_2 + d_3+ C_2^{\bs{a}} +  C_2^{\bs{c}})
\end{align}

{\rm $\Gamma$ $= (0, \, 0, \, 0)$} $-$ The little co-group elements of the unitary subgroup of $\mathbf{G}_{M}$ at $\Gamma$ has all the elements $E,d_1, C_2^{\bs{b}}$ and $\Ps$. We may construct the matrix representation of these elements for the magnons and find the characters. These are computed to be
\begin{align}
& \chi_{\Gamma}^{M}(d_1)=0~~\chi_{\Gamma}^{M}(C_2^{\bs{b}})=0~~\chi_{\Gamma}^{M}(\Ps)=0 \nonumber \\
& \implies \Gamma^M = 2\left( \Gamma_1^+  + \Gamma_1^-  +\Gamma_2^+  +\Gamma_2^-   \right)
\label{eq:GammaDecomp}
\end{align}
where the factor of two comes about because the diagonalization gives modes at $\bs{k}$ and $-\bs{k}$ which are identical here because there is inversion symmetry.

We now include the effect of the anti-unitary elements by finding
\beq
\sum_{\alpha'} \chi_{\Gamma}^{p}\left( \element{R'_\alpha}{\bs{t}'_\alpha}^2  \right) = 4
\eeq 
for each irrep in Eq.~\ref{eq:GammaDecomp} which, according to the criterion Eq.~\ref{eq:AUTest}, gives only class (a) coreps. We therefore find four distinct bands at $\Gamma$.  In order of increasing energy these are $\Gamma_1^+, \Gamma_2^+, \Gamma_1^-,\Gamma_2^- $. 


{\rm $Y$ $= (0, \, -1, \, 0)$} $-$ Similarly to $\Gamma$ the little co-group has elements $E,d_1, C_2^{\bs{b}}$ and $\Ps$, the characters of the non-identity elements are all zero and the irreps are 
\beq
 Y^M = 2\left( Y_1^+  + Y_1^-  +Y_2^+  + Y_2^-   \right)
\eeq
only now $\sum_{'} \chi_{Y}^{p}\left( \element{R'_\alpha}{\bs{t}'_\alpha}^2  \right) = 0$ so the coreps are of type (c). The coreps therefore bind unitary irreps into pairs and the pattern of pairing is $DY_1 = (Y_1^+,Y_1^-)$ and $DY_2 = (Y_2^+,Y_2^-)$ so there are two doubly degenerate bands at $Y$.

It follows from the results at $\Gamma$ and $Y$, the ordering of the $\Gamma$ irreps in energy and compatibility relations that a crossing of two bands between $\Gamma$ and $Y$ occurs in this model. There are two such Weyl points to be consistent with fermion doubling on opposite sides of $\Gamma$ and, since the ordering of the $\Gamma$ point energies is crucial to the existence of the point, it is evidently an accidental crossing. This calculation therefore exposes the symmetry origin of the Weyl point between bands $2$ and $3$ shown in Fig.~\ref{fig:magnons111}.

{\rm $T$ $= (0, \, -1, \, -1)$} $-$ Like $Y$ and $\Gamma$, the little co-group at $T$ contains all the elements $E,d_1, C_2^{\bs{b}},\Ps$ and the characters of the nontrivial elements are zero giving $T^M = 2(T_1 + T_1)$ where $T_1$ is a 2D irrep. Inclusion of anti-unitary elements reveals that the coreps belong to class (a) so there are two two-fold degenerate bands.

{\rm $Z$ $= (0, \, 0, \, -1)$} $-$ From elements $E,d_1, C_2^{\bs{b}},\Ps$ at $(0,0,-1)$ we obtain two copies of the 2D irrep $Z_1$ and class (a) coreps so there are two two-fold degenerate magnon modes at this point.

{\rm $L$ $= (\textstyle \frac{1}{2}, \, \frac{1}{2}, \, \frac{1}{2})$} $-$ This point has only $E$ and $\Ps$ elements, zero character for the inversion symmetry and type (a) coreps leading to four distinct modes.

{\rm $B$ $= (0, \, u, \, -1)$} $-$ This is the line $[Z T]$ with unitary symmetry elements $E$ and $C_2^{\bs{b}}$. The character for $C_2^{\bs{b}}$ is zero and therefore $B^M = 2(B_1 + B_1 + B_2 + B_2)$. The coreps are of type (a) leading to four 1D modes.

{\rm $H$ $= (0, \, -1, \, u)$} $-$ We focus on the line $[Y T]$. The unitary symmetry elements along the line are $E$ and $d_1$ and the character for $d_1$ is zero. Then $H^M = 2(H_1 + H_1 + H_2 + H_2)$ and the coreps are of type (c) meaning that there is a two-fold degeneracy of the magnon modes where each mode has bound $(H_1,H_2)$ irreps.

{\rm $A$ $= (u, \, 0, \, -1)$} $-$ We focus now on the line $[Z A]$ (and equivalently on $[Y A_1]$). As with the $[Y T]$ line, the unitary elements along these lines are $E$ and $d_1$, there are four 1D irreps that are bound into two copies of $(A_1, A_2)$ so the lines both have two two-fold degenerate bands. 

{\rm $\Sigma$ $= (u, \, 0, \, 0)$} $-$ This is the line $[\Gamma X]$ (and equivalently $[X_1 T]$), which is different to $[Z A]$ and $[Y T]$ although the unitary element are identical leading to four 1D irreps. This is because the anti-unitary elements lead to corep criterion $\chi(E)+\exp(i \bs{k}\cdot(0,1/2,1/2))=2$  meaning that the coreps are type (a) and the modes are therefore singly degenerate. 
Similar argument hold for line $\Delta = [\Gamma Y] = (0, \, u, \, 0)$ and $\Lambda = [\Gamma Z] = (0, \, 0, \, u)$.

We have accounted for all the degeneracies in the magnon spectrum from the enhanced symmetry group. If we had instead taken the symmetry group of the underlying magnetic structure neglecting the pure spin space transformations available to the Kitaev-Heisenberg coupled spins then the representation theory would have predicted a degeneracy along $[Z A]$ and {\it nowhere else}. 

\section{Informal Arguments for Band Touching}
\label{app:111FMInformal}

The analysis given in the main text based on representation theory supplied the underlying symmetry constraints on the observed robust degeneracies in the magnon spectra. It is occasionally possible and certainly more illuminating to give a direct correspondence between a magnetic symmetry and any degeneracy. For example, we have encountered time reversal glide symmetries, $\Ts d_i$, in various cases. The action of the glide is given in Eqs.~\ref{eq:d1xyz}, \ref{eq:d2xyz} and \ref{eq:d3xyz}. For concreteness we consider, $\Ts d_2$ acting on a magnon state at a given momentum $\bs{k}=k_1\bs{b}_1+k_2\bs{b}_2+k_3\bs{b}_3$
\begin{widetext}
\beq
\label{eq:d1xyz2}
  \Ts d_2 : \ket{k_1, k_2, k_3, l} = \begin{cases}
                     			\ket{k_2-k_3, k_2, k_2 - k_1, 3}^*\,,            & l=1\\
                     			\ket{k_2-k_3, k_2, k_2- k_1, 4}^*\,,            & l=2\\
                     		e^{2\pi i k_2}	\ket{k_2-k_3, k_2, k_2 - k_1, 1}^*\,,  & l=3 \\
                     			e^{2\pi i k_2} \ket{k_2-k_3, k_2, k_2 - k_1, 2}^*\,,  & l=4\,.
            		      \end{cases}
\eeq
\end{widetext}
From this $(\Ts d_2)^2 = \exp (\pm 2\pi i k_2) = -1$ when $k_2=\pm 1/2$. This has the implication that at invariant momenta satisfying $k_2 = \pm1/2$ and $k_1 - k_3 = \pm 1/2$ $-$ along a line at the Brillouin zone boundary $-$ Kramers theorem enforces a degeneracy. For the Kitaev-Heisenberg ferromagnet in $[111]$ field, the magnetic symmetry includes $\Ts d_2$ and $\Ts d_3$ and the presence of both relies on the existence of spin and space transformations. The above straightforward argument leads to the presence of crossing zone boundary nodal lines. Similar arguments were employed to understand magnon nodal lines in the canted zig-zag order in a magnetic field \cite{Choi2019}.

\bibliography{Magnon_Sym}

\end{document}